\documentclass[a4paper,fleqn,usenatbib]{mnras}

\usepackage[T1]{fontenc}
\usepackage{ae,aecompl}

\usepackage{graphicx}	
\usepackage{amsmath}	
\usepackage{amssymb}	

\usepackage{mathptmx}
\usepackage{txfonts}

\defcitealias{GH2005}{GH2005}

\title[A $\Gamma$- \& $Z$-dependent Wolf-Rayet mass-loss]{Driving classical Wolf-Rayet winds: A $\Gamma$- and $Z$-dependent mass-loss}

\author[A. A. C. Sander et al.]{
Andreas A. C. Sander,$^{1}$\thanks{E-mail: Andreas.Sander@armagh.ac.uk}
J. S. Vink,$^{1}$
and W.-R. Hamann$^{2}$
\\
$^{1}$Armagh Observatory and Planetarium, College Hill, Armagh BT61 9DG, Northern Ireland, UK\\
$^{2}$Institut f{\"u}r Physik \& Astronomie, Universit{\"a}t Potsdam, Karl-Liebknecht-Str.\ 24/25, 14476 Potsdam, Germany
}

\date{Accepted 2019 October 28. Received 2019 October 28; in original form 2019 August 06}

\pubyear{2019}

\begin{document}
\label{firstpage}
\pagerange{\pageref{firstpage}--\pageref{lastpage}}
\maketitle

\begin{abstract}
Classical Wolf-Rayet (WR) stars are at a crucial evolutionary stage for constraining the fates of massive stars. 
The feedback of these hot, hydrogen-depleted stars dominates their surrounding by tremendous injections 
of ionizing radiation and kinetic energy. 
The strength of a WR wind decides the eventual mass of its remnant, likely a massive black hole.
However, despite their major influence and importance for gravitational wave detection statistics, WR winds are particularly 
poorly understood.
In this paper, we introduce the first set of hydrodynamically consistent stellar atmosphere models
for classical WR stars of both the carbon (C) and nitrogen (N) sequence, i.e. WC and WN stars,
as a function of stellar luminosity-to-mass ratio (or Eddington Gamma), and metallicity.
We demonstrate the inapplicability of the CAK wind theory for classical WR stars and confirm earlier 
findings that their winds are launched at the (hot) iron (Fe) opacity peak. 
For $\log Z/Z_\odot > -2$, Fe is also the main accelerator throughout the wind.
Contrasting previous claims of a sharp lower mass-loss limit for WR stars, we obtain a smooth transition to optically thin winds.
Furthermore, we find a strong dependence of the mass-loss rates on Eddington $\Gamma$,
both at solar and sub-solar metallicity. Increases in WC carbon and oxygen abundances turn out to 
slightly reduce the predicted mass-loss rates.
Calculations at subsolar metallicities indicate that below the metallicity of the SMC, WR mass-loss rates 
decrease much faster than previously assumed, potentially allowing for high black hole masses even in the local universe.
\end{abstract}




\section{Introduction}
  \label{sec:intro}

Among the highly influential massive stars, Wolf-Rayet (WR) stars are a rare
but particularly important class. Their outstanding spectral appearance with huge emission
lines dominating their optical spectrum allowed not only their discovery back in the 19th 
century \citep{WR1867}, but also led to the idea that these stars have strong stellar winds \citep{Beals1929},
long before this was discovered to be inherent to the overwhelming majority of hot and massive stars.

By definition, WR stars are a spectroscopic category of stars with strong emission lines. 
As this can include very different types of objects, the
term ``classical Wolf-Rayet'' (cWR) stars has been introduced in the more recent years, describing the
hydrogen-depleted WR stars, associated with the end-points of the evolution of the 
most massive stars \citep[e.g.][]{Crowther2007}. Even today it is not yet known whether  
cWR stars produce supernovae (SNe) Ibc \citep[e.g.][]{Yoon+2012,Groh+2013}, or if they collapse directly into a black hole (BH), i.e. without a visible display.
In the latter case, SNe Ibc might generally be associated with lower mass
binary `stripped' helium (He) stars \citep[e.g.][]{Eldridge+2013} which may or may not have
a WR-type spectral appearance depending on their mass-loss rate $\dot{M}$. 
Even in this case, stellar wind mass loss during the He-burning phase is the key to removing 
the remaining hydrogen (H), in order to produce H-free Ibc SNe \citep[e.g.][]{Gilkis+2019}. 

The final phases of massive star evolution over a wide mass range -- for both single and binary  
evolution -- heavily rely on the assumed He star mass-loss rates. The need for a proper understanding of He star mass loss
has become even more prevalent after the discovery of gravitational 
waves (GWs) from merging BHs, such as GW 150914. \citet{Abbott+2016}
showed that the evolutionary scenarios towards collapse depend heavily on the 
assumed WR mass loss, and only if a metallicity ($Z$) dependence \citep{VdK2005} is accounted for, the `heavy' nature of GW 150914 becomes possible 
\citep[][]{Belczynski+2010,Abbott+2016}. 

The $Z$-dependent mass-loss rates of \citet{VdK2005} which have for example been used to
constrain the history of the massive BHs, were only computed for two `prototypical' late-type
WR stars with relatively low effective temperatures (on the order of 50\,000 K). So far, a general theoretical 
prescription for WR stars does not exist.
Therefore, most stellar evolutionary models \citep[e.g.][]{Georgy+2012,Chen+2015,Grassitelli+2016,Renzo+2017,LC2018} still rely 
on empirical $\dot{M}$ recipes, such as \citet{NL2000}. Whilst these recipes 
might provide meaningful constraints inside the regime for which they were derived, they 
may fail substantially outside \citep{Vink2017}. 

Our aim is thus to alleviate current shortcomings by computing the up to date most-sophisticated 
hydro-dynamically consistent cWR wind models. Until now, only one
such model has been constructed for a WC5 star by \citet[][hereafter GH2005]{GH2005}, highlighting the 
challenge in this task. In recent years, \citet{Sander+2015,Sander+2017,Sander+2018} developed a new 
hydrodynamical version of the PoWR stellar atmosphere 
code \citep{HK1998,GKH2002} which is ideally suited for this task. 
One of the key advantages of this method is that it not only includes multi-line scatterings
\citep[as do Monte Carlo methods, e.g.][]{AL1985,Vink+1999,Noebauer+2015} but 
it also performs the radiative transfer in the co-moving frame (CMF), without any reliance 
on the Sobolev approximation.

Traditionally, WR winds are believed to be radiatively driven, making them similar to OB star winds. Still, the concepts successfully applied in the regime of OB supergiants, especially the so-called CAK theory \citep*[named after][]{Castor+1975} and its later extensions and modifications (`mCAK') \citep[e.g.][]{FA1986,Pauldrach+1986,KP2000} turned out to be insufficient in describing the more dense WR winds \citep[e.g.][]{LL1993}, presumably due to the inherent restriction to single scattering. While the fact that the luminosities and terminal wind velocities of OB and WR stars are quite similar was one of the arguments for radiative driving in the first place, their empirical $\dot{M}$ measured from radio and/or deduced from stellar atmosphere models are about an order of magnitude higher \citep[e.g.][]{SHW1989,NL2000,PVN2008}. Over the years, insight has grown thanks to the understanding of WR ionization stratification \citep{LA1993}, 
photon trapping \citep{Owocki1994}, and multi-line scattering \citep[e.g.][]{Gayley+1995}.

By studying the conditions at the sonic point, \citet{NL2002} concluded that WR winds with their high $\dot{M}$ could potentially be launched in the optically thick regime due to an iron opacity bump. The idea of launching WR winds by starting the acceleration in the deep layers well below of what an observer would describe as the photosphere goes back to \citet{KI1992}. While their model had a continuum-driven wind with an artificially enhanced opacity, the release of the OPAL opacities then let \citet{PE1995} to follow up on this concept without the principal need of an artificial opacity enhancement.
\citetalias{GH2005} for the first time developed a model with a consistent flux-weighted mean opacity including the \ion{Fe}{ix-xvi} ions from Opacity Project data \citep[e.g.][]{TOPBASE2001}. They demonstrated, that the launch of the wind in the deep layers implies a smaller radius than commonly assumed when analysing WR stars with models using a prescribed $\beta$-law \citep[cf. the results for WR\,111 in][]{Sander+2012,Sander+2019}. As a consequence, the population of \ion{Fe}{ix-xvi}, the so-called M-shell ions, is significantly enhanced, providing sufficient acceleration for launching a WR wind. The corresponding opacity is essentially forming the `(hot) iron bump' in the OPAL tables by \citet{OPAL1996}. 

The empirical constraint of WR radii is cumbersome as WR spectra tend to form solely in the wind. This has led to discrepancies between empirical results and hydrostatic radii predicted by stellar evolution models. WR radii are an important important constrain in massive star evolution e.g.\ to determine the progenitor systems of double black hole binaries \citep[e.g.][]{Marchant+2016}. In this work, we investigate compact WR atmosphere models, motivated by the result from \citetalias{GH2005} hinting that the larger radii in empirical models might just be a consequence of their simplified (i.e.\ $\beta$-law) treatment of the density and velocity stratification. While hydrostatic stellar structure calculations \citep[e.g.][]{GOV2012,McE2016} yield so-called `inflated' hydrostatic radii for WR stars, recent hydrodynamic structure calculations by \citet{Grassitelli+2018} support and generalize the picture of He stars with a compact subsonic structure and an extended pseudo-photosphere at large radii. Such models demonstrate that optically thick winds compatible with the empirical $\dot{M}$ for WNE stars can be launched by the hot iron bump, consistent with the results from the hydrodynamic atmosphere model for WR\,111 by \citetalias{GH2005}.

In the last decade it has further become clear that the most massive stars in the Universe are not O-type stars, but WR stars with H, termed WNh stars \citep{Hamann+2006,Martins+2008,Liermann+2010} with up to 200-300 solar masses \citep{Crowther+2010}. These very massive stars \citep[VMS;][]{Vink+2015} are predicted to have strong, $\Gamma_\text{e}$-dependent $\dot{M}$ \citep[][]{Vink2006,GH2008}, which is confirmed  by empirical results \citep{Graefener+2011,Bestenlehner+2014}. $\Gamma_\text{e}$ denoted the ratio of acceleration due to Thomson scattering and gravity, i.e. $\Gamma_\text{e} \propto q_\text{ion} \cdot L/M$ \citep[cf.~e.g.~Eq.\,22 in][]{Sander+2015}. While for canonical O-type stars \citet{Vink+2011} predicted a relatively shallow $\Gamma_\text{e}$-dependence in relatively good agreement with CAK, for optically denser winds where the wind efficiency parameter
\begin{equation}
  \label{eq:etadef}
	\eta := \frac{\dot{M}\varv_\infty}{L/c}
\end{equation} 
crosses unity \citep[][]{VG2012}, $\dot{M}$ shows a much steeper $\Gamma_\text{e}$-dependence, resulting in a `kink' in the derived $\dot{M}(\Gamma_\text{e})$-relation, seemingly in disagreement with the CAK theory. Given that VMS/WNh winds are only marginally optically thick, while cWR winds are usually very optically thick, we will explore a regime beyond the applicability of CAK, where a new generation of hydro-dynamically consistent atmospheres becomes a necessity for a profound understanding of WR and He star winds in general.

The paper is organized as follows: after we introduce the stellar atmosphere models in Sect.\,\ref{sec:powr}, the following Sect.\,\ref{sec:arad} comprises a detailed discussion of the radiative acceleration in WR atmospheres. In Sect.\,\ref{sec:results}, we then discuss the sets of calculated models and their implication for wind driving and mass-loss of cWR stars. The main insights from the paper are comprised in the concluding Sect.\,\ref{sec:conclusions}. In the appendix, we furthermore demonstrate the breakdown of a CAK force multiplier approach for cWR stars (Sect.\,\ref{asec:cakfail}) due the complex, non-monotonic opacity structure, and discuss the problem of finding an analytic description for the radiative acceleration (Sect.\,\ref{asec:aradfit}).

\section{Stellar atmosphere models}
  \label{sec:powr}
	
For our study we use the PoWR model atmopshere code \citep[e.g.][]{GKH2002,HG2003,Sander+2015} with the option to solve to hydrodynamically consistent stratifications introduced in \citet{Sander+2017}. The models assume a spherically symmetric star with an expanding, but stationary outflow.	To properly account for the non-LTE conditions of hot star atmospheres, the population numbers for all levels (cf.~Table\,\ref{tab:datom}) are calculated by solving the high-dimensional system of statistical equilibrium equations. The radiative transfer is treated in the comoving frame (CMF) to avoid any major simplifications \citep[cf.~e.g.][]{Mihalas+1975}. Together with the calculation of the temperature stratification, the solution of the statistical equations and the radiative transfer are performed iteratively until a consistent model for the full atmosphere is obtained. Hydrodynamic stratification updates, meaning new calculations of the velocity and density structure from the hydrodynamic equation of motion, are performed and applied every time the corrections to the populations numbers have dropped below a certain limit and the flux is reasonably well conserved (usually on the order of a few percent). Once the overall iteration cycle including the stratification updates is converged, the emergent spectrum of the star is calculated from the formal integral in the observer's frame. A more detailed description of the concepts of these `next-generation' PoWR models is given in \citet{Sander+2017}.

The inner boundary of our models, where we assume the diffusion approximation to be valid, is defined by the stellar radius $R_\ast$, which we define at a Rosseland continuum optical depth of $\tau_\text{Ross,cont} = 20$. This quantity is conserved during all stratification updates, thus ensuring our input parameters $T_\ast$ and $L$, connected by the Stefan-Boltzmann law
	\begin{equation}
	  \label{eq:lrt}
	  L = 4 \pi R_\ast^2 \sigma_\text{SB} T_\ast^4\text{,}
	\end{equation}
remain consistent. $L$ denotes the luminosity of the star, while $T_\ast$ is the effective temperature corresponding to $R_\ast$, meaning $T_\ast := T_\text{eff}(R_\ast) = T_\text{eff}(\tau_\text{Ross,cont} = 20)$. As this is deep in the optically thick regime, this effective temperature is not identical to the electron temperature $T_\text{e}$. In fact, $T_\text{e}(R_\star)$ is typically on the order of $\approx\,2\dots2.5\,T_\ast$ for classical Wolf-Rayet stars. The considerable scatter is a consequence of $T_\text{e}(R_\star)$ depending on the total optical depth including lines $\tau_\text{Ross}$, i.e.
\begin{equation}
  T_\text{e}(R_\star) \approx T_\ast \sqrt[4]{\frac{3}{4}\tau_\text{Ross}(R_\ast)+\frac{1}{2}}\text{,}
\end{equation}
instead of just $\tau_\text{Ross,cont}$. Choosing a fixed $\tau_\text{Ross,cont}$ as a boundary value is numerically favourable, but as the ratio between $\tau_\text{Ross}(R_\ast)$ and $\tau_\text{Ross,cont}(R_\ast)$ varies for different model parameters, this comes at the cost of the described scatter in $T_\text{e}(R_\star)$.

Another common definition for the effective temperature of a star is to refer to the radius where the Rosseland optical depth is equal to $2/3$. We define this value as $T_{2/3} = T_\text{eff}(\tau_\text{Ross} = 2/3)$ in our work and will discuss the role of its corresponding radius $R_{2/3}$ later in Sect.\,\ref{sec:arad}. Further input parameters for each PoWR model are the stellar mass $M_\ast$, the mass-loss rate $\dot{M}$, the clumping stratification $D(r)$, and an initial approximation for the velocity $\varv(r)$. The density structure is implied via
the equation of continuity
\begin{equation}
  \label{eq:conteq}
	 \dot{M} = 4 \pi r^2\,\rho(r)\,\varv(r)\text{.}
\end{equation}
In all models presented in this work, $\varv(r)$ is iteratively adjusted, while the mass-loss rate $\dot{M}$ can either be updated as well or kept fixed. In the latter case, another fundamental stellar parameter of course has to be changed instead. For the stratification update in our code we allow to either
\begin{itemize}
  \item keeping $L$ and $M_\star$ fixed to determine the consistent $\dot{M}$,
  \item fixing $L$ and $\dot{M}$ to deduce the consistent $M_\star$, or
  \item keeping instead $M_\star$ and $\dot{M}$ fixed to obtain the consistent $L$.
\end{itemize}
The general numerical implementation is very similar that described in \citet{Sander+2017}, but the flexibility to switch between these different methods allows us to use in each case the correction mechanism that is of interest for our particular study and numerically most efficient. As the CMF radiative transfer is usually the bottleneck in terms of calculation time, choosing the method that leads to corrections least affecting the radiative transfer is a significant help to keep the computational effort manageable. It is important to keep in mind that this choice of what is a fixed input and what is updated is purely for computational reasons. Test calculations were performed to ensure that the different methods do not lead to conflicting results. Hence, the finally obtained sets of $L$, $M$, and $\dot{M}$ are consistent between the different methods and can always be read as the mass-loss rate $\dot{M}$ following from a set of given stellar parameters $L$ and $M$, regardless of the actual computational method.

\begin{figure}
  \includegraphics[width=\columnwidth]{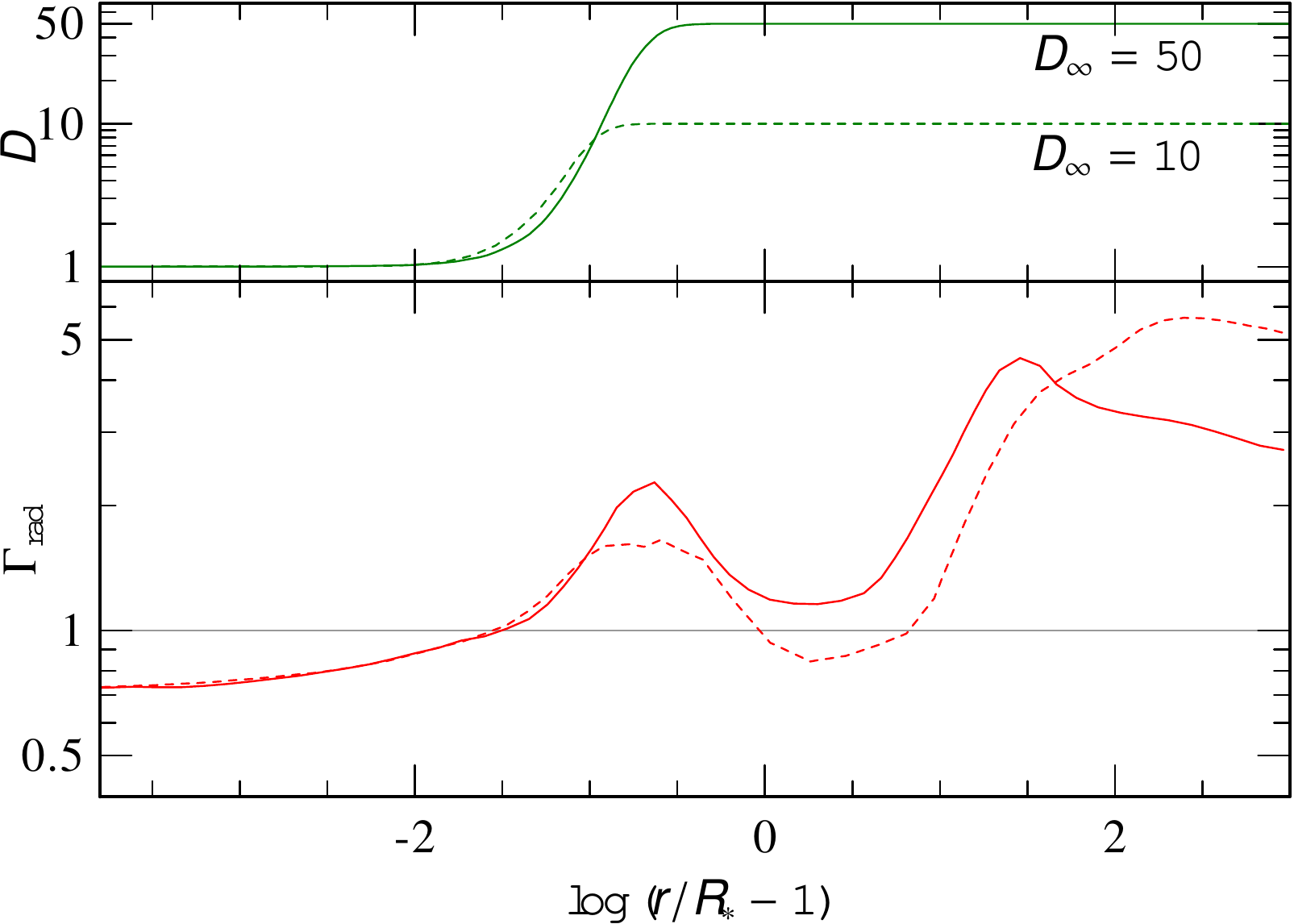}
  \caption{Comparison of the radiative acceleration normalized to gravity ($\Gamma_\text{rad} = a_\text{rad}/g$, lower panel) and the clumping stratification (upper panel) for WN models with $D_\infty = 10$ (dashed lines) and $D_\infty = 50$ (solid lines).}
  \label{fig:cmp-arad-clumping}
\end{figure}

We choose a depth-dependent clumping stratification with a maximum clumping factor of $D_\infty = 50$. With a void interclump medium, this corresponds to a volume filling factor $f_\text{V} = D^{-1} = 0.02$. While this value is in the range of what has been used for several OB star analyses \citep[e.g.][]{Bouret+2012,Mahy+2015}, we are aware that it is considered to be more towards the drastic end of what is actually considered to be reasonable, while $D \approx 10$ or $f_\text{V} = 0.1$ is more commonly applied and usually more consistent with what is inferred from electron scattering wings. However, the two main reasons to keep this value are to be consistent with the only previously existing consistent early-type WR model by \citetalias{GH2005} and to particularly avoid $\Gamma_\text{rad} < 1$ in the wind which would lead to a non-monotonic velocity profile, which then would need to be modified in order to work within our CMF radiative transfer framework. Since we are using a depth-dependent approach for $D(r)$, the actual effect on the derived $\dot{M}$ for a given set of $T_\ast$, $L$, and $M_\ast$ is really small ($\approx 0.03\,$dex) between $D_\infty = 10$ and $D_\infty = 50$ as long as the same recipe for $D(r)$ is used. The reason for this is illustrated in Fig.\,\ref{fig:cmp-arad-clumping}, where one can see that the radiative acceleration equals gravity (i.e. $\Gamma_\text{rad} = 1)$ already in a region where $D(r)$ has not yet reached its maximum value (and $\Gamma_\text{rad}$ also drops back below $1$ for the $D_\infty = 10$-model). Instead the -- empirically highly uncertain -- onset of clumping can play an important role. However, since this would add at least one more dimension, we limit this work to models using the depth-dependent descriptions via $\tau$ and $\varv$ explained in \citet{Sander+2017} with $\tau_\text{cl} = 10$ or $\varv_\text{cl} = 100\,\mathrm{km}\,\mathrm{s}^{-1}$ and point out the need for a future, more in-depth study of the imprint of the clumping stratification on the driving of WR winds.
Our choice of $D_\infty = 50$ affects of course the derived terminal velocities $\varv_\infty$. Test calculations in regimes where solutions for both $D_\infty = 50$ and $D_\infty = 10$ could be obtained, e.g.\ for very dense winds with $\log \dot{M} \approx -4.4$, reveal that $\varv_\infty$ is about $30\%$ larger for $D_\infty = 50$ compared to a similar calculation with $D_\infty = 10$.
 
As introduced in \citet{Sander+2015} and discussed in \citet{Sander+2017}, we can write the hydrodynamic equation of motion as
  \begin{align}
	  \label{eq:stdhydro}
     \varv \left( 1 - \frac{a^2}{\varv^2} \right) \frac{\mathrm{d} \varv}{\mathrm{d} r} & = a_\text{rad} - g + 2 \frac{a^2}{r} - \frac{\mathrm{d} a^2}{\mathrm{d} r} \\
	  \label{eq:stdhydroG}
		                                                        & = \frac{GM}{r^2} \left(\Gamma_\text{rad} - 1\right) + 2 \frac{a^2}{r} - \frac{\mathrm{d} a^2}{\mathrm{d} r}
  \end{align}
with $a_\text{rad}$ denoting the total radiative acceleration which we will discuss in Sect.\,\ref{sec:arad} in more detail. The quantity $a(r)$ denotes the isothermal sound speed $a_\text{s}$ corrected for a micro-turbulence term, i.e.
  \begin{equation}
    a^2(r) := \frac{k_\text{B} T(r)}{\mu(r)\,m_\text{H}} + \frac{1}{2} \varv_\text{mic}^2 \equiv a_\text{s}^2 + \frac{1}{2} \varv_\text{mic}^2
  \end{equation}
with $\varv_\text{mic}$ assumed to have a constant value of $30\,\mathrm{km\,s^{-1}}$ in this work. While such a constant micro-turbulence likely does not reflect the true situation in a WR atmosphere, the potential parameters which can be varied in a model atmosphere are manifold. Thus, we decided to focus on the more fundamental parameters in this work, leaving the finer contributions such as $\varv_\text{mic}$ or a detailed discussion of the clumping stratification $D(r)$ to be discussed in a future work.

The hydrodynamic equation of motion in the form of Eq.\,(\ref{eq:stdhydro}) or Eq.\,(\ref{eq:stdhydroG}) has a critical point at $R_\text{crit} := R\left(\varv = a\right)$, that is the radius where the wind speed $\varv$ is equal to the turbulence-adjusted sound speed. Due to the parametrization of $a_\text{rad}$ as a function of radius, this critical point is not identical to the CAK critical point, which would be located further out. Since we include a micro-turbulence term in $a(r)$, we refrain from calling it the `sonic point' to avoid any confusion with the radius where $\varv = a_\text{s}$, which is slightly further in due to $\varv_\text{mic} > 0$. A more detailed study of the influence of $\varv_\text{mic}$ is beyond the scope of the present paper, but one has to be aware that the origin of turbulence is commonly attributed to sub-surface convention zones \citep[e.g.][]{Cantiello+2009,Grassitelli+2015,Jiang+2015}, which would likely be incompatible with a wind launching mechanism at the hot iron bump \citep[e.g.][]{RM2016}. However, the value of $30\,\mathrm{km\,s^{-1}}$ chosen in this work has only a very limited influence on the location of the critical point. We illustrate this in Fig.\,\ref{fig:checkopaross}, highlighting both $R_\text{sonic}$ and $R_\text{crit}$ for a typical WNE model at $Z_\odot$ where both radii differ by only $0.002\,R_\ast$. Thus, even the complete absence of turbulence would lead to only a small change of our results derived in this work.

\section{The radiative acceleration}
  \label{sec:arad}

The strength, but also the complexity of our locally consistent hydrodynamical approach is rooted in the detailed treatment of the radiative transfer, which is performed in the CMF to keep the opacities and emissivities isotropic despite the expansion. From the CMF radiative transfer we also obtain the radiative acceleration which we can write in 1D as 
\begin{align}
  \label{eq:arad}
	a_\text{rad}(r) & = \frac{4\pi}{c} \frac{1}{\rho(r)} \int\limits_{0}^{\infty} \kappa_\nu(r) H_\nu(r)\,\mathrm{d}\nu \\
	\label{eq:aradvar}
	                & = \frac{4\pi}{c} \int\limits_{0}^{\infty} \varkappa_\nu(r) H_\nu(r)\,\mathrm{d}\nu\text{.}
\end{align}
By not assuming any explicit analytical expression for $a_\text{rad}$, we implicitly account for various affects such as line overlaps and multiple scattering.
The two flavors of Eq.\,(\ref{eq:arad}) depend on whether the opacity is written in the form of an absorption coefficient $\kappa_\nu$ that has the units of an inverse length (i.e. cm$^{-1}$), or as a mass attenuation coefficient $\varkappa_\nu$ in cm$^2$\,g$^{-1}$ with $\kappa_\nu(r) = \rho(r)\varkappa_\nu(r)$. As mentioned in \citet{Sander+2017}, this opacity includes all line and continuum opacities, i.e. bound-bound, bound-free, and free-free transitions as well as Thomson electron scattering. Especially the so-called `true continuum' of bound-free and free-free transitions, which is usually negligible for OB star winds, can become important for WR star winds at some depths and thus cannot be neglected a priori.

Using $a_\text{rad}$ in the form of Eq.\,(\ref{eq:aradvar}) allows to eliminate the explicit dependence of the density. Replacing the Eddington flux $H_\nu$ with $F_\nu = 4 \pi H_\nu$ and introducing the flux-weighted mean opacity
\begin{equation}
  \varkappa_{F}(r) = \frac{1}{F(r)} \int\limits_{0}^{\infty} \varkappa_\nu(r) F_\nu(r)\,\mathrm{d}\nu
\end{equation}
with $F = \int F_\nu\,\mathrm{d}\nu$ it is often written as
\begin{align}
  \label{eq:aradvk}
  a_\text{rad}(r) = \frac{1}{c} \varkappa_{F}(r) F(r) = \frac{\varkappa_{F}(r)L}{4 \pi r^2 c}\text{.}
\end{align}
In the last equation, we used $L(r) = 4 \pi r^2 F(r)$ to express the integrated flux $F$ by  the stellar luminosity $L$. For very dense winds, $L$ is not exactly constant as the energy removed for driving the wind could in theory become large enough to reduce the observed luminosity. However, even in our Wolf-Rayet calculations this is a minor effect. In this work, we yield $\Delta \log L/L_\odot < 0.05$ for the most extreme (i.e.\ high $\dot{M}$ and high $Z$) considered cases.

\begin{figure}
  \includegraphics[width=\columnwidth]{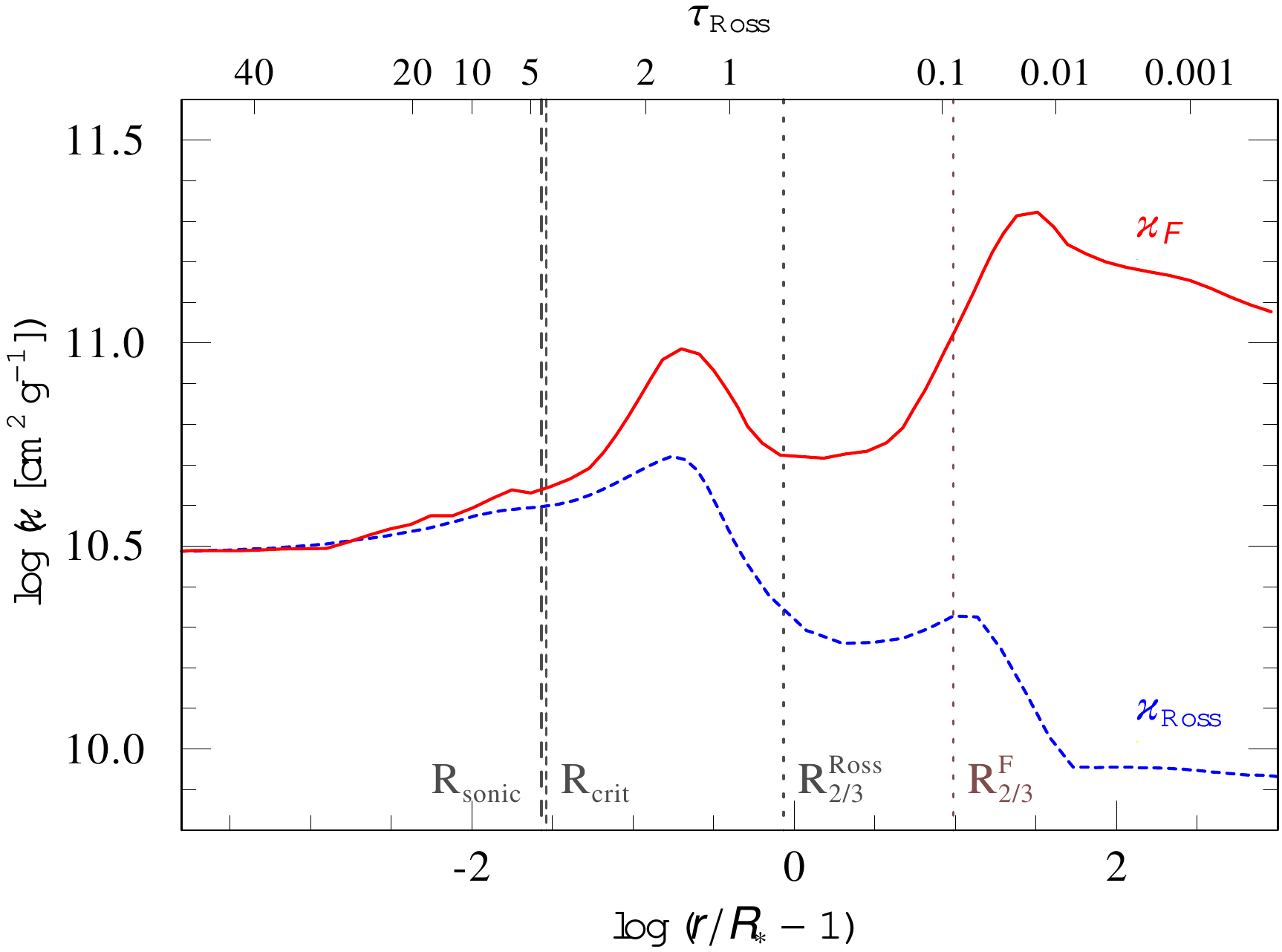}
  \caption{Comparison of the flux-weighted mean opacity (red, solid) to the Rosseland mean opacity (blue, dashed) in an atmosphere model for a hydrogen-free WN star (see Table\,\ref{tab:wcwnesolcmp} for parameters) at solar metallicity.}
  \label{fig:checkopaross}
\end{figure}

Instead of $a_\text{rad}$, it is often more favourable to discuss the ratio
\begin{equation}
  \label{eq:gammarad}
	\Gamma_\text{rad}(r) := \frac{a_\text{rad}(r)}{g(r)} = \frac{\varkappa_{F}(r)}{4\pi c G}\frac{L}{M}\text{.}
\end{equation}
Since the mass of the stellar atmosphere compared to the total mass is negligible and also $L$ is approximately constant, all radial dependency of $\Gamma_\text{rad}$ is due to the flux-weighted mean opacity $\varkappa_{F}$. The quantity $\varkappa_{F}$ should not be mixed up with the Rosseland opacity 
\begin{equation}
  \label{eq:opaross}
  \varkappa_\text{Ross}^{-1}(r) := \frac{\int_{0}^{\infty} \varkappa_\nu^{-1}(r) \frac{\partial B_\nu}{\partial T}\,\mathrm{d}\nu}{\int_{0}^{\infty} \frac{\partial B_\nu}{\partial T}\,\mathrm{d}\nu}\end{equation}
	that is for example used in stellar structure calculations or wind driving studies based on the grey OPAL opacity tables \citep[e.g.][]{NL2002,RM2016,Graefener+2017,Sanyal+2017,Grassitelli+2018,Ro2019}. In the deeper layers of the atmosphere where the diffusion approximation is valid, $\varkappa_{F} \approx \varkappa_\text{Ross}$ holds as illustrated in Fig.\,\ref{fig:checkopaross}, but the difference between the two quantities becomes considerable around and above the critical point with the two quantities being off by more than an order of magnitude in the outer wind. Thus, the construction of a consistent wind stratification cannot be achieved without the detailed calculation of $\varkappa_{F}$, as only this quantity gives a proper handle on the radiative acceleration and the driving of a stellar wind.

\begin{figure}
  \includegraphics[width=\columnwidth]{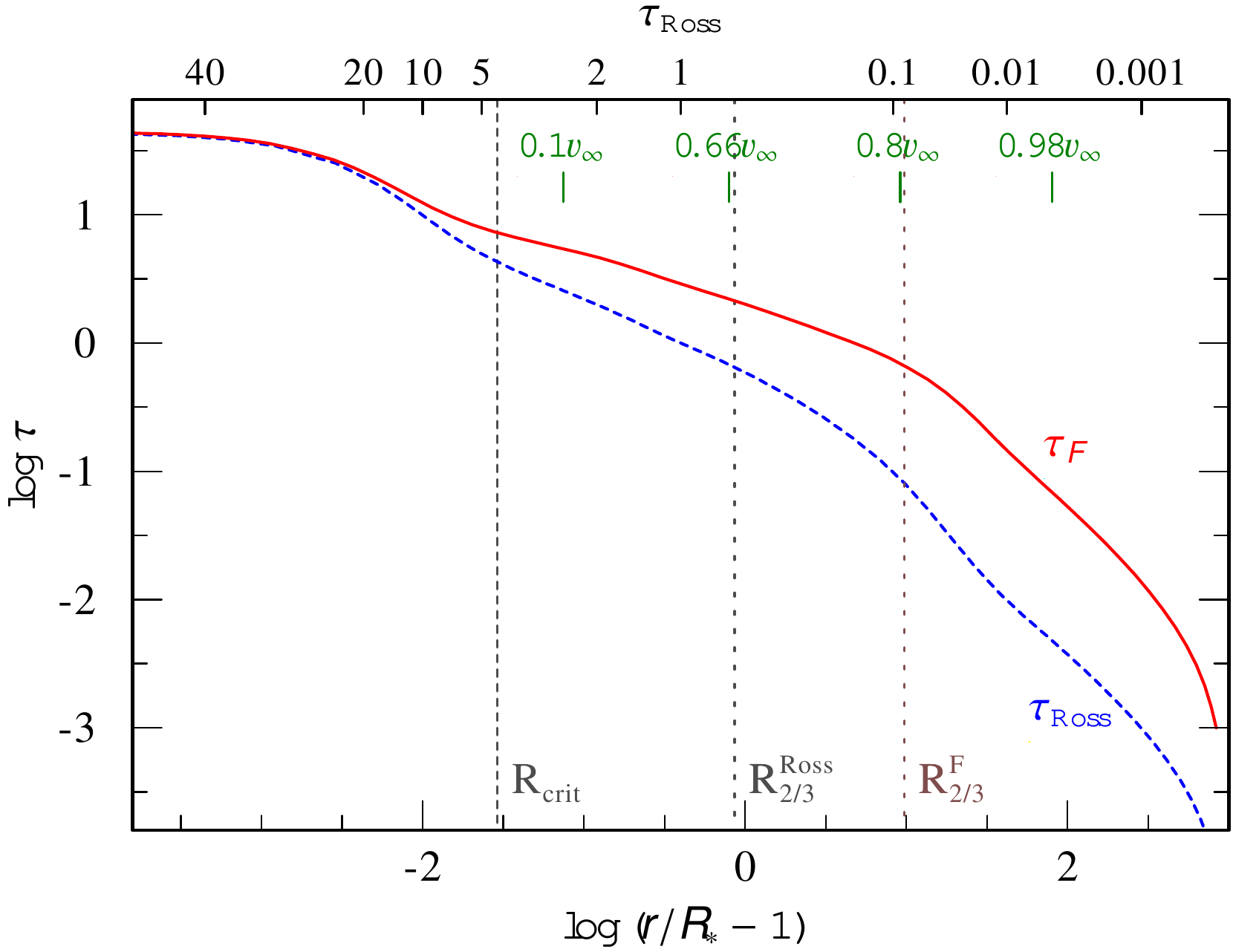}
  \caption{Comparison of the flux-weighted optical depth (red, solid) to the Rosseland optical depth (blue, dashed) in an atmosphere model for a hydrogen-free WN star (see Table\,\ref{tab:wcwnesolcmp} for parameters) at solar metallicity.}
  \label{fig:checktauscales}
\end{figure}

In analogy to the Rosseland optical depth, we can define the flux-weighted optical depth 
\begin{equation}
  \label{eq:tauf}
	\tau_{F}(r) := \int\limits^\infty_{r} \varkappa_{F}(r^\prime)\,\rho(r^\prime)\,\mathrm{d}r^\prime
\end{equation}
which can serve as an optical depth scale directly reflecting the wind driving situation. A comparison with the more common $\tau_\text{Ross}$ for a typical hydrogen-free WN model at $Z_\odot$ is depicted in Fig.\,\ref{fig:checktauscales}. While in the innermost region both scales approach the same total value, their difference is quite remarkable in the outer layers. In particular the region where the wind is optically thick ($\tau_{F} > 2/3$) is by an order of magnitude larger than what one would infer from the Rosseland scale. At this distance of $\approx 11\,R_\ast$, the wind has already reached more than $80\%$ of its terminal velocity. To distinguish the corresponding radius for the two different optical depth scales we denote them as $R_{2/3}^\text{Ross}$ and $R_{2/3}^{F}$ respectively. Wherever a quantity is labeled only $R_{2/3}$, it has the meaning of $R_{2/3}^\text{Ross}$.

\begin{figure}
 \includegraphics[width=\columnwidth]{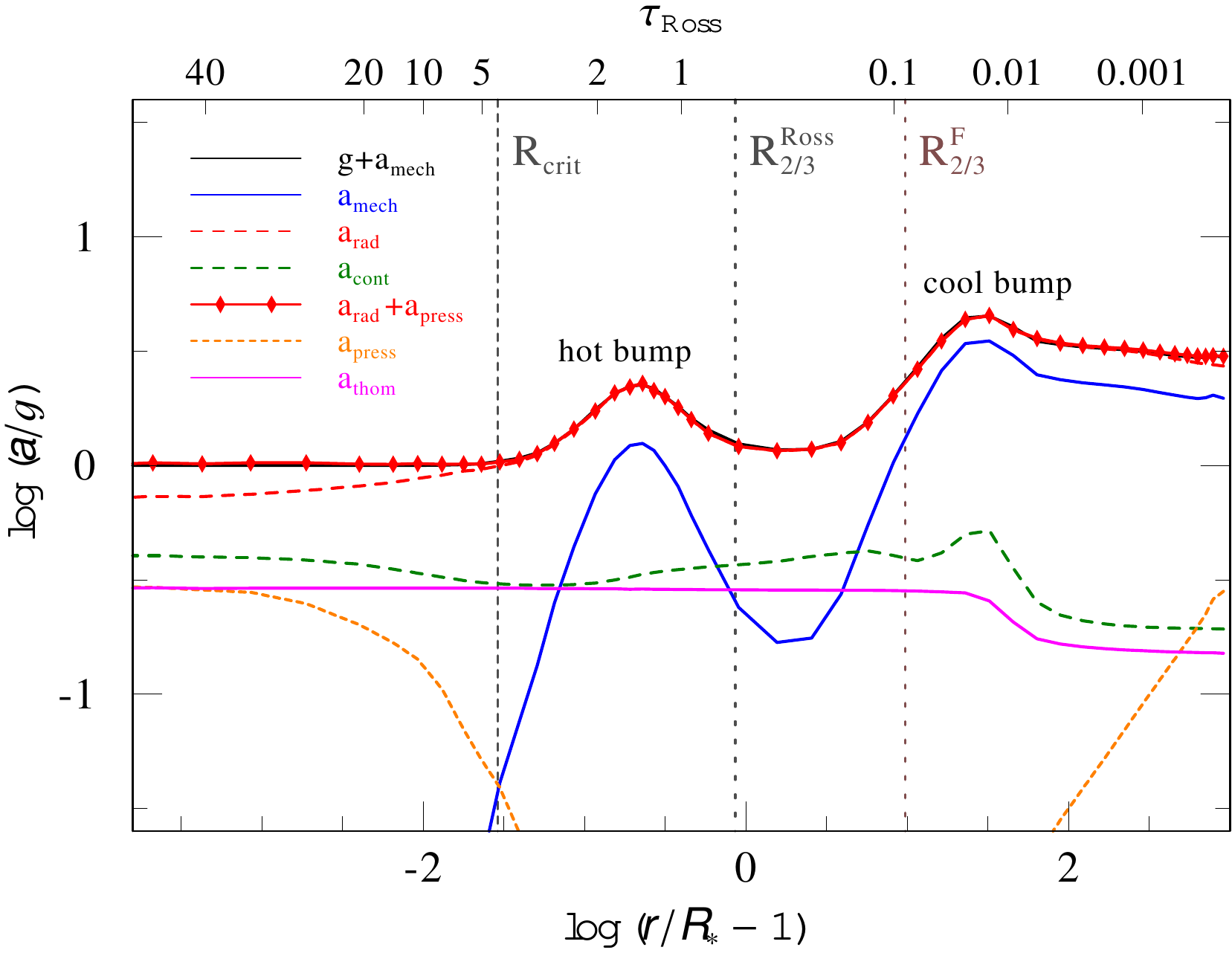}
  \caption{Acceleration stratification for a hydrodynamically consistent WNE model with $Z = Z_\odot$ (see Table\,\ref{tab:wcwnesolcmp} for parameters): the sum of gravity and inertia (black solid line) matches the total wind acceleration (red diamond line). The contributions to these two terms are indicated by different line styles and colors as indicated in the plot. The so-called hot and cool bump stemming from the line acceleration contribution are also labelled. The particular ions contributing ions to the bumps are shown in Fig.\,\ref{fig:leadions-wne}.}
  \label{fig:aradillu}
\end{figure}

For our hydrodynamically consistent models, we demand that the sum of radiation (following from $\varkappa_{F}$, cf.\ Eq.\,\ref{eq:aradvk}) and gas pressure (including turbulence) to be equal to the sum of gravity and inertia throughout the stellar atmosphere. This is illustrated in Fig.\,\ref{fig:aradillu} where show the acceleration balance for an example WNE model. The figure also shows that the contributions to $a_\text{rad}$ from free electron (Thomson) scattering and the total continuum which includes the opacities from bound-free and free-free transitions on top of the Thomson contribution.
Although numerically crucial to get the proper critical point, the actual contribution of the gas and turbulent pressure gradients is small and becomes important only closer to the boundaries: in the deeper subsonic layers, where the contributions from the line acceleration get smaller and smaller, it raises to about $30\%$. In the far outer wind, its relative contribution grows to the fact that there is a $1/r$-term in $a_\text{press}$ compared to the $1/r^2$-terms of the other forces which decline more rapidly. 

\begin{figure}
  \includegraphics[width=\columnwidth]{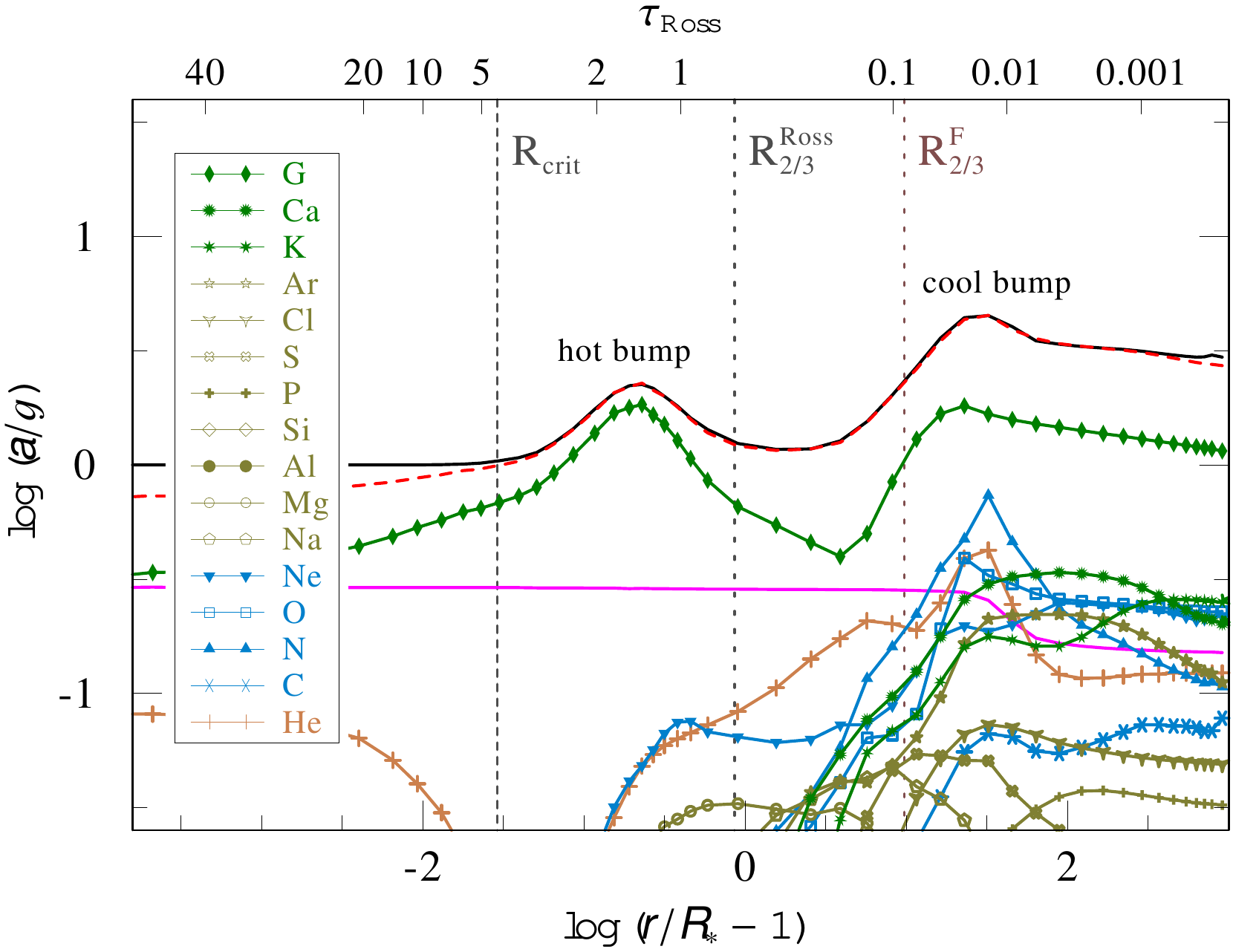}
  \caption{Contributions to the radiative acceleration by different elements for the same models as depicted in Fig.\,\ref{fig:aradillu}: The sum of gravity and inertia (black solid line) and the total radiative acceleration (red dashed line) are added for comparison, as well as the contribution due to free electron scattering (purple solid line). The element `G' stands for a generic element representing the whole iron group \citep[see][]{GKH2002}.}
  \label{fig:aradelemillu}
\end{figure}

From Fig.\,\ref{fig:aradillu}, we can infer that the main contribution to $a_\text{rad}$ is due to the line acceleration. The two `bumps', already present in the $\varkappa_{F}$-curve in Fig.\,\ref{fig:checkopaross}, are termed the  `hot' and `cool' bump and can be understood when breaking $a_\text{rad}$ down into the contributions of the different elements (Fig.\,\ref{fig:aradelemillu}). It is evident, that both bumps are mainly caused by iron group opacities (treated as the generic element $G$) although there are additional significant contributions to $a_\text{rad}$, especially in the outer wind. The `cool' bump is further increased by N, O, and He contributions, with the latter stemming from bound-free opacities. In models with dense winds like our WNE example here, \ion{He}{iii} is no longer the leading ionization stage, but instead most of the He recombines into \ion{He}{ii}. As a consequence of less available free electrons, the Thomson contribution decreases.

In addition to the iron group elements, several other elements contribute to the acceleration in the outer wind. It is noteworthy that this is not limited to the CNO elements, but instead the more heavy and electron-rich Ca, K, and Ar provide important contributions here. Moreover, Ne is on the same order as oxygen, while the influence of C is much lower due to the depleted abundance in WN stars. Elements such as Si, P, or S, which are important to consider for O or B stars \citep[e.g.][]{Vink+1999,Noebauer+2015,Sander+2017,Sander+2018}, provide only minor or even negligible contributions to the wind driving in a WNE. A full list of all considered elements and ions is listed in the appendix Table\,\ref{tab:datom}.

\begin{figure}
  \includegraphics[width=\columnwidth]{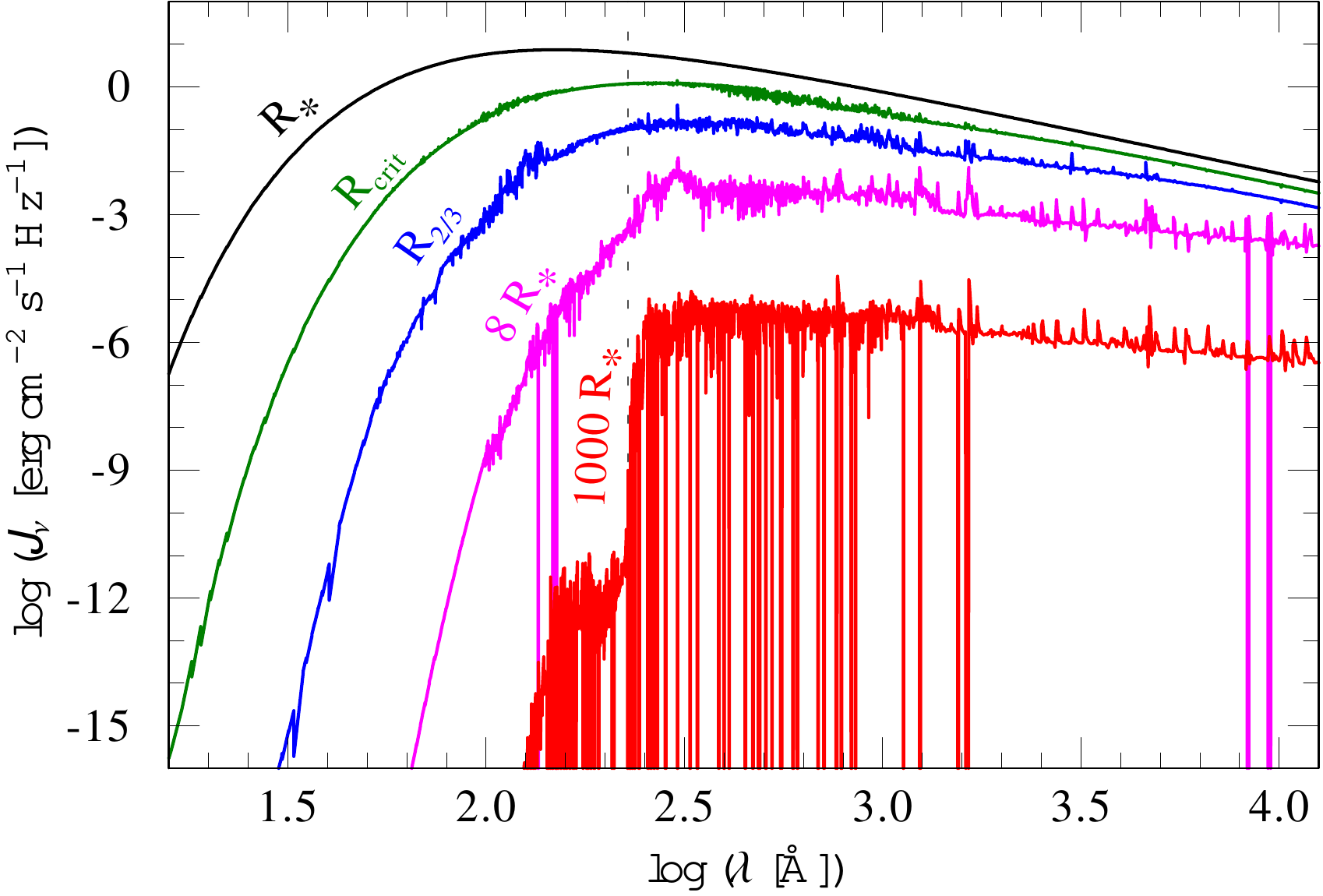}
  \caption{The radiation field $J_\nu(r,\lambda)$ in the comoving frame at different distances $r$ from the inner boundary $R_\ast$, corresponding to a Rosseland continuum optical depth of $\tau_\text{Ross,cont} = 20$.  Note that the curves in this figure are plotted on the coarse frequency grid used in the calculation of the radiative rates, while the radiative transfer itself is performed on a finer grid. For computational and memory reasons, the fine-grid results are not stored permanently. The dashed vertical line denotes the \ion{He}{ii} ionization edge. (WNE model at $Z_\odot$, see Table\,\ref{tab:wcwnesolcmp} for a complete set of parameters.)}
  \label{fig:jnueillu}
\end{figure}

\begin{figure}
  \includegraphics[width=\columnwidth]{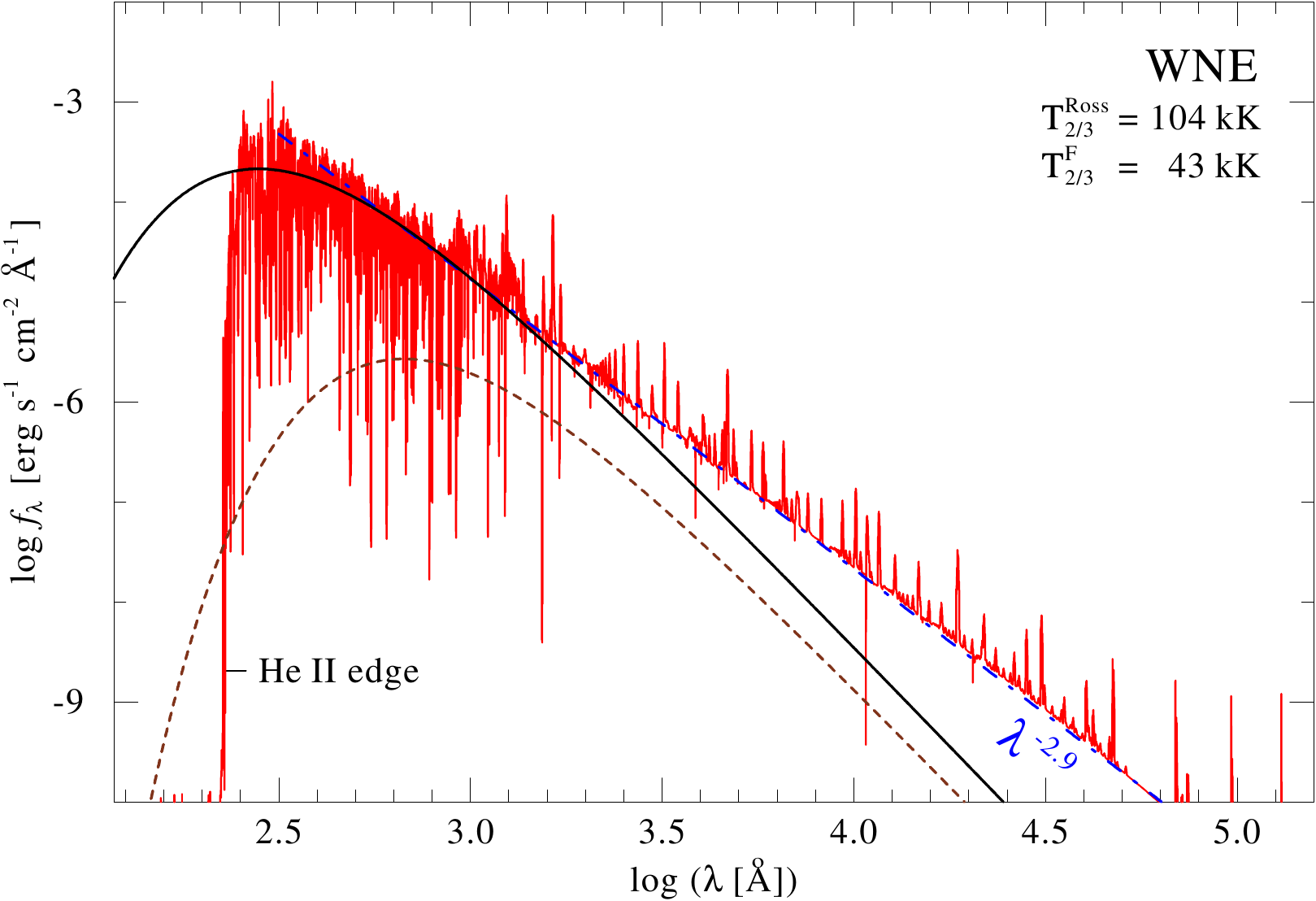}
  \caption{Emergent spectral energy distribution (red) of a WNE (same as in Fig.\,\ref{fig:jnueillu}) compared to black bodies with effective temperatures obtained from $\tau_\text{Ross} = 2/3$ (black) and $\tau_F = 2/3$ (maroon, dashed). The blue dot-dashed line represents the slope of the spectrum redwards of the flux peak in the UV. The EUV flux drop corresponds to the $\ion{He}{ii}$ ionization edge.}
  \label{fig:blackbody-hdwne}
\end{figure}

In Fig.\,\ref{fig:jnueillu}, the angle-integrated intensity $J_\nu$ is shown as a function of wavelength $\lambda$ at different distances from $R_\ast$. At the inner boundary of our stellar atmosphere we are in an optically very thick regime ($\tau_\text{Ross,cont} = 20$, $\tau_\text{Ross} \approx 44$), where we can safely assume the diffusion approximation to be valid. Thus, the resulting $J_\nu$ is an almost perfect blackbody. This changes outwards as depicted in Fig.\,\ref{fig:jnueillu}, but the departures from a blackbody are not large as long as we remain in the optically thick regime, which includes our critical point at $\tau_\text{Ross} \approx 4$. 
In the wind however, the shape of $J_\nu$ is significantly altered, especially in the outer wind where most of the \ion{He}{ii} ionizing photons are absorbed. While not obvious from Fig.\,\ref{fig:jnueillu}, the photons are re-emitted at longer wavelengths, leading to an emergent flux $\propto \lambda^{-2.9}$ as illustrated in Fig.\,\ref{fig:blackbody-hdwne}, considerably flatter than the $\lambda^{-4}$-slope one would infer from the Rayleigh-Jeans part of a blackbody. Defining an effective temperature for a WR star is not straight-forward due to the fact that for these dense winds the whole spectrum is formed in the wind. In Fig.\,\ref{fig:blackbody-hdwne}, we use the effective temperature corresponding to $R(\tau_\text{Ross} = 2/3)$ and $R(\tau_F = 2/3)$ for comparison. While the flux-weighted optical depth $\tau_F$ is important in terms of wind driving, $T_\text{eff}(\tau_F = 2/3)$ does not provide any useful representation. Instead, the Rosseland optical depth here represents where the continuum part of the spectrum is formed and we thus define $T_{2/3} := T_\text{eff}(\tau_\text{Ross} = 2/3)$. The corresponding blackbody touches the actual slope of the emergent spectrum in the UV, but cannot reproduce the other main features. It is clear that a blackbody is not a good representation of the emergent flux of a WR star. Even when adding a cutoff at the \ion{He}{ii} ionization edge, any blackbody representation would significantly underestimate the flux at longer wavelengths. Due to the different slope, the amount of underestimation is wavelength-dependent, but easily surpasses an order of magnitude already in the near-IR, getting larger ever after. This effect is more pronounced for denser winds with the slope going back to Rayleigh-Jeans in the thin-wind limit. Even then, for a star with $T_\ast \approx 140\,$kK the emergent flux at short (e.g. EUV) wavelengths can still deviate from a blackbody by an order of magnitude.

\section{Results}
  \label{sec:results}

In this work we calculate different series of models, anchored by the prototypical model for WR\,111 by \citetalias{GH2005}. We limit our study to models with $T_\ast = 141\,$kK, reflecting early-type, He-burning stars with compact $R_\ast$ \citep[as recently also motivated by][]{Grassitelli+2018}. Corresponding to the WC model with $\log \dot{M} = -5.15$, we calculated a WNE model with the same $\dot{M}$. These were then used as anchors for different model sets. In the first set, we calculated a series of models with fixed $\dot{M}$ and flexible $\Gamma_\text{e}$ or, respectively, $L/M_\ast$-ratio, where we vary the metallicity $Z$ ranging from $10$ times solar down to $10^{-5}$ times solar. In the second and third series, we fix $Z$ to $1\,Z_\odot$ and $0.1\,Z_\odot$, while exploring the $\dot{M}(\Gamma_\text{e})$ domain.

A detailed set of input and output parameters for the anchoring WNE and WC models at $Z_\odot$ with $\log \dot{M} = -5.15$ is given in Table\,\ref{tab:wcwnesolcmp}. (The third model with $\log \dot{M} = -6.1$ refers to the WC model with a thin wind and is shown for comparison. The different implications in regards to driving for optically thin solutions are discussed further below and, with specific reference to this model, in Sect.\,\ref{asec:cakfail} of the appendix.) The chemical composition given in Table\,\ref{tab:wcwnesolcmp} is identical for all other models at $Z_\odot$ in this work. Throughout the rest of this paper, we refer to the metallicity $Z$ in the sense of non self-enriched elements. Thus for WC models, this excludes C and O, which are kept at $X_\text{C} = 0.6$ and $X_\text{O} = 0.05$ for all $Z$ unless otherwise noted. In the WNE models, the CNO elements are mixed due to the bottleneck in the CNO cycle, leading to a nitrogen enhancement. These mixed abundances are then scaled for WN models at different $Z$.

$T_\ast$, $\varv_\text{mic}$ and $D_\infty$ are kept fixed throughout this work. The choice of $T_\ast = 141\,$kK implies that our solutions will likely not be applicable for late-type WR stars. For these types, the discrepancy between the empirically deduced values for $T_\ast$ and the effective temperatures inferred from hydrostatic stellar evolution calculations are largest. Via Eq.\,(\ref{eq:lrt}), this temperature discrepancy can also be viewed as a radius discrepancy and has been termed the `WR radius problem' \citep[e.g.][]{Grassitelli+2018}. The prototypical study by \citetalias{GH2005} has shown that for the WC5 star WR\,111 this discrepancy could be removed by using a hydrodynamically consistent stratification, where $T_\ast$ changed from $85\,$kK to $140\,$kK and $R_\ast$ is reduced from $\approx 2.5\,R_\odot$ to $0.9\,R_\odot$, in line with a star essentially sitting on the He zero age main sequence (He-ZAMS). In light of this result, we calculate such compact, hydrodynamically consistent models in this work and leave a broader study of the $T_\ast$-domain for future work.

Comparing our anchoring models reveals that when we force both models to have the same $\dot{M}$, the WNE star has a higher stellar mass. In turn, this implies a WNE star with the same $M_\ast$ (and $L$) as a WC star can drive a wind with a higher $\dot{M}$. While the higher C and O abundances in a WC star do not lead to an increase of its mass-loss rate compared to a WNE star as their ionization stages are too high at the high electron temperatures around $R_\text{crit}$ to provide substantial line driving, they provide additional opacity in the wind, leading to higher terminal velocities. This also holds when we fix $M_\ast$: In a test case with $M_\ast = 11\,M_\odot$ and $\log L/L_\odot = 5.45$, we obtain $\log \dot{M} = -4.8$ and $\varv_\infty = 2052\,\mathrm{km}\,\mathrm{s}^{-1}$ for the WNE, while we get $\log \dot{M} = -4.9$ and $\varv_\infty = 2159\,\mathrm{km}\,\mathrm{s}^{-1}$ for the WC star. While the difference in $\varv_\infty$ is a direct result of enhanced line-driving, the inverse difference in $\dot{M}$ is not. As Table\,\ref{tab:wcwnesolcmp} illustrates, the WNE has a higher value for the ionization parameter $q_\text{ion}$ and thus a higher $\Gamma_\text{e} \propto q_\text{ion} \cdot L/M$ \citep[cf.~Eq.\,22 in][]{Sander+2015}. For fixed $L/M$, it is therefore only the larger He mass fraction in the WNE making the difference. More He instead of C and O leads to more free electrons compared the WC stars, thus increasing the Thomson contribution and therefore the mass-loss rate. The only elements that can provide substantial line opacities directly affecting $\dot{M}$ are those with complex electron structures such as iron, while others only have an indirect effect via their contribution to the free electron budget.

\begin{table}
  \caption{Hydrodynamically consistent WNE and WC models with $Z_\odot$}
  \label{tab:wcwnesolcmp}

  \centering
  \begin{tabular}{lccc}
      \hline
           & WNE & WC & WC (thin) \\
      \hline  
      $T_\ast$\,[kK] &  \multicolumn{3}{c}{$141$}  \\
      $R_\ast$\,[R$_\odot$] &  \multicolumn{3}{c}{$0.9$}  \\
      $\log\,(L~[\mathrm{L}_\odot])$ &  \multicolumn{3}{c}{$5.45$}  \\
      $\log\,(\dot{M}~[\mathrm{M}_\odot\,\mathrm{yr}^{-1}])$ &  $-5.15$ & $-5.15$ & $-6.1$ \\
			$\varv_\text{mic}$\,[km\,s$^{-1}$] & \multicolumn{3}{c}{$30$} \\
      \smallskip
      $D_\infty$ &  \multicolumn{3}{c}{50}  \\

			$X_\text{H}$   &   \multicolumn{3}{c}{--}  \\
			$X_\text{He}$  &    $0.9787$            &   $0.6182$   &   $0.6182$   \\
			$X_\text{C}$   &    $4.0\cdot10^{-4}$   &   $0.6$      &   $0.6$      \\
			$X_\text{N}$   &    $0.015$             &    --        &    --        \\
			$X_\text{O}$   &    $1.0\cdot10^{-3}$   &   $0.05$     &   $0.05$     \\
			$X_\text{Ne}$  &   \multicolumn{3}{c}{$1.3\cdot10^{-3}$}  \\
			$X_\text{Na}$  &   \multicolumn{3}{c}{$2.7\cdot10^{-6}$}  \\
			$X_\text{Mg}$  &   \multicolumn{3}{c}{$6.9\cdot10^{-4}$}  \\
			$X_\text{Al}$  &   \multicolumn{3}{c}{$5.3\cdot10^{-5}$}  \\
			$X_\text{Si}$  &   \multicolumn{3}{c}{$8.0\cdot10^{-4}$}  \\
			$X_\text{P}$   &   \multicolumn{3}{c}{$5.8\cdot10^{-6}$}  \\
			$X_\text{S}$   &   \multicolumn{3}{c}{$3.1\cdot10^{-4}$}  \\
			$X_\text{Cl}$  &   \multicolumn{3}{c}{$8.2\cdot10^{-6}$}  \\
			$X_\text{Ar}$  &   \multicolumn{3}{c}{$7.3\cdot10^{-5}$}  \\
			$X_\text{K}$   &   \multicolumn{3}{c}{$3.1\cdot10^{-6}$}  \\
			$X_\text{Ca}$  &   \multicolumn{3}{c}{$6.1\cdot10^{-5}$}  \\
      \medskip
			$X_\text{Fe}$  &   \multicolumn{3}{c}{$1.6\cdot10^{-3}$}  \\

			\multicolumn{3}{l}{\textit{Results:}} \\
      $\varv_\infty$\,[km\,s$^{-1}$] \rule[0mm]{0mm}{3mm} & $1805$ &  $2132$  &  $4268$ \\
      $M_\ast$\,[M$_\odot$]                               & $14.9$ &  $13.9$  &  $19.4$ \\
			$R_\text{crit}$\,[R$_\ast$]                         & $1.03$ &  $1.03$  &  $1.04$ \\[0.1em]
			$R_{2/3}^\text{Ross}$\,[R$_\ast$]                   & $1.86$ &  $1.60$  &  $1.03$ \\[0.2em]
			$R_{2/3}^{F}$\,[R$_\ast$]                           & $10.7$ &  $7.74$  &  $1.19$ \\
			$\tau_\text{Ross}(R_\text{crit})$                   & $5.38$ &  $3.53$  &  $0.26$ \\
      $T_\text{e}(R_\text{crit})$\,[kK]                   &  $196$ &   $188$  &   $147$ \\
      $T_\text{eff}(R_\text{crit})$\,[kK]                 &  $139$ &   $139$  &   $139$ \\
      $T_{2/3}$\,[kK]                                     &   $95$ &   $112$  &   $140$ \\
      $\left<\mathcal{A}\right>$                          & $4.06$ &  $7.17$  &  $7.17$ \\
      $q_\text{ion}$                                      & $0.43$ &  $0.35$  &  $0.39$ \\
      $\Gamma_\text{e}$                                   & $0.29$ &  $0.25$  &  $0.18$ \\
      $\eta$                                              & $2.18$ &  $2.61$  &  $0.60$ \\
    \hline
  \end{tabular}
\end{table}

Comparing the WC parameters in Table\,\ref{tab:wcwnesolcmp} to the results from \citetalias{GH2005}, we see only minor discrepancies. This is a bit surprising given the fact that \citetalias{GH2005} used way fewer elements and a much larger value of $\varv_\text{Dop} = 250\,$km s$^{-1}$ in the CMF calculations. While necessary at that time to speed up the calculations, this also leads to an additional broadening of the iron-line cross sections compared to our more standard estimate of $\varv_\text{Dop} = 100\,$km s$^{-1}$. We thus also calculated additional models with both values of $\varv_\text{Dop}$ using their original chemical composition and clumping stratification, revealing that the various differences between their and our approach partly cancel each other, leading to the good agreement. The calculations further revealed that the terminal velocity $v_\infty$ in these kind of models is very sensitive to all kinds of changes, including the precise method of temperature correction. Thus we need to assume an uncertainty for $\varv_\infty$ of up to $10\%$ in all our models presented in this work.

\subsection{The driving role of Carbon and Oxygen in WC stars}
  \label{sec:wcabu}

\begin{figure}
  \includegraphics[width=\columnwidth]{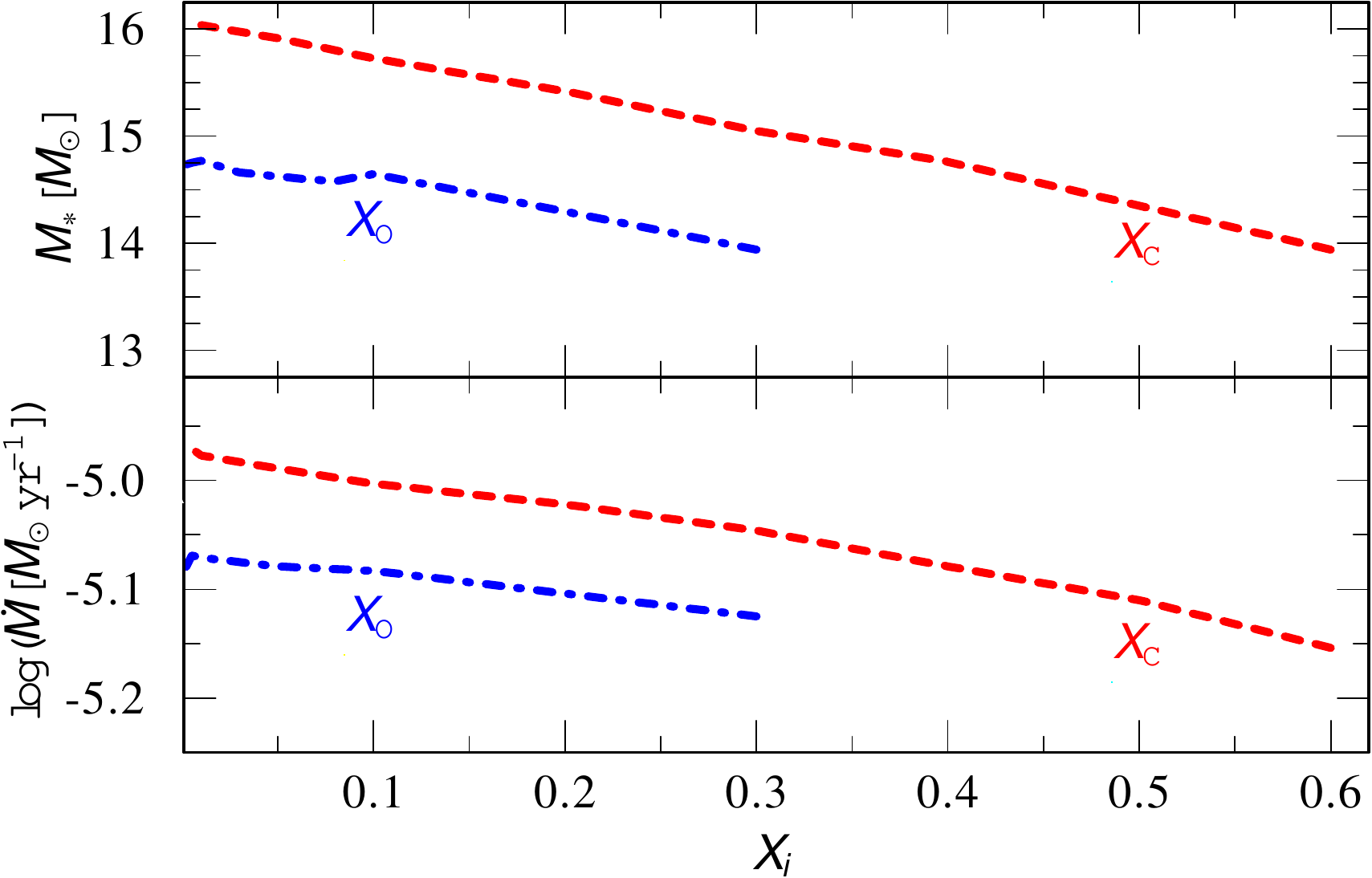}
  \caption{Upper panel: Stellar masses for hydrodynamically consistent WC star models with $\log L/L_\odot = 5.45$, $\log \dot{M} = -5.15\,[M_\odot \mathrm{yr}^{-1}]$, and different mass fraction of $C$ (red, dashed, $X_\text{O} = 0.05$) and $O$ (blue, dashed-dotted, $X_\text{C} = 0.4$). All other metals have solar abundances. Lower panel: Similar calculation, but now with fixed $M_\ast = 13.9$ and flexible $\dot{M}$.}
  \label{fig:wc-co-influence}
\end{figure}

For WC stars, the abundances of carbon and oxygen have an influence on the derived results. We investigate this with two series of models: One, where we fix $L$ and $\dot{M}$ to study the derived stellar masses, and a second where we fit $L$ and $M_\ast$ to investigate the influence on $\dot{M}$. The results are shown in Fig.\,\ref{fig:wc-co-influence}, where we see that for both elements higher abundances are counter-productive in regards to wind launching. The same $\dot{M}$ can be driven by a WC star with $X_\text{C} = 0.2$ that is about $1.5\,M_\odot$ or $10\%$ more massive than a WC star with $X_\text{C} = 0.6$. This means that for the same $L$ and $M_\ast$, the WC star with the lower $X_\text{C}$ would have a larger $\dot{M}$, in line with our result above that a WNE star with the same $L$ and $M$ as a WC star has the larger $\dot{M}$. Again, less C and O lead to a higher Thomson opacity. In actual stellar evolution this would of course go the other way round, as the amounts of carbon and oxygen are increasing with time. The resulting decrease of $\dot{M}$ due to the increased mixing of He-burning products in the atmosphere however, is then also competing with the increase of $\dot{M}$ due to the increase of $\Gamma_\text{e} \propto L/M$ during the life of a WC star resulting from the actual loss of mass itself.
	
Our second series depicted in the lower panel Fig.\,\ref{fig:wc-co-influence} confirms that -- with the exception of $X_\text{O} < 0.005$, higher $X_\text{C}$ and $X_\text{O}$ decrease $\dot{M}$. The isolated dependence of a particular mass fraction on $\dot{M}$ is exponential and a linear fit of $\log \dot{M}$ as a function of $X_i$ yields
	\begin{align}
  \log \dot{M} & = -4.971 (\pm 0.003) - 0.285 (\pm 0.010) X_\text{C}~\text{~~~and} \\
  \log \dot{M} & = -5.072 (\pm 0.002) - 0.165 (\pm 0.017) X_\text{O}\text{,}
\end{align}
revealing that a higher $X_\text{C}$ seems to have about twice the impact of a higher $X_\text{O}$. Based on these findings and taking the empirically derived fractions for $X_\text{C}$ and $X_\text{O}$, the uncertainty in $\dot{M}$ of observed WC stars due to unsure abundances in C and O is about $0.2$\,dex.

\subsection{Dependencies on luminosity and mass}
  \label{sec:mdotlmdep}

In order reach a particular value for $\Gamma_\text{e}$ or $L/M_\ast$ in our calculations, we usually fix the luminosity $L$ while varying the stellar mass $M_\ast$. This numerically favourable technique allows to calculate models for a good number of $L/M_\ast$ values. As not all of these particular combinations of $L$ and $M_\ast$ will be realized in nature \citep[see, e.g.,][]{L1989}, this of course prompts the question whether different combinations of $L$ and $M_\ast$ leading to the same $\Gamma_\text{e}$ might affect the derived wind parameters such as $\dot{M}$, or only $\Gamma_\text{e}$ matters here.

\begin{figure}
  \includegraphics[width=\columnwidth]{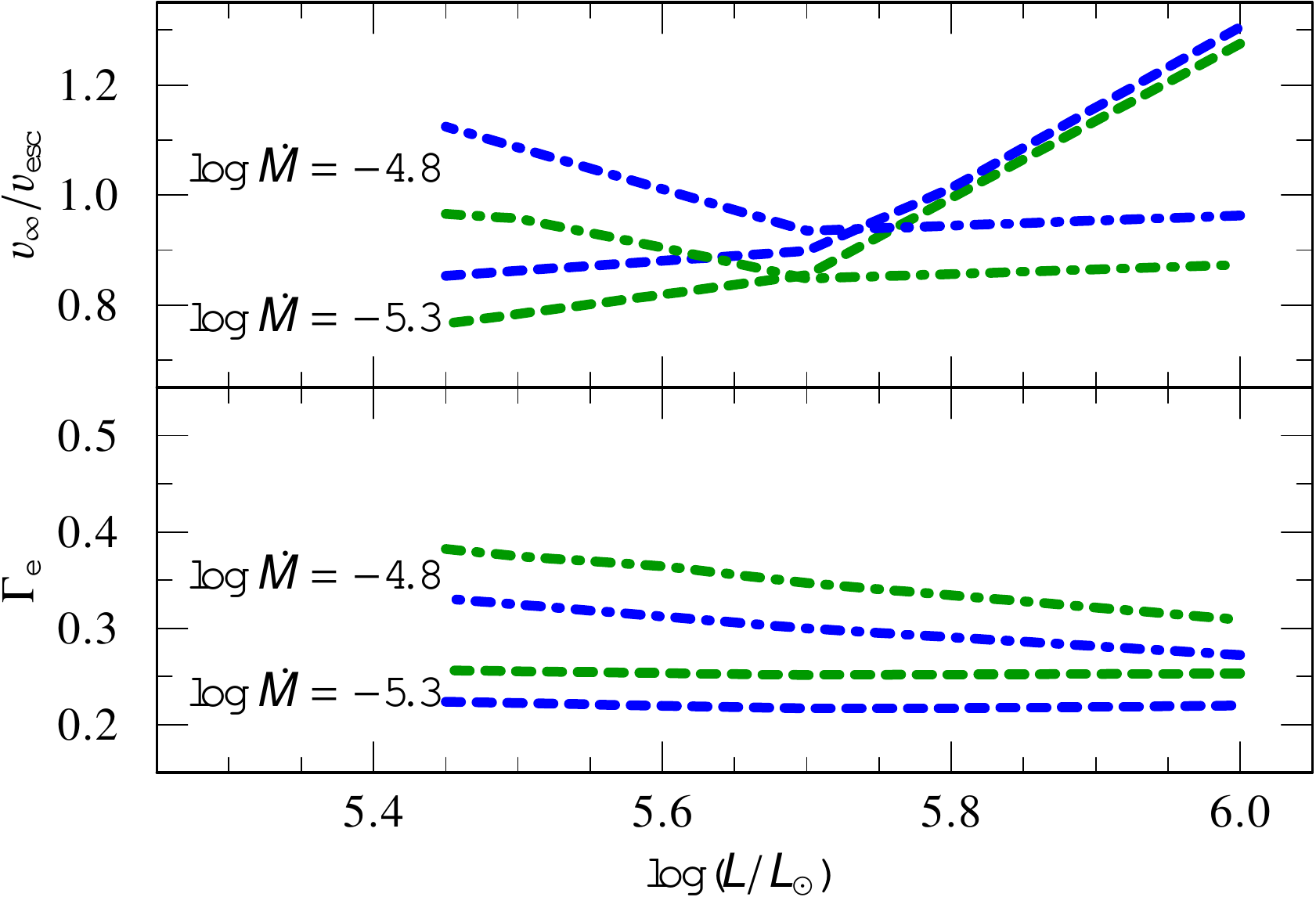}
  \caption{Eddington parameter $\Gamma_\text{e}(R_\star)$ (lower panel) and $\varv_\infty/\varv_\text{esc}$-ratio (upper panel) for consistent WN (green) and WC star models (blue) with different $L$ and $M$, but identical $\dot{M}$.}
  \label{fig:wn-gamma-vratio}
\end{figure}

To investigate this, we calculated for two different mass-loss rates each a small series of WC and WN models with different luminosities. The derived values of $\Gamma_\text{e}$ can be seen in the lower panel of Fig.\,\ref{fig:wn-gamma-vratio}. For the lower value of $\dot{M}$, $\Gamma_\text{e}$ stays more or less constant, meaning that $\dot{M}$ depends only on the ratio of $L$ and $M$, but not the individual values themselves. For the larger $\dot{M}$ however, we see a decrease of $\Gamma_\text{e}$ with higher $L$. 
While a dependency on $\Gamma_\text{e}$ is expected, the absence of a further $L$-dependence would be in contrast to relations for OB and WNh stars, which all include such a dependence \citep[e.g.][]{Vink+2000,GH2008,Vink+2011}, meaning $\dot{M} = \dot{M}(\Gamma_\text{e},L,T_\ast,X_i)$. For He-ZAMS stars, one could in principle eliminate either $M$ or $L$ from the recipe by using mass-luminosity relations \citep[e.g.][]{L1989,Graefener+2011}, but as mentioned above this is not the intension of this paper which is purely based on atmosphere calculations. A dependence of $\dot{M} = \dot{M}(\Gamma_\text{e},T_\ast,X_i)$ would reduce the problem by one dimension and allow a simple scaling of the resulting $\dot{M}$ to be e.g.\ easily used in stellar evolution calculations. The results for $\log \dot{M} = -4.8$ show that this is only approximately fulfilled. Fortunately, the remaining dependence on $L$ does appear to be weak with e.g. $\log L/M_\ast \propto -0.155 \log L$ for our WC calculations. 

In the upper panel of Fig.\,\ref{fig:wn-gamma-vratio}, we look at the outer wind layers, which seem to be more complex. Neither $\varv_\infty$, nor its ratio with the
 escape velocity $\varv_\text{esc} := \sqrt{2 G M_\ast R_\ast^{-1}}$
remain constant for different $L$, but similar $\Gamma_\text{e}$. Including the factor $\sqrt{1 - \Gamma_\text{e}(R_\ast)}$ in the definition of $\varv_\text{esc}$ changes the absolute values, but not the observed trends. 

While the derived values for the terminal velocities have to be taken with care, the mass-loss rates appear to be more robust. In the following calculations, we will stick to $\log L/L_\odot = 5.45$. Our results can be approximately scaled to other luminosities (and masses) as long as $L/M_\ast$ is kept constant. The small, but non-vanishing impact of $L$ in our test models with higher $\dot{M}$ means that such scaling introduces an uncertainty. Our $\dot{M}(\Gamma_\text{e})$-relations (cf.\ Sect.\,\ref{sec:mdotfit}) thus need future follow-up calculations for different luminosity ranges.

\subsection{Metallicity trends}
  \label{sec:zdep}

A major uncertainty for our understanding of earlier generations of stars and galaxies is their mass-loss history. Line-driven mass loss requires a considerable abundance of elements with a sufficient number of lines covering the spectral energy distribution, especially in the UV and EUV. The iron group elements with in particular Fe itself are a key ingredient here. Thus, a lower abundance of Fe and other elements should considerably reduce the stellar mass loss.

\subsubsection{Proximity to the Eddington Limit}
  \label{sec:eddprox}

\begin{figure}
  \includegraphics[width=\columnwidth]{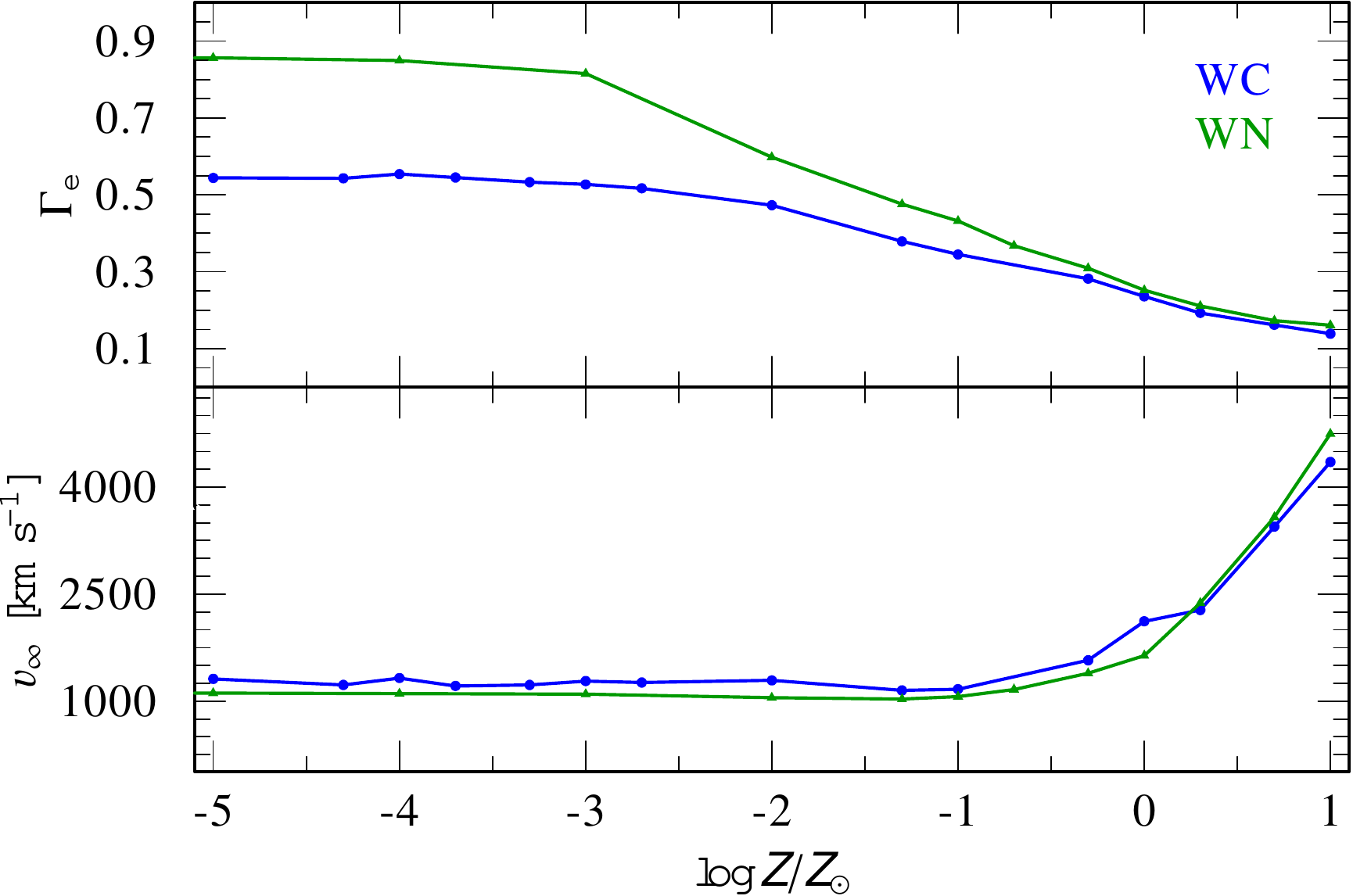}
  \caption{$\Gamma_\text{e}(Z)$ (upper panel) and $\varv_\infty(Z)$  (lower panel) for our WNE and WC sample (cf.~Table\,\ref{tab:wrgeddzmodels}) with the same luminosity and mass-loss rate}
  \label{fig:cmp-vinf-gedd-z}
\end{figure}

To study how a certain amount of mass loss could be driven in different metallicity regimes, we calculate a series of models with constant $\dot{M}$, but different $Z$ and consequently different $L/M$-ratios. An overview of the results from $10\,Z_\odot$ down to $10^{-5}\,Z_\odot$ is given in the Appendix Table\,\ref{tab:wrgeddzmodels} for WNE and WC stars respectively. The trend for $\dot{M}(Z)$ is also depicted in the upper panel of Fig.\,\ref{fig:cmp-vinf-gedd-z}, where we see that WNE stars would need to get much closer to the Eddington limit at lower $Z$ than WC stars due to the self-enriched carbon abundance in WC stars. A closer look at the low-$Z$ models reveals that this contribution is not only from spectral lines, but to a considerable part ($\approx 30..40\%$ depending on the layers) from the bound-free and free-free opacities of \ion{C}{iii} and \ion{C}{iv}. For the WNE stars, where all non-He abundances scale with $Z$, the contribution from the true continuum is also significant at very low $Z$, but all of this can be attributed to He.

The true continuum is crucial for both, launching the wind and maintaining the acceleration. The obtained terminal velocities $\varv_\infty$ would be considerably lower without taking these opacities into account. As they do not vanish at very low $Z$, the obtained values of $\varv_\infty$ tend to stay essentially constant below $0.1\,Z_\odot$ as depicted in the lower panel of Fig.\,\ref{fig:cmp-vinf-gedd-z}. In fact, there is a small increase of $\varv_\infty$ with lower $Z$ after a minimum is reached around $0.1\,Z_\odot$. 
As a direct consequence of the almost constant $\varv_\infty$ and the fixed $\dot{M}$, the wind efficiency parameter $\eta$ (cf.~Eq.\,\ref{eq:etadef})
does not change either at low $Z$ as we vary our $\Gamma_\text{e}$ by changing $M_\ast$ while keeping $L$ fixed.

\begin{figure}
  \includegraphics[width=\columnwidth]{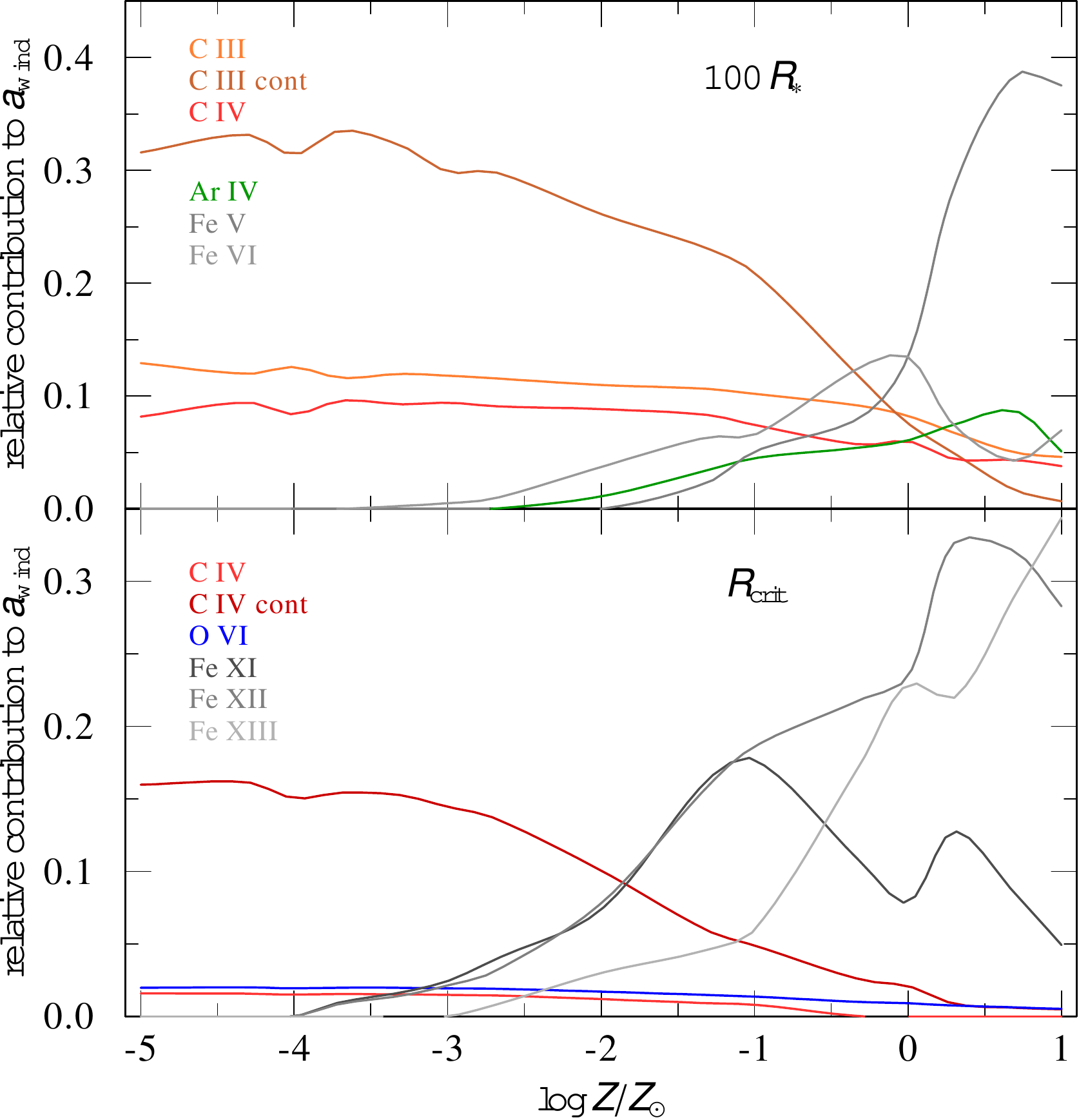}
  \caption{Lead line and true continuum contributions to the wind driving at $R_\text{crit}$ (lower panel) and $100\,R_\ast$ (upper panel) for a WC star with $T_\ast = 141$\,kK and $\log \dot{M} = -5.15$ at different $Z$}
  \label{fig:wc-lead-driving-ions-z}
\end{figure}

In Fig.\,\ref{fig:wc-lead-driving-ions-z} we show the most important contributors for a WC star with a given mass-loss rate at different metallicities. Given an evolutionary scenario that could produce these objects, the self-enriched carbon surpasses the role of Fe for launching the wind at the critical point below $\log Z/Z_\odot \la -2$ due to the contributions of the \ion{C}{iv} continuum transitions. \citet{VdK2005} found that $\dot{M}$ tends to settle for $\log Z/Z_\odot < -3$, which would agree with our finding that Fe continues to shrink in importance, but still plays a role until $\log Z/Z_\odot \approx -4$ where the line contributions from \ion{C}{iv} and \ion{O}{vi} are more important, although on an overall weak level of contribution. 
In the outer wind, a transition in the relative contributions happens already at higher $Z$ than at the base of the wind: At $100\,R_\ast$, the importance of the \ion{C}{iii} continuum already surpasses iron below $\log Z/Z_\odot \la 0.5$, i.e.\ below LMC metallicity.
%

\begin{figure}
	\includegraphics[width=0.9\columnwidth]{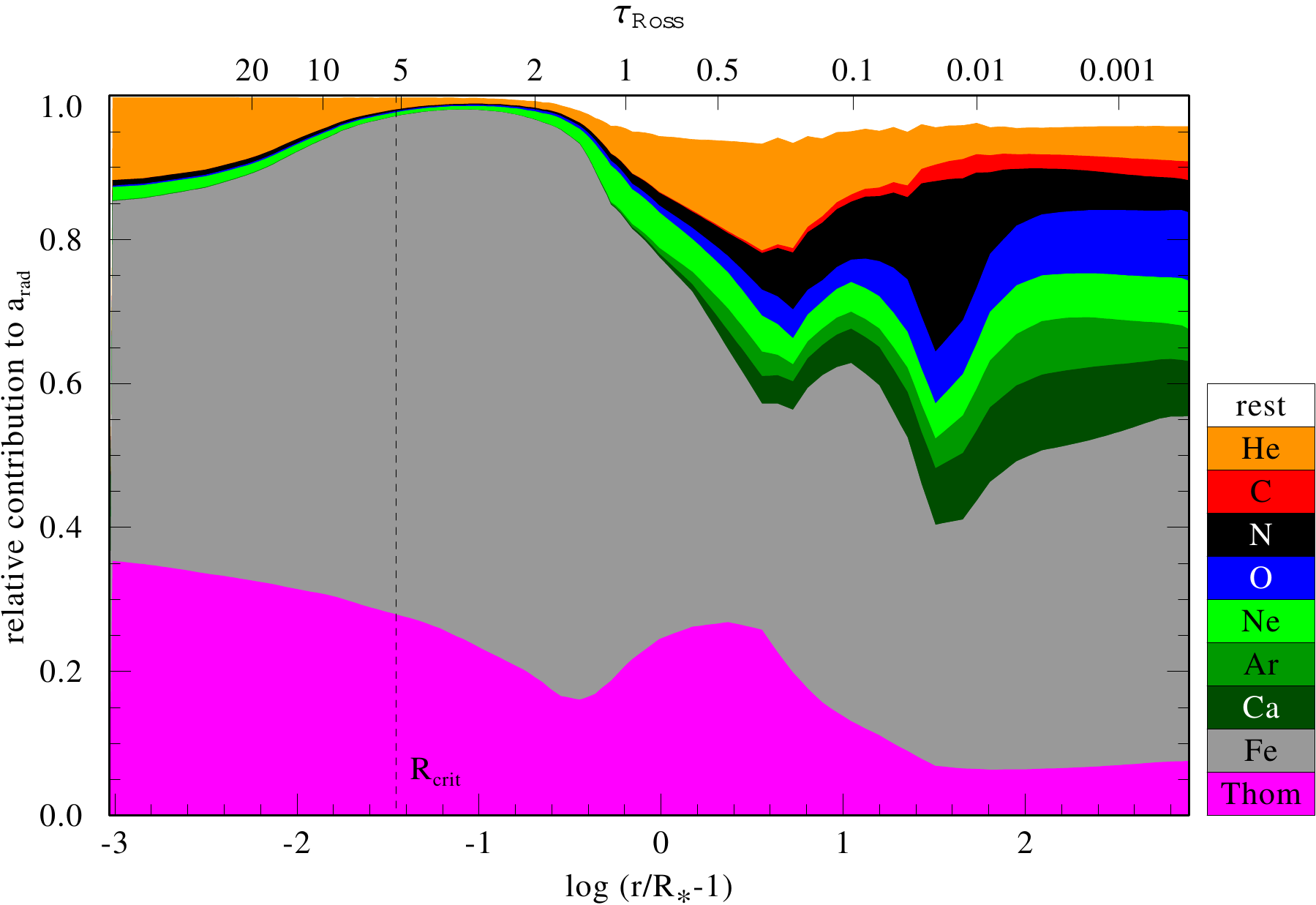}\\[0.5em]
  \includegraphics[width=0.9\columnwidth]{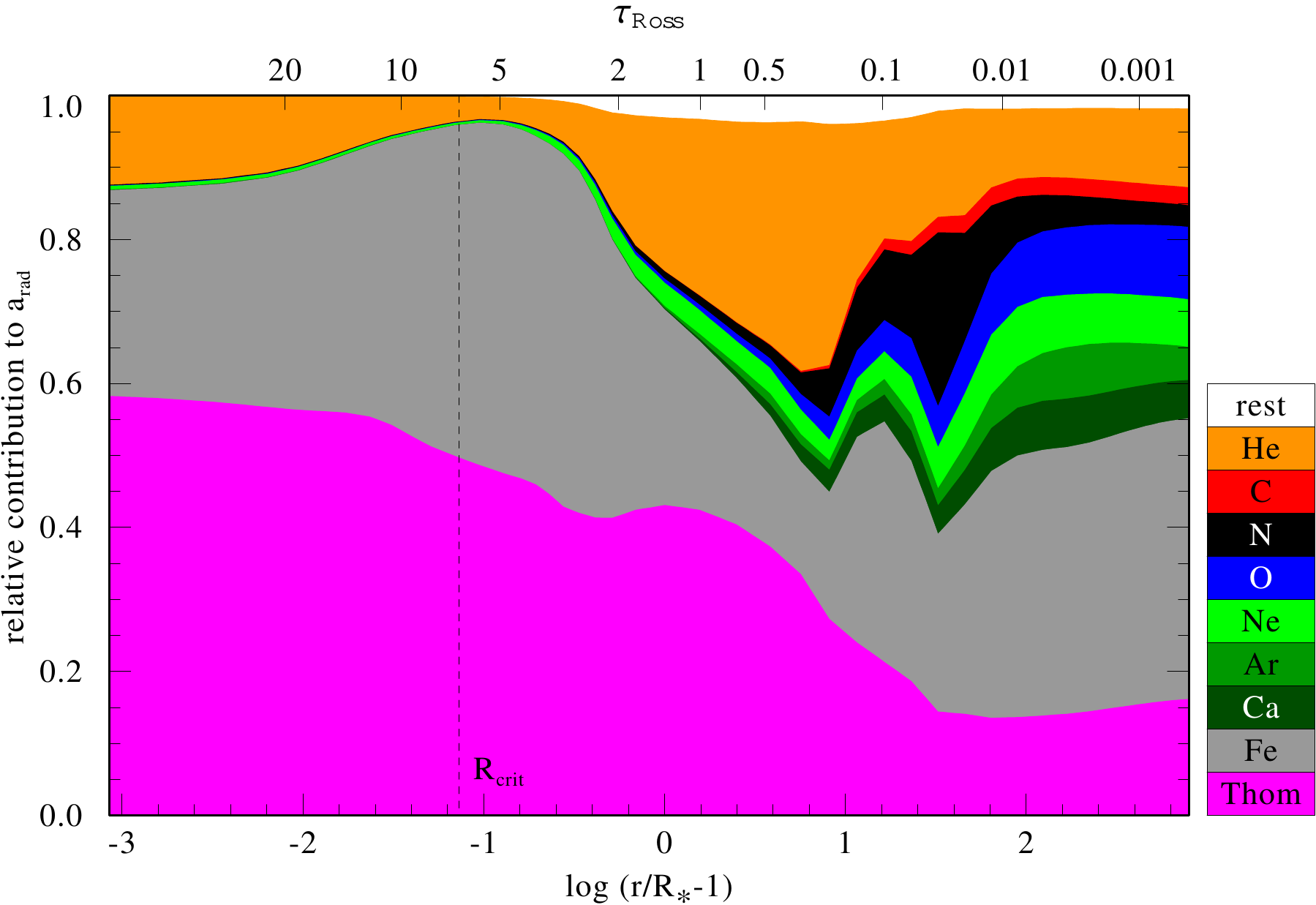}\\[0.5em]
	\includegraphics[width=0.9\columnwidth]{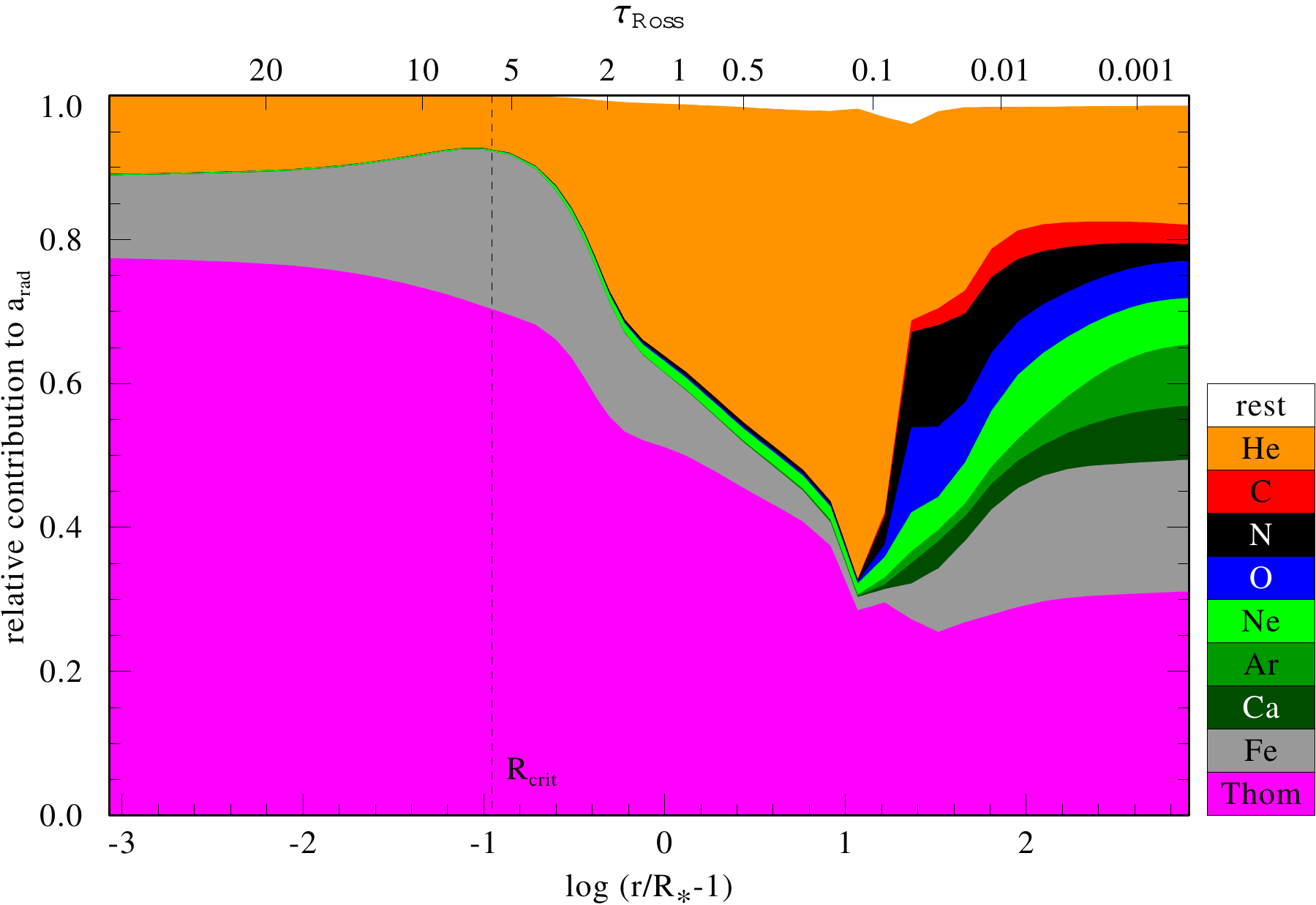}
  \caption{Relative contributions to $a_\text{rad}$ for a WNE star at $Z = 1$, $0.1$, and $0.01\,Z_\odot$ (top to bottom panels)}
  \label{fig:driving-relelemcontrib-wne-zsol}
\end{figure}
 
For a WN model with $\log \dot{M} = -5.15$, we show the full radial dependency of the elements contributing to the radiative acceleration at three different $Z$ in Fig.\,\ref{fig:driving-relelemcontrib-wne-zsol}.
Similar to what we saw in our discussion for the WC stars, Fe is the key element to launching the winds of classical WR stars at high metallicity. For lower $Z$ this role weakens and the stars have to get closer to the Eddington limit to reach a similar mass loss. For $Z = 10^{-1} Z_\odot$, the role of free electron scattering is already a bit larger than the iron contribution. At very low $Z$, the line contributions for WN stars essentially disappear and winds at e.g.\ $10^{-3} Z_\odot$ are driven by continuum opacities.
As long as the metallicity is not negligible, which tends to happen below $10^{-2}\,Z_\odot$, lower ionization stages are eventually reached in the outer wind and contributions from a variety of elements come into play. Major contributors here are N, O, Ne, Ar, and Ca with roughly equal impact despite their different abundances. Although the mass fractions of Ar and Ca are on the order of a few times $10^{-5}$ at $Z_\odot$, their transitions seem to fill up `gaps' in frequency space, while elements of similar abundance like Al or Na do not.

\begin{figure*}
  \includegraphics[width=0.8\textwidth]{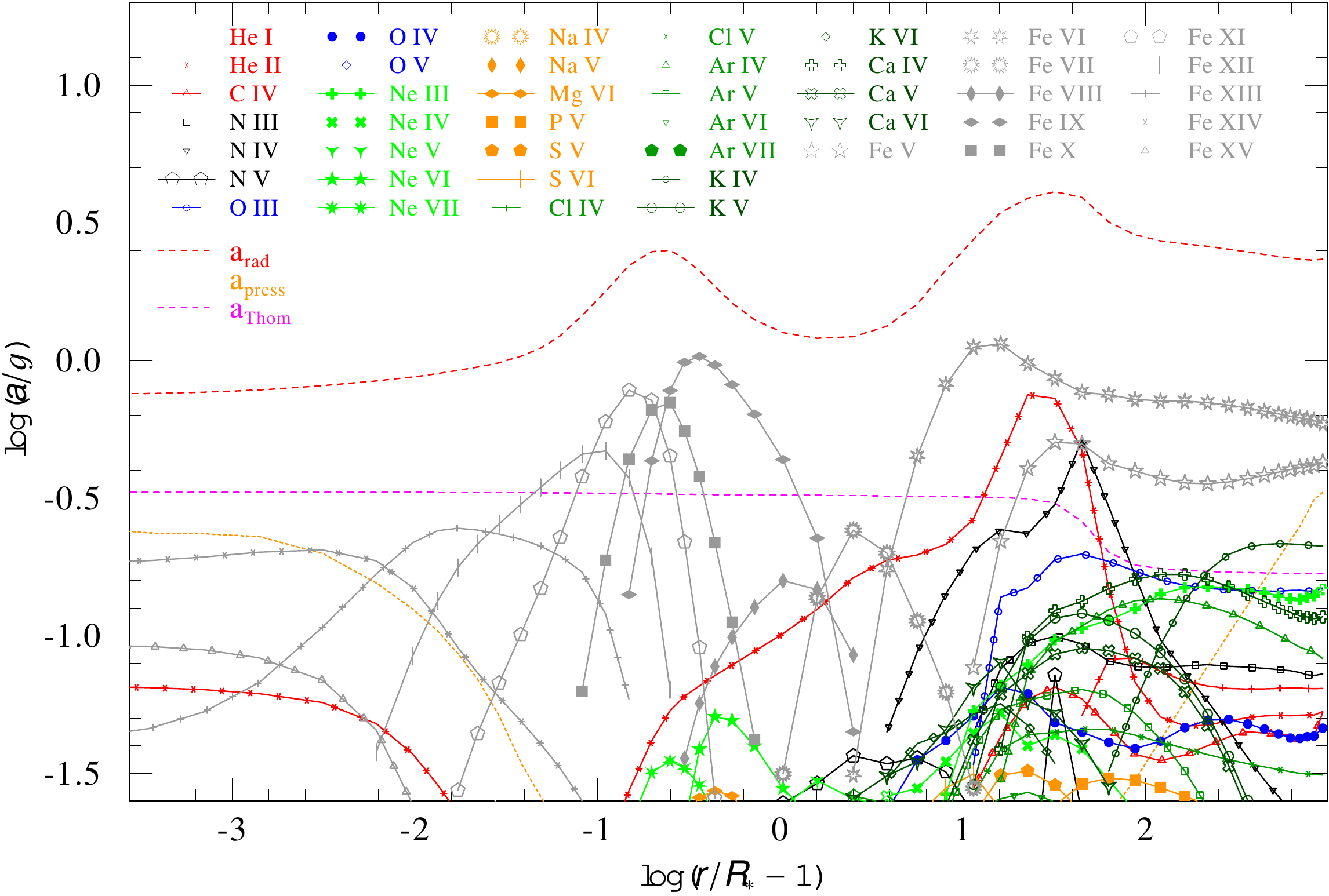}
  \caption{Contributions of the different ions to the radiative acceleration of a hydrodynamically consistent WN model with $T_\ast = 141\,$KK, $\log L/L_\odot = 5.45$, $M_\ast = 13\,M_\odot$, and $\log \dot{M} = -5.0$. All contributions with more then $\Gamma_\text{ion} = a_\text{ion}/g > 0.02$ are shown with different ions indicated by a combination of different color and symbol. For comparison, the total radiative acceleration ($a_\text{rad}$), the Thomson acceleration from free electrons, and the contribution from gas (and turbulence) pressure are also shown.}
  \label{fig:leadions-wne}
\end{figure*}

As briefly mentioned in Sect.\,\ref{sec:arad} and illustrated in Fig.\,\ref{fig:leadions-wne}, Si, P, and S do not provide a major contribution despite their abundance. On the contrary, the abundance of carbon is important, as it is depleted in favour of nitrogen due to the CNO cycle equilibrium. With $X_\text{C}$ being about two orders of magnitude lower than $X_\text{N}$, its low contribution in WN winds is in sharp contrast to its significant role in WC winds. All of these results illustrate that the role an element plays for wind driving can neither be easily guessed from the abundance, nor from the general position in the periodic table of elements, but detailed consistent calculations of the non-LTE population and the radiative transfer are crucial.

\subsubsection{Breakdown of mass loss at low Z}
  \label{sec:mdotz}

\begin{figure}
  \includegraphics[width=\columnwidth]{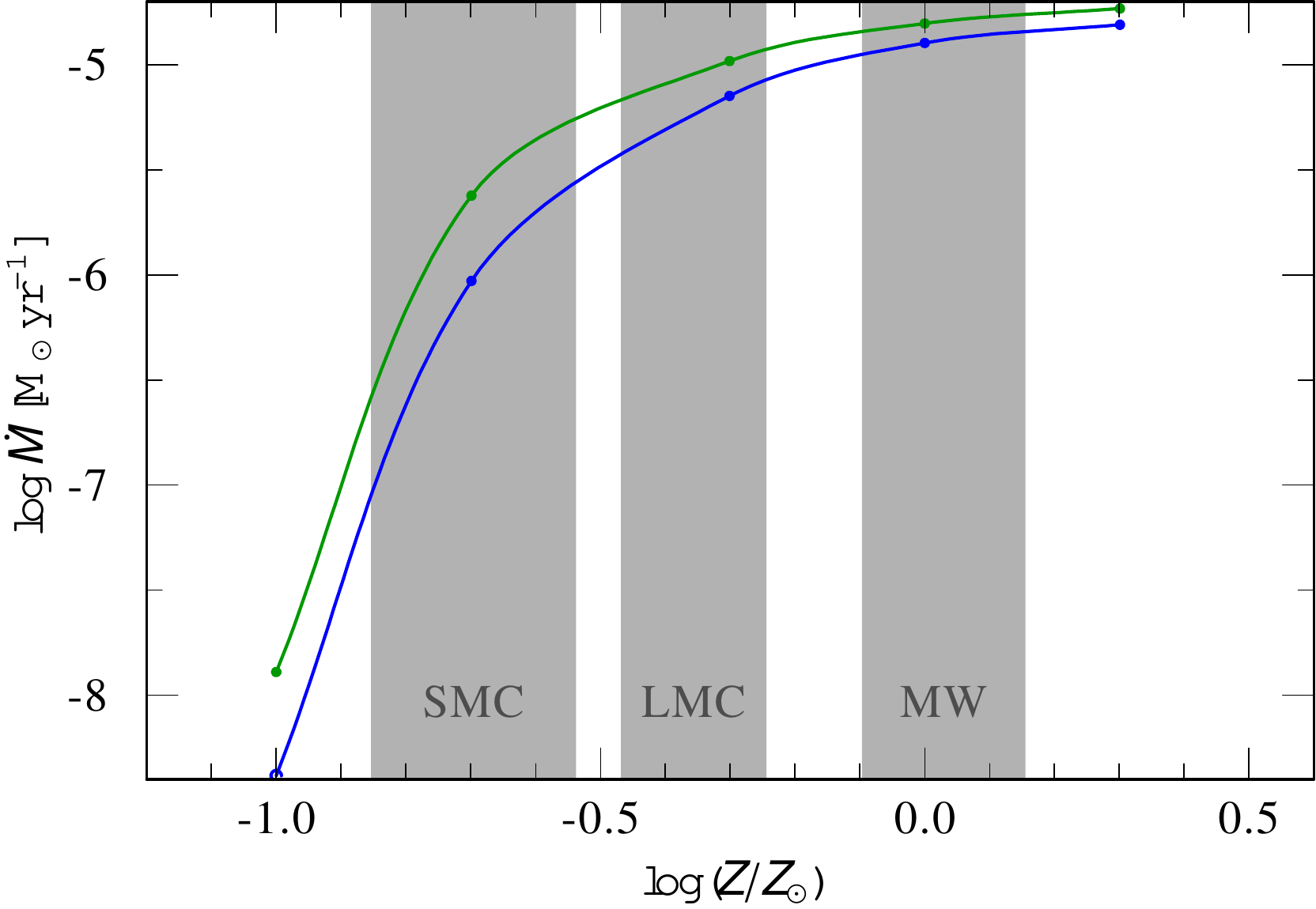}
  \caption{$\dot{M}(Z)$ for hydrodynamically consistent WN (green) and WC (blue) stars with fixed $\log L/L_\odot = 5.45$, $M_\ast = 11.5\,M_\odot$, and $T_\ast = 141\,$kK.}
  \label{fig:cmp-mdot-z}
\end{figure}

While we previously fixed $\dot{M}$ and looked at the resulting $L/M_\ast$-ratio, we now consider the opposite approach and look at a WN and WC model with fixed $L$ and $M_\ast$ at different $Z$. The derived mass-loss rates in Fig.\,\ref{fig:cmp-mdot-z} reveal a dramatic decrease of $\dot{M}$ with $Z$ for both WR types. While the slope is modest between $0.5$ and $2\,Z_\odot$ where $\dot{M}$ varies by $< 0.5\,$dex, the difference to the SMC regime is already an order of magnitude. At $0.1\,Z_\odot$, we get value of $\log \dot{M} \approx -8.4$, two orders of magnitude lower than for the SMC. Given the low metal abundance, such stars -- if they exist -- would have optically thin winds and provide a significant \ion{He}{ii} ionizing flux. 

A steep dependence of $\dot{M}(Z)$ is known from theoretical modelling of massive OB star winds \citep[e.g.][]{Vink+2001} and also for the most massive stars which have WN-type spectra \citep{GH2008}. Monte-Carlo calculations for late-type WN and WC stars by \citet{VdK2005} gave steep slopes of $\dot{M} \propto Z^{0.86}$ for WN and $\dot{M} \propto Z^{0.66}$ for WC stars in the regime between $Z_\odot$ and $0.1\,Z_\odot$. Empirically constrained results for WN stars by \citet{Hainich+2015} yield an even steeper $\dot{M} \propto Z^{1.2}$. Our CMF-based models for early subtypes here point to an even steeper drop with lower $Z$ that does not obey a simple power law. Calculations for He stars at $0.1\,Z_\odot$ and below turn out to be challenging, giving the values at $0.1\,Z_\odot$ a higher uncertainty. Nonetheless, the breakdown below $Z_\text{SMC}$ is reproduced in various additional models that lead to sequences presented in Fig.\,\ref{fig:cmp-mdot-z}.

\begin{table}
  \caption{Parameters of the WR models calculated for the $\dot{M}(\Gamma_\text{e})$ sequences ($T_\ast = 141\,$kK, $\log L/L_\odot = 5.45$)}
  \label{tab:tabwrhd-mddep}

  \centering
  \begin{tabular}{cccccc}
      \hline 
         $\log L/M$ &
         $\Gamma_\text{e}(R_\ast)$  &  
         $\log \dot{M}$ & 
         $\varv_\infty$ &
         $T_{2/3}^\text{Ross}$ &
         $T_\text{e}(R_\text{crit})$ \\[0.5mm]
      \hline
		\multicolumn{6}{c}{\textit{WN, solar abundances, CNO mixed}} \\
  4.55 & 0.55 & -4.39 & 2138 &  41 & 270  \\
  4.47 & 0.45 & -4.59 & 2132 &  54 & 243  \\
  4.40 & 0.38 & -4.79 & 2128 &  68 & 220  \\
  4.33 & 0.33 & -5.00 & 2049 &  87 & 203  \\
  4.28 & 0.29 & -5.16 & 2043 & 104 & 192  \\
  4.22 & 0.26 & -5.29 & 2061 & 117 & 182  \\
  4.19 & 0.24 & -5.50 & 1970 & 132 & 173  \\
  4.17 & 0.23 & -5.70 & 2812 & 138 & 161  \\
  4.15 & 0.22 & -5.89 & 3560 & 139 & 155  \\
	\medskip
  4.13 & 0.21 & -6.09 & 4341 & 139 & 150  \\

  \multicolumn{6}{c}{\textit{WN, $0.1$ solar abundances, CNO mixed}} \\
	4.67 & 0.72 & -4.39 & 1212 &  33 & 253  \\
  4.59 & 0.61 & -4.59 & 1421 &  39 & 230  \\
  4.54 & 0.54 & -4.79 & 1335 &  44 & 210  \\
  4.52 & 0.51 & -4.89 & 1317 &  49 & 201  \\
  4.50 & 0.49 & -5.00 & 1316 &  55 & 193  \\
  4.49 & 0.48 & -5.09 & 1299 &  63 & 186  \\
  4.48 & 0.46 & -5.20 & 1264 &  72 & 179  \\
  4.47 & 0.45 & -5.29 & 1166 &  81 & 173  \\
  4.46 & 0.45 & -5.39 & 1148 &  90 & 168  \\
  4.45 & 0.44 & -5.50 & 1015 &  98 & 163  \\
  4.45 & 0.44 & -5.59 & 1044 & 107 & 158  \\
  4.44 & 0.43 & -5.79 & 1088 & 123 & 151  \\
  4.44 & 0.42 & -6.00 & 1192 & 135 & 145  \\
  4.43 & 0.42 & -6.20 & 1320 & 137 & 139  \\
  4.43 & 0.41 & -6.39 & 1556 & 138 & 132  \\
	\medskip
  4.43 & 0.41 & -6.59 & 1754 & 138 & 129  \\

	\multicolumn{6}{c}{\textit{WC, solar abundances, except for CNO}} \\
  4.62 & 0.51 & -4.39 & 2243 &  39 & 268  \\
  4.52 & 0.40 & -4.59 & 2310 &  63 & 238  \\
  4.44 & 0.33 & -4.79 & 2366 &  78 & 218  \\
  4.39 & 0.30 & -4.89 & 2164 &  84 & 210  \\
  4.36 & 0.28 & -5.00 & 2341 &  97 & 200  \\
  4.31 & 0.25 & -5.15 & 2132 & 112 & 188  \\
  4.27 & 0.22 & -5.29 & 2184 & 124 & 180  \\
  4.22 & 0.20 & -5.50 & 2394 & 136 & 167  \\
  4.20 & 0.19 & -5.70 & 2906 & 138 & 160  \\
  4.18 & 0.18 & -5.89 & 3647 & 139 & 153  \\
  4.16 & 0.18 & -6.09 & 4268 & 140 & 147  \\
  4.15 & 0.17 & -6.29 & 4772 & 140 & 141  \\
  4.13 & 0.16 & -6.50 & 5264 & 140 & 138  \\
	\medskip
  4.12 & 0.16 & -6.70 & 6050 & 140 & 127  \\

		\multicolumn{6}{c}{\textit{WC, $0.1$ solar abundances, except for CNO}} \\
  4.71 & 0.63 & -4.39 & 1724 &  41 & 250  \\
  4.67 & 0.57 & -4.50 & 1689 &  45 & 236  \\
  4.61 & 0.50 & -4.70 & 1643 &  49 & 214  \\
  4.59 & 0.47 & -4.79 & 1562 &  55 & 205  \\
  4.56 & 0.44 & -4.89 & 1524 &  62 & 197  \\
  4.54 & 0.42 & -5.00 & 1460 &  72 & 189  \\
  4.53 & 0.41 & -5.09 & 1281 &  80 & 182  \\
  4.52 & 0.40 & -5.16 & 1223 &  83 & 178  \\
  4.50 & 0.39 & -5.29 & 1268 &  93 & 170  \\
  4.49 & 0.37 & -5.50 & 1612 & 111 & 156  \\
  4.47 & 0.36 & -5.70 & 1290 & 131 & 146  \\
  4.47 & 0.36 & -5.89 & 1705 & 138 & 140  \\
  4.46 & 0.35 & -6.09 & 1978 & 138 & 131  \\
  4.46 & 0.35 & -6.29 & 2005 & 139 & 126  \\
  4.45 & 0.34 & -6.50 & 2022 & 139 & 119  \\
  4.45 & 0.34 & -6.70 & 2248 & 139 & 122  \\
    \hline
  \end{tabular}	
\end{table}

\begin{figure}
  \includegraphics[width=\columnwidth]{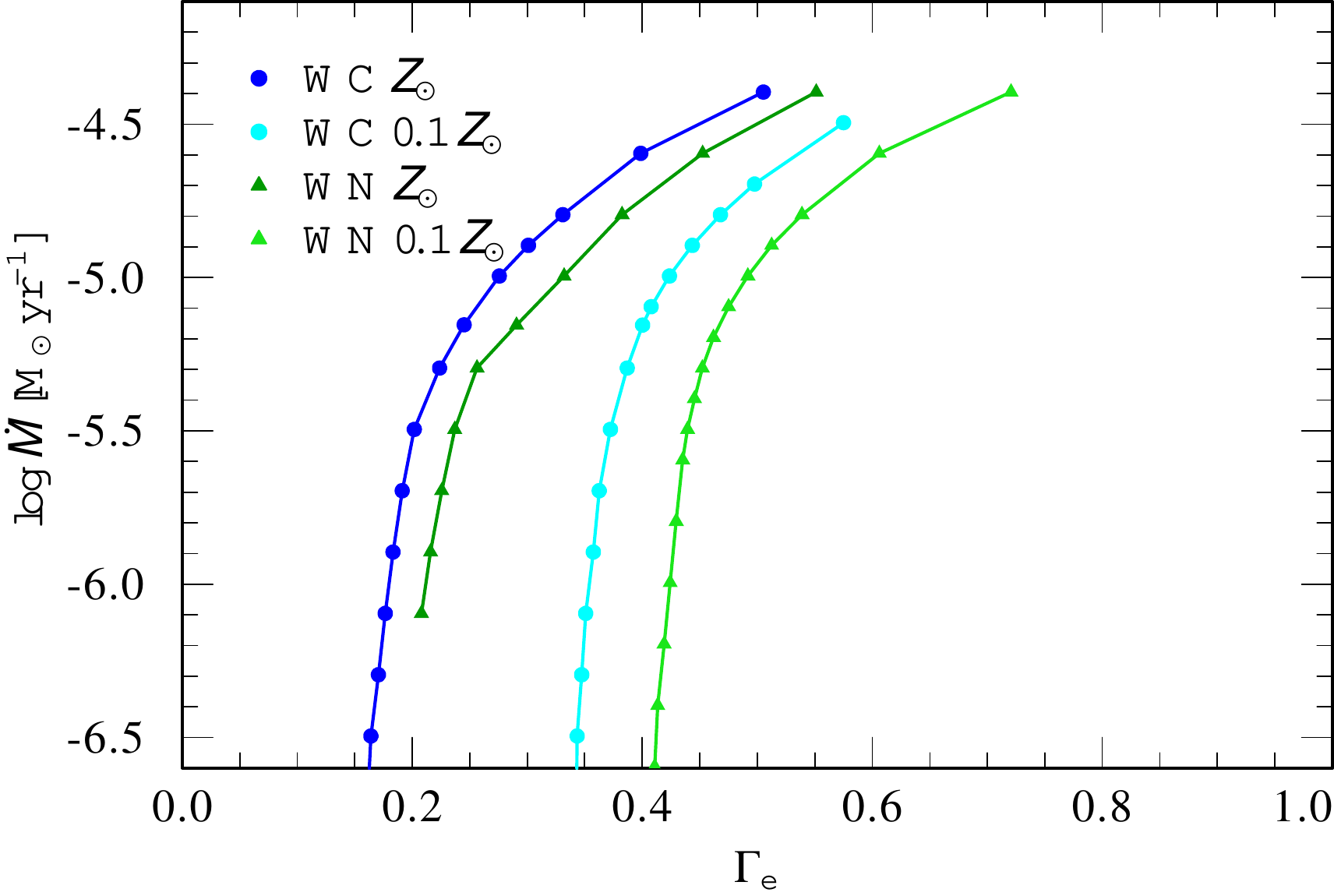}
  \caption{$\dot{M}(\Gamma_\text{e})$ and for different hydrodynamically consistent models at solar metallicity and at $0.1\,Z_\odot$. All models use $T_\ast = 141\,$kK.}
  \label{fig:cmp-mdot-gedd}
\end{figure}

\begin{figure*}
  \includegraphics[width=0.8\textwidth]{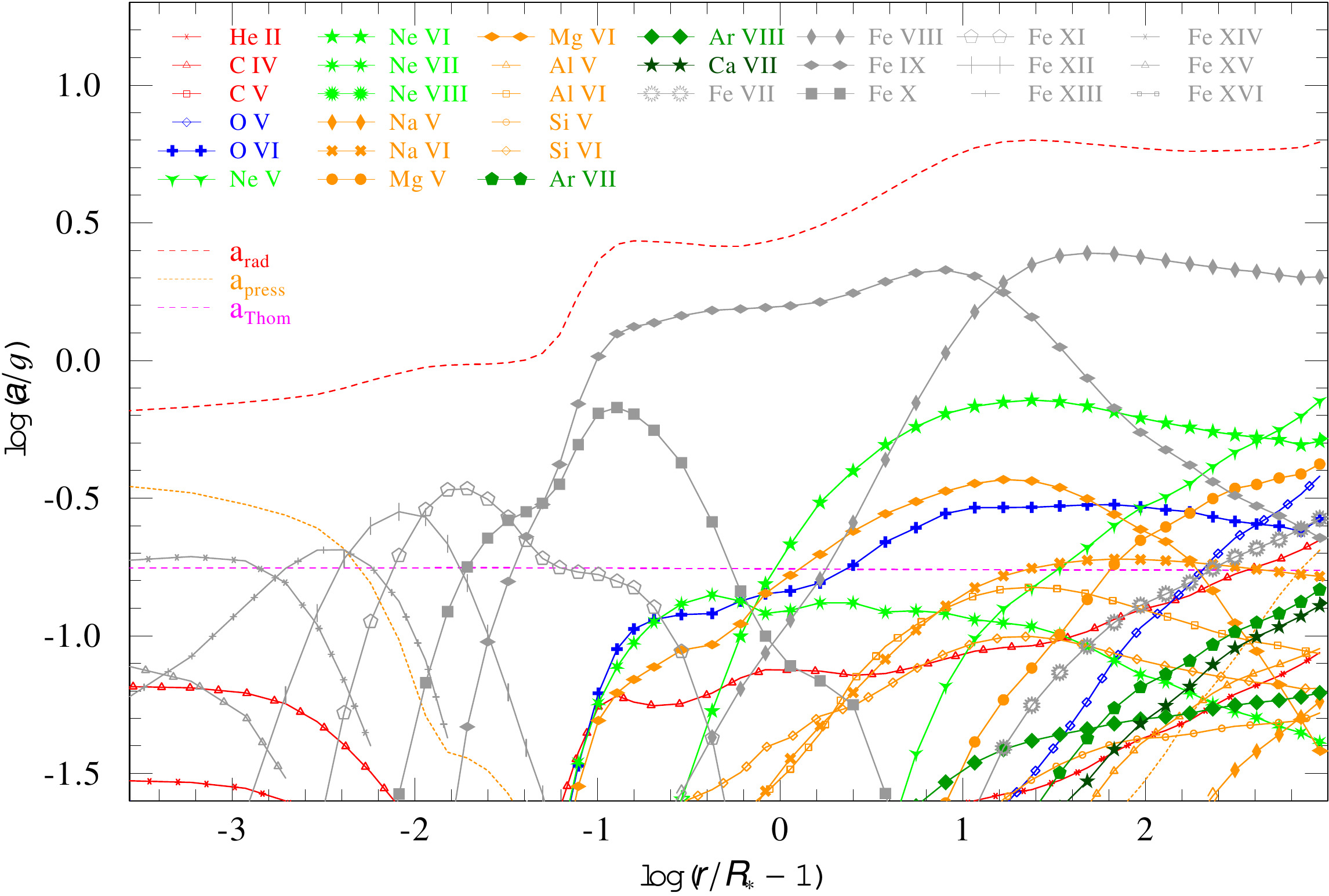}
  \caption{Contributions of the different ions to the radiative acceleration of a hydrodynamically consistent WC model with $T_\ast = 141\,$KK, $\log L/L_\odot = 5.45$, $M_\ast = 19.4\,M_\odot$, and $\log \dot{M} = -6.1$. All contributions with more then $\Gamma_\text{ion} = a_\text{ion}/g > 0.02$ are shown with different ions indicated by a combination of different color and symbol. For comparison, the total radiative acceleration ($a_\text{rad}$), the Thomson acceleration from free electrons, and the contribution from gas (and turbulence) pressure are also shown.}
  \label{fig:leadions-wc-thin}
\end{figure*}

\subsection{Mass loss vs.\ Gamma}
  \label{sec:mdotgamma}
	
For our investigation of $\dot{M}(\Gamma_\text{e})$, we calculated four sequences of hydrodynamically consistent models. For $Z_\odot$ and $0.1\,Z_\odot$, we each calculated a series of WN and WC models. The obtained parameters are listed in Table\,\ref{tab:tabwrhd-mddep} and the results in the $\dot{M}$-$\Gamma_\text{e}$-plane are shown in Fig.\,\ref{fig:cmp-mdot-gedd}. As expected, there is a shift to higher $\Gamma_\text{e}$ for the models with $0.1\,Z_\odot$, while the overall behaviour of the four curves looks very similar. When the winds become thinner, implying lower $\dot{M}$, the gradient of all curves are increasing significantly. The curves at $0.1\,Z_\odot$ are steeper than the ones at $Z_\odot$. 

\subsubsection{Transition to the optically thin wind regime}
  \label{sec:thintrans}

For lower $\dot{M}$, the wind configuration changes significantly as illustrated in Fig.\,\ref{fig:leadions-wc-thin}, where we show the contributing ions for a WC model with a rather thin wind corresponding to a mass-loss rate of $\log \dot{M} = -6.1$. The comparison with Fig.\,\ref{fig:leadions-wne} shows a substantial change of the ionization stages with \ion{Fe}{ix} now being the leading contributor from the onset of the wind far out until transitioning to \ion{Fe}{viii}. The `dip' between the two bumps is essentially gone as \ion{Fe}{viii} is a now a significant driving contributor which was not the case in the thick wind situation. This goes along with a higher electron temperature in the outer regions of the thin wind model.

\begin{figure}
  \includegraphics[width=\columnwidth]{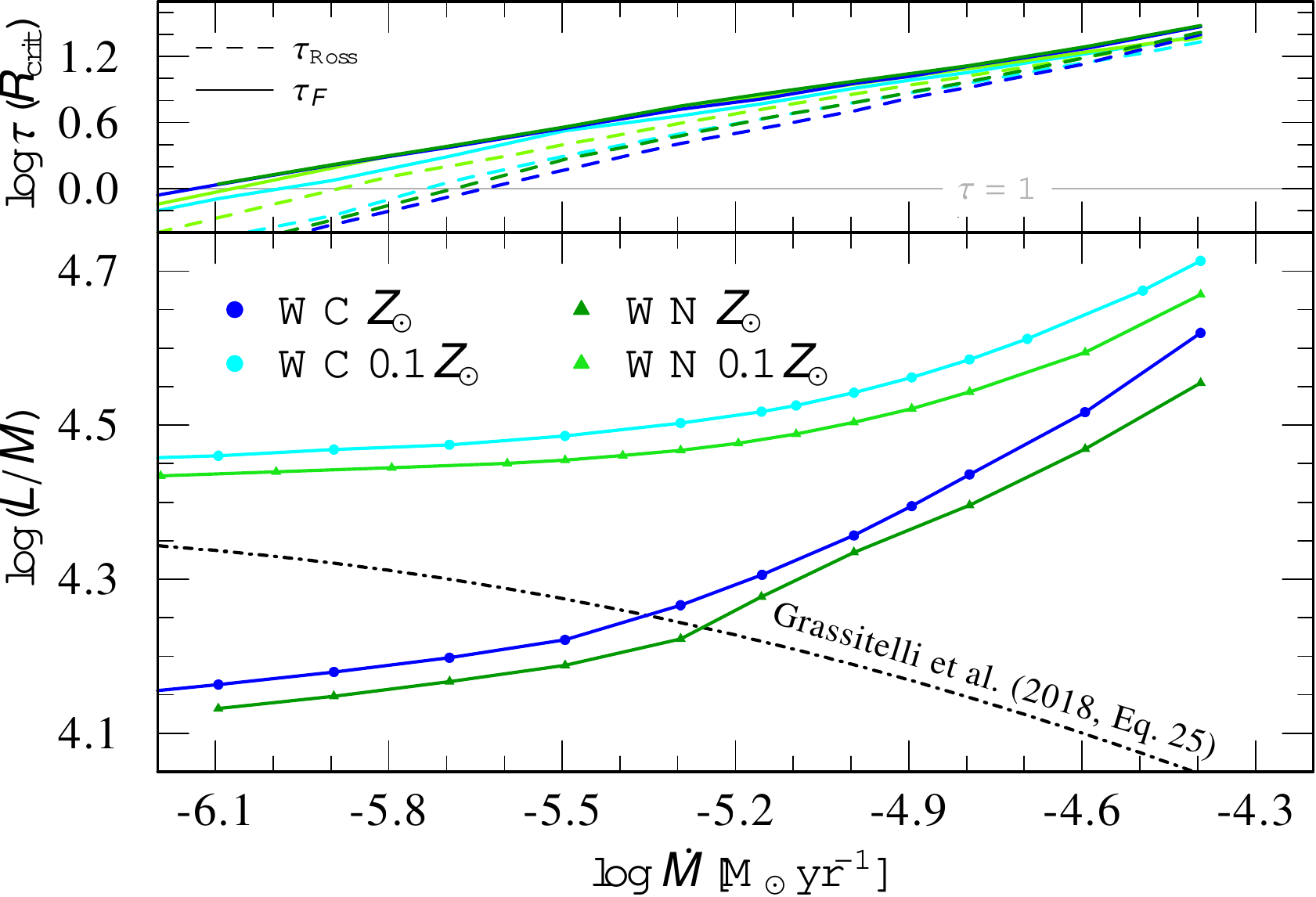}
  \caption{Lower panel: $L/M_\ast$ versus $\dot{M}$ for different hydrodynamically consistent models at solar metallicity and $0.1\,Z_\odot$ (indicated by lighter colors). The prediction for the minimum $L/M_\ast$ for a given $\dot{M}$ by \citet{Grassitelli+2018} in the case of a wind being driven by the hot iron bump at $Z_\odot$ is drawn as a dashed-dotted line.
	         Upper panel: Wind optical depth $\tau_F(R_\text{crit})$ and $\tau_\text{Ross}(R_\text{crit})$ for each model series as a function of the mass-loss rate $\dot{M}$.}
  \label{fig:cmp-mdot-lm}
\end{figure}

\begin{figure}
  \includegraphics[width=\columnwidth]{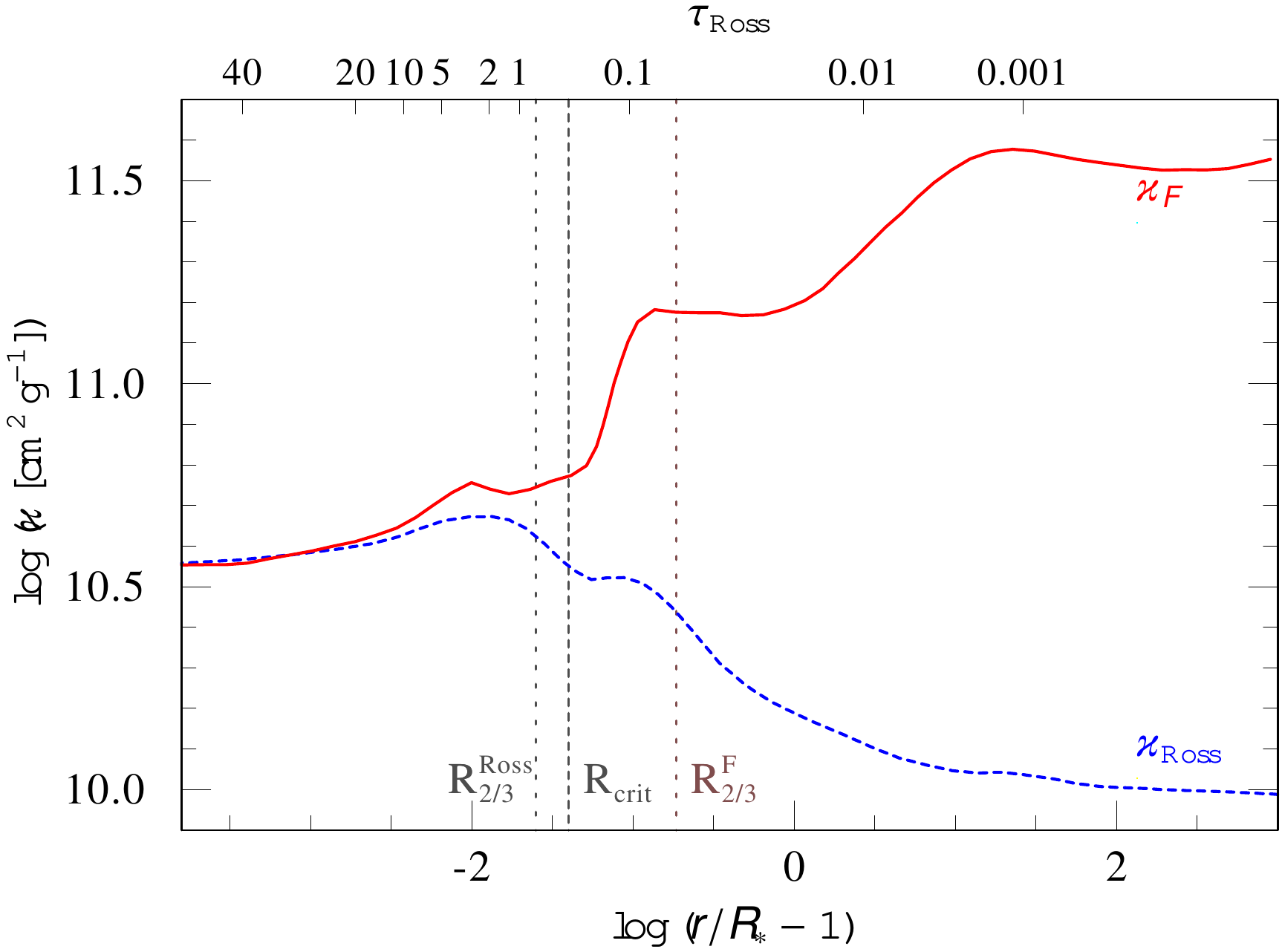}
  \caption{Comparison of the flux-weighted mean opacity (red, solid) with the Rosseland mean opacity (blue, dashed) for the WC model with the driving stratification depicted in Fig.\,\ref{fig:leadions-wc-thin} with $\log \dot{M} = -6.1$ and $Z = Z_\odot$.}
  \label{fig:checkopaross-thin}
\end{figure}

The finding of thin-wind solutions launched at the hot iron bump could be seen as a contradiction of recent predictions based on stellar structure calculations by \citet{Grassitelli+2018} and semi-analytic models based on OPAL tables by \citet{Ro2019}. Both works concluded that below a certain $\dot{M}$ the hot iron bump would lead to an inflation of the envelope while only becoming supersonic way further out. Instead, our WN and WC model sequences at $Z_\odot$ pass smoothly through this precipitated limit as depicted in Fig.\,\ref{fig:cmp-mdot-lm}. An inspection of the model stratifications show that our findings for the model with $\log \dot{M} = -6.1$ are representative for all sequences with a rather smooth transition to different ionization stages in the wind. Given the very flat slope of $\log L/M_\ast$ vs.\ $\log \dot{M}$ at the low $\dot{M}$ end in Fig.\,\ref{fig:cmp-mdot-lm}, however, the observed change in $\dot{M}$ for a minor step in $L/M_\ast$ is quite drastic. Thus, for the values of $L$ and $M_\ast$ realized in nature there could indeed be an abrupt transition in $\dot{M}$.

As shown in the upper panel of Fig.\,\ref{fig:cmp-mdot-lm}, the optical depth of the wind denoted by $\tau_F(R_\text{crit})$ is smoothly decreasing with lower $\dot{M}$, indicating no abrupt regime change. Instead, we obtain an essentially linear relation between $\log \tau_F(R_\text{crit})$ and $\log \dot{M}$.
The regime of the limit predicted by \citet{Grassitelli+2018}, corresponding to $\log{M} \approx -5.3$ to $-5.5$, does not exactly coincide with $\tau_{F}(R_\text{crit})$ passing unity, but there is a switch from \ion{Fe}{vi} to \ion{Fe}{vii} as the leading driver in the outer wind. The transition to an optically thin wind seems to happen rather gradually than abrupt. As illustrated in Fig.\,\ref{fig:checkopaross-thin}, where we show the two mean opacities similar to Fig.\,\ref{fig:checkopaross}, now for our previously discussed thin-wind WC model, the critical point is no longer in the optically thick regime. Therefore the assumption of $\varkappa_\text{Ross} = \varkappa_F$ and thus the use of OPAL opacity tables is no longer valid at the base of the wind, explaining why we obtain solutions in contrast to \citet{Grassitelli+2018} and \citet{Ro2019}. We do not have a comprehensive set of WR models with LMC metallicity in this work, but our findings indicate that if there is no abrupt minimum $\dot{M}$ in terms of hydrodynamics for classical WR stars, the suggested decrease of the clumping factor by \citet{Ro2019} for the early-type WR stars in LMC is probably not needed. Regarding their spectroscopic appearance, all of the models at $\log \dot{M} = -6.1$ still show major optical emission lines, although weaker than usual for early WR subtypes. Nonetheless, they would be spectroscopically classified as WR stars. Only models with a much lower $\log \dot{M} \approx -6.8$ would probably qualify for a transition type classification.

The switch to \ion{Fe}{vii} (and then \ion{Fe}{viii} at even lower $\dot{M}$) as the main driver in the outer wind coincides with the wind -- unlike at higher $\dot{M}$ -- also being no longer optically thick in the EUV range for hundreds of stellar radii. The lower wind densities allow the \ion{He}{ii} ionizing photons to escape, making helium stars with thin winds huge contributors of \ion{He}{ii} ionizing flux with $\log Q_{\ion{He}{ii}} = 48.3$ in the case of our models with $\log L/L_\odot = 5.45$ and $\log \dot{M} = -6.1$. While this mass-loss rate might be too low for the typical WN and WC star in our Galaxy, stripped He stars of lower masses -- e.g. resulting from binary evolution -- will have similar or even lower values of $\Gamma_\text{e}$ and thus provide an enormous amount of hard ionizing flux. Recently, \citet{Goetberg+2018} have calculated a series of models for stripped He stars. Their most massive model assumes $\log \dot{M} = -5.8$ and has $\log L/M_\ast = 4.07$, which is slightly below our calculated regime. Given our results, their assumption for $\dot{M}$ would be too high \citep[see also][]{Vink2017}, meaning that their \ion{He}{ii} ionizing flux, which is about two orders of magnitude lower than their \ion{H}{i} ionizing flux, is still underestimated by an order of magnitude. However, their value for $T_\ast$ is a bit lower than ours and more tailored models will be required to further constrain the wind and ionizing flux situation for stripped helium stars of a lower luminosity (and mass) range.

Given that $T_\ast$ and $L$ are kept fixed in the calculations of our sequences, lower values of $\Gamma_\text{e}$ imply higher masses and thus higher values of $g_\ast = G M_\ast R_\ast^{-2}$. As a consequence, the location of the critical point $R_\text{crit}$ is moving further out as more gravity has to be overcome. Looking back at the slope of $\Gamma_\text{rad}$ for the high-density models (see, e.g., Figs.\,\ref{fig:cmp-arad-clumping}, \ref{fig:aradillu} or \ref{fig:leadions-wne}), one might expect that below a certain $\dot{M}$ we could enter the regime of multiple critical points. This means as $R_\text{crit}$ would eventually `climb up' the hot iron bump, it could lead to regions with $\Gamma_\text{rad} < 1$ further out. In this case one would encounter a situation with $\Gamma_\text{rad} < 1$ in the innermost layers, transitioning to an acceleration region with $\Gamma_\text{rad} > 1$, then followed by a deceleration region with $\Gamma_\text{rad} < 1$, until eventually transitioning back to an accelerating wind with $\Gamma_\text{rad} > 1$. Neither physical, nor mathematical constraints forbid this and it might very well be realized in some WR winds in nature.
Multiple locations with $\Gamma = 1$ imply either an inflated envelope \citep[see e.g.\ the discussion in][]{Grassitelli+2018} or a deceleration region in the wind. Both would coincide with a non-monotonic $\varv(r)$, which can happen in our hydrodynamic solution, but would then be inhibited to avoid problems in the following CMF radiative transfer. Indeed some calculations run into this situation, depending on the chosen start approximation. However, when calculating our models with $T_\ast = 141\,$kK in sequence, always using the last model as a start, we find our previously described solutions with a monotonic $\varv(r)$. While $R_\text{crit}$ gets slightly larger with lower $\dot{M}$, the models are also not (significantly) inflated as the radiative acceleration changes its slope when the wind becomes optically thin. Thus, the whole solution topology remains in a situation with just one critical point. Future research will have to check whether the avoidance of multiple critical points is limited to a narrow parameter range (e.g.\ the selected $T_\ast$- and $L$-regime) or a more generally occurring phenomenon.

\subsubsection{Fit relations for the mass-loss}
  \label{sec:mdotfit}

\begin{figure}
  \includegraphics[width=\columnwidth]{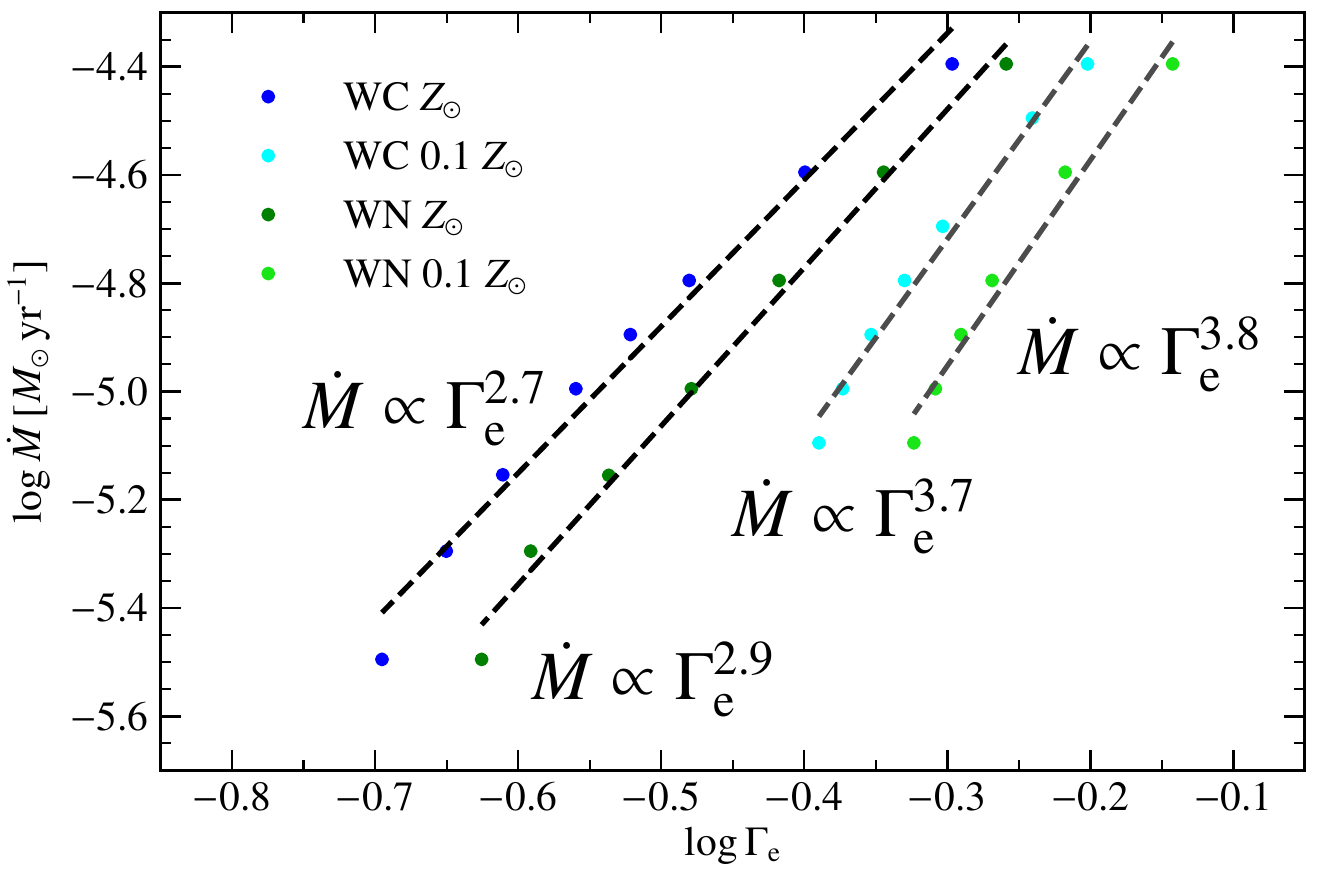}
  \caption{Linear fit to $\log \dot{M}$ versus $\log \Gamma_\text{e}$ for the upper part of the data depicted in Fig.\,\ref{fig:cmp-mdot-gedd}.}
  \label{fig:fit-mdot-gedd-lin}
\end{figure}

\begin{figure}
  \includegraphics[width=\columnwidth]{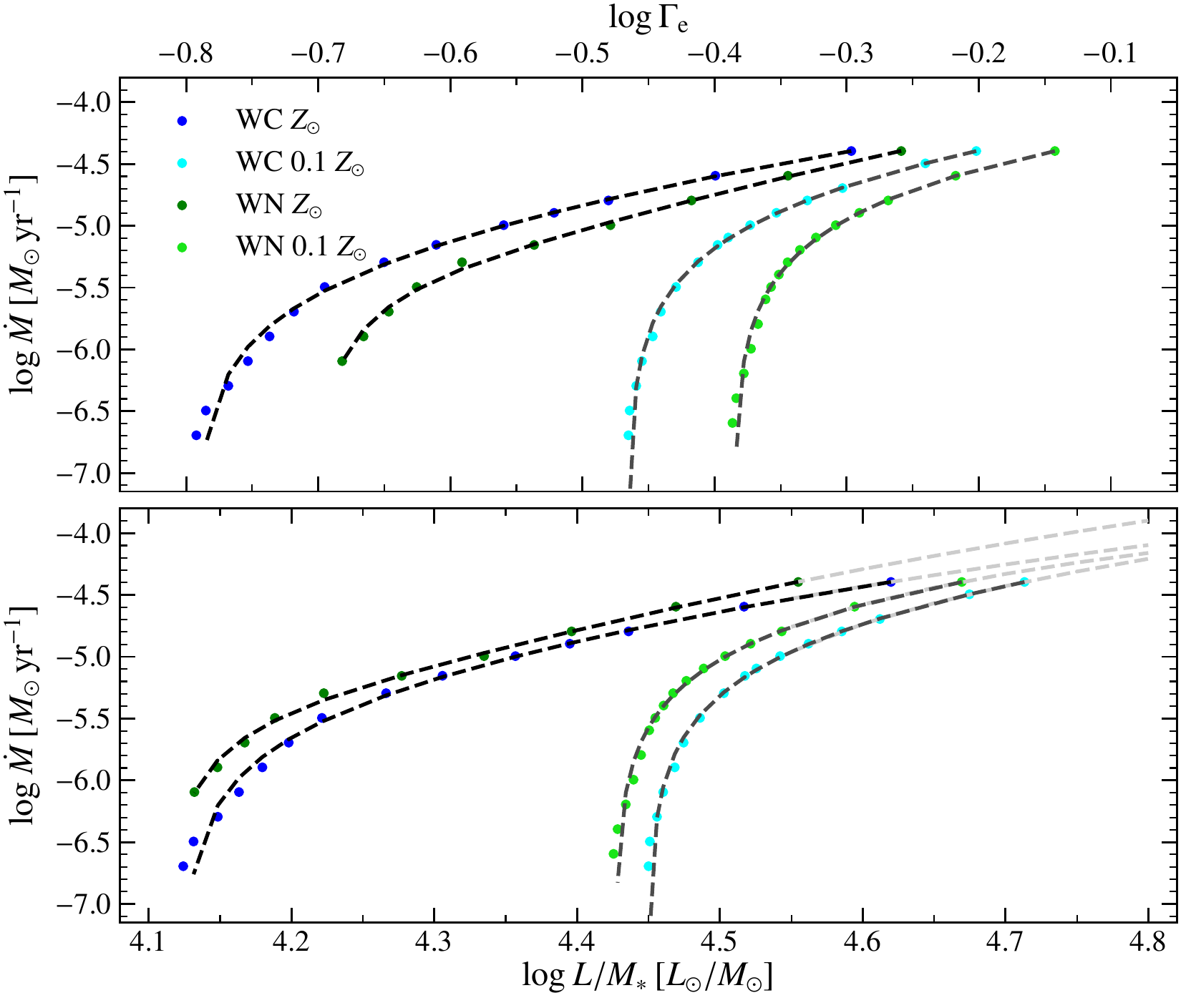}
  \caption{Nonlinear fit to $\log \dot{M}$ versus $\log \Gamma_\text{e}$ (upper panel) and $\log L/M_\ast$ (lower panel) for the WN and WC models from Table\,\ref{tab:tabwrhd-mddep} with extrapolation denoted in grey.}
  \label{fig:fit-mdot-gedd-ldm}
\end{figure}

For the classical WR stars, we can try to derive simple relations of the type $\dot{M} \propto \Gamma_\text{e}^{x}$ by forcing a linear fit in the log-log-Domain. This is depicted in Fig.\,\ref{fig:fit-mdot-gedd-lin}, where we obtain very similar slopes for the WC and WN stars at $Z_\odot$ with the WNs being potentially slightly steeper. For the sample at $0.1\,Z_\odot$, we get steeper slopes than for $Z_\odot$. The data points also clearly indicate that the actual trend for $\dot{M}(\Gamma_\text{e})$ is non-linear and we need to consider higher orders if we want to obtain a recipe for a broader range of mass-loss rates.

The dependency on $\Gamma_\text{e}$ is so steep that a good fit could not be obtained in a double-logarithmic plane. Instead we derived a 3rd-order polynomial representation of $\dot{M}$ vs.\ $\log \Gamma_\text{e}$, i.e.
\begin{equation}
  \dot{M} = a_0 + a_1 \log \Gamma_\text{e} + a_2 \left(\log \Gamma_\text{e}\right)^2 + a_3 \left(\log \Gamma_\text{e}\right)^3
\end{equation}
with the coefficients given in Table\,\ref{tab:mdgamfitparams}. While this formula provides a decent description within the range of covered parameters, simple extrapolations are not encouraged, in particular towards higher $\Gamma_\text{e}$, as illustrated by the gray dashed lines in Fig.\,\ref{fig:fit-mdot-gedd-ldm}. The extrapolated range would predict a crossover of the relations for the WN and WC stars at $Z_\odot$, which is unlikely given the similar wind launching mechanism. As discussed at the beginning of Sect.\,\ref{sec:results}, the higher values of $\dot{M}$ for the WC stars compared to the WN stars of the same $\Gamma_\text{e}$ are an effect of the ionization parameter $q_\text{ion}$, which -- in the inner part -- is $\approx 0.39$ for our WC and $\approx 0.49$ for our WN models.

\begin{table}
  \caption{Analytic fit for the mass-loss rate dependency}
  \label{tab:mdgamfitparams}

  \centering
  \begin{tabular}{ccrrrr}
      \hline
         & \multicolumn{1}{c}{\!$Z$\!}   &  \multicolumn{1}{c}{$a_0$}  &   \multicolumn{1}{c}{$a_1$} &  \multicolumn{1}{c}{$a_2$} &  \multicolumn{1}{c}{$a_3$} \\
				 & \multicolumn{1}{c}{\!$[Z_\odot]$\!} & \multicolumn{1}{c}{\![$10^{-4}$]\!} & \multicolumn{1}{c}{\![$10^{-4}$]\!} & \multicolumn{1}{c}{\![$10^{-4}$]\!} & \multicolumn{1}{c}{\![$10^{-4}$]\!} \\
		  \hline
		      \multicolumn{6}{c}{\emph{$\dot{M}$ vs.\ $\log \Gamma_\text{e}$}}              \\
           WN & $1.0$      & $1.334$ &  $5.293$ &  $7.569$ &  $3.900$  \\       									
           WN & $0.1$      & $0.718$ &  $2.301$ &  $0.261$ & $-2.255$  \\       									
           WC & $1.0$      & $1.166$ &  $3.555$ &  $3.717$ &  $1.373$  \\
      \medskip  WC & $0.1$ & $0.943$ & $3.182$  &  $2.411$ & $-0.156$  \\
		     \multicolumn{6}{c}{\emph{$\dot{M}$ vs.\ $\log L/M_\odot$}}              \\
           WN  & $1.0$    &   $-284.175$       &     $203.817$  &  $-48.819$   &     $3.905$   \\    									
           WN  & $0.1$    &    $247.954$       &    $-157.529$  &   $32.964$   &    $-2.266$   \\								
           WC  & $1.0$    &    $-89.923$       &  	  $66.773$  &  $-16.582$   &     $1.377$   \\							
           WC  & $0.1$    &     $62.349$       &     $-32.002$  &    $4.749$   &    $-0.159$   \\								
      \hline
	\end{tabular}
	
\end{table}

To illustrate the effect of higher mass-loss rates for WC stars compared to WN stars with the same $L$ and $M$, as well as to avoid the need for requiring $q_\text{ion}$ in order to get to the fundamental stellar parameters, we also derive relations for $\dot{M}$ as a function of $\log L/M_\ast$, again using a 3rd-order polynomial, i.e.
\begin{equation}
  \dot{M} = a_0 + a_1 \log \frac{L}{M_\ast} + a_2 \left(\log \frac{L}{M_\ast}\right)^2 + a_3 \left(\log \frac{L}{M_\ast}\right)^3
\end{equation}
with $L$ and $M_\ast$ being given in solar units and $\dot{M}$ in $M_\odot\,\text{yr}^{-1}$.
The coefficients $a_i$ are given in the lower part of Table\,\ref{tab:mdgamfitparams} and the curves are illustrated in the lower panel of Fig.\,\ref{fig:fit-mdot-gedd-ldm}. The relations derived here are a bit more robust with higher orders having smaller coefficients and the extrapolations yield more reasonable results than in the $\Gamma_\text{e}$-description.

\begin{figure}
  \includegraphics[width=\columnwidth]{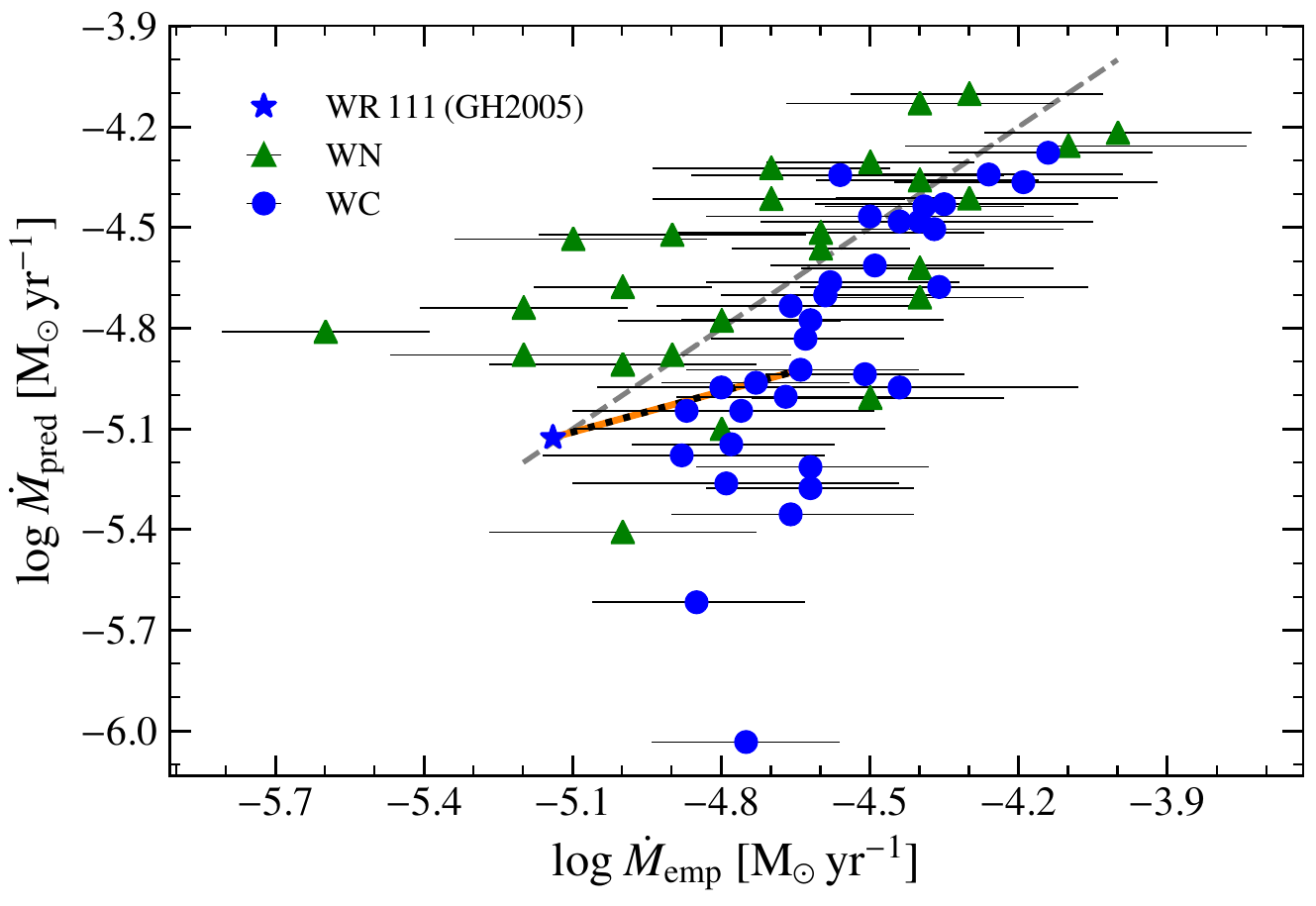}
  \caption{Comparison of the empiricially derived $\dot{M}$ for hydrogen-free WR stars from \citet{Sander+2019} and \citet{Hamann+2019} with predictions for $\dot{M}(L/M_\ast)$ using the coefficients from Tab.\,\ref{tab:mdgamfitparams}. The hydrodynamically consistent modelling of WR\,111 from \citetalias{GH2005} is also shown, using the updated $L$ (and resulting $M_\ast$) based on Gaia DR2 distance in the fit formula. The black dotted line on an orange background connects this result with the the empirical solution from \citet{Sander+2019}.}
  \label{fig:cmp-mdot-emp-pred}
\end{figure}

A first glimpse of how our relations for $\dot{M}(\Gamma_\text{e})$ perform in comparison to observations is shown in Fig.\,\ref{fig:cmp-mdot-emp-pred}. Here, we compare a sample of hydrogen-free WR stars using the recent empirical results from \citet{Hamann+2019} and \citet{Sander+2019} where the luminosities and mass-loss rates have been updated based on Gaia DR2 parallaxes. For comparison, we also show the result for WR\,111 from \citetalias{GH2005}. The value we obtain with our formula and the $L/M_\ast$ based on \citet{Sander+2019} agrees nicely with their tailored result. This is not a surprise as our model series was motivated on their parameter set, but given all the differences in the details it is also a nice sanity check. The remaining sample in Fig.\,\ref{fig:cmp-mdot-emp-pred} is based on purely empirical approaches where simply $\beta$-laws were adopted for the velocity stratification. 

Given the fact that our comparison data stems from a grid analysis, we estimate a systematic uncertainty of $0.15\,$dex -- reflecting the difference between two grid models -- for the empirical mass-loss rates, which we add to the errors from the distance uncertainty. As we see in Fig.\,\ref{fig:cmp-mdot-emp-pred}, even with these rather generous errors, several stars do not touch the identity relation. The two strongest outliers are the weak-lined and peculiar WN3 star WR\,46, which might not be a single star as discussed in \citet{Hamann+2006}, potentially explaining the apparently low empirical $\dot{M}$, and the WC4 star WR\,52, for which we predict an $\dot{M}$ that is an order of magnitude too low. Although not a drastic as for WR\,46, several weak-lined WNs get much higher predictions for $\dot{M}$ than observed, hinting that either our assumptions in this work are not valid for this type of stars, or their lines might be diluted by a yet undetected companion.
WR\,52 has a very high wind efficiency with $\eta \approx 24 $ in \citet{Sander+2019}, more than twice the value that WR\,111 has in the same study. One solution could be that WR\,52 is more evolved and the mass of $8.5\,M_\odot$ deduced from the luminosity in \citet{Sander+2019} is overestimated. However, for the shift of about one magnitude in $\dot{M}$, the mass would need to go down by about a factor two, which is hard to justify. An underestimation of the luminosity would be more plausible, but the rather well constrained parallax in Gaia DR2 also puts doubts on this. Here, and also for the scatter of the other objects, we need to keep in mind that in this work we only have models for one $T_\ast$ and $\dot{M}$ will not only change with $L$, $M$, and $X_i$, but also $T_\ast$. Nonetheless, the overall agreement is promising and hints that at least a good part of the Galactic WR population could be compact He stars with a dense, complex wind structure. Still, the difference between the empirically deduced result for WR\,111 and the tailored hydrodynamical model (illustrated by a black line connecting the data points in Fig.\,\ref{fig:cmp-mdot-emp-pred}) makes it clear that even in this case there is probably no simple `correction formula'. In other words, parameters derived from an HD analysis will differ in varying magnitude between different WR stars, indicating that really a new generation of atmosphere models is needed to properly constrain the stellar parameters of classical WR stars.

\subsection{The velocity profile}
  \label{sec:velo}

\begin{figure*}
  \includegraphics[width=0.32\textwidth]{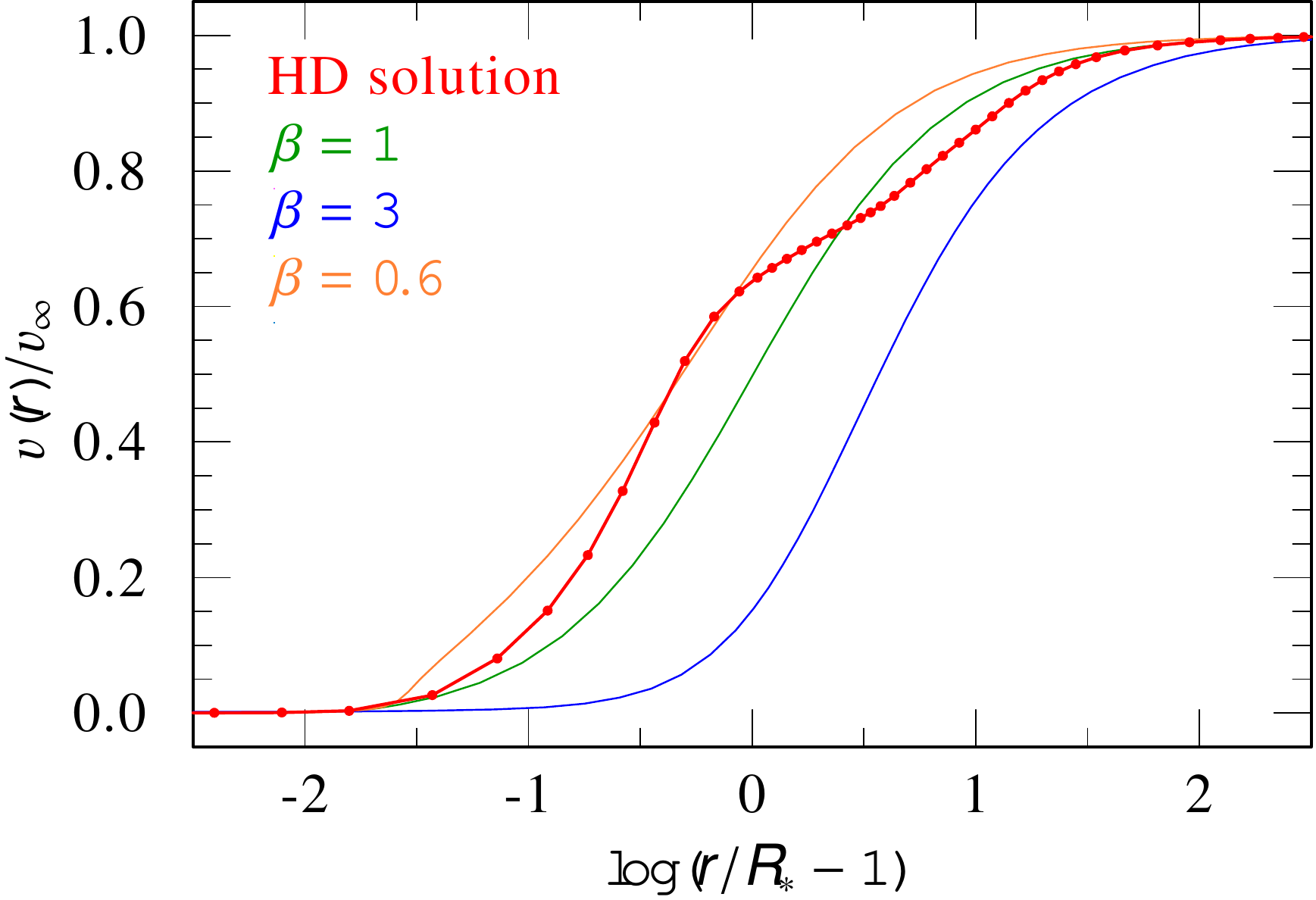} \hfill
  \includegraphics[width=0.32\textwidth]{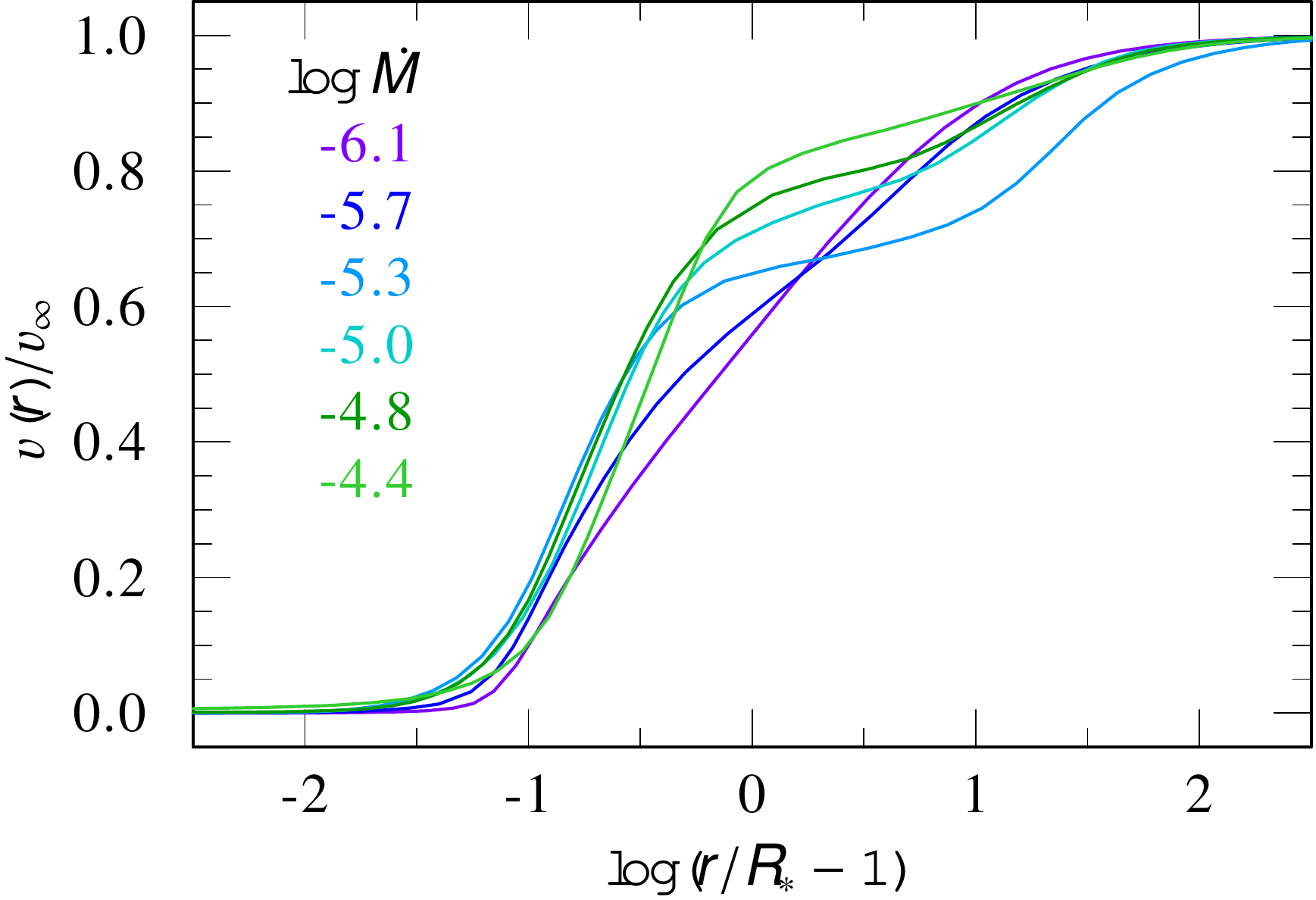} \hfill
  \includegraphics[width=0.32\textwidth]{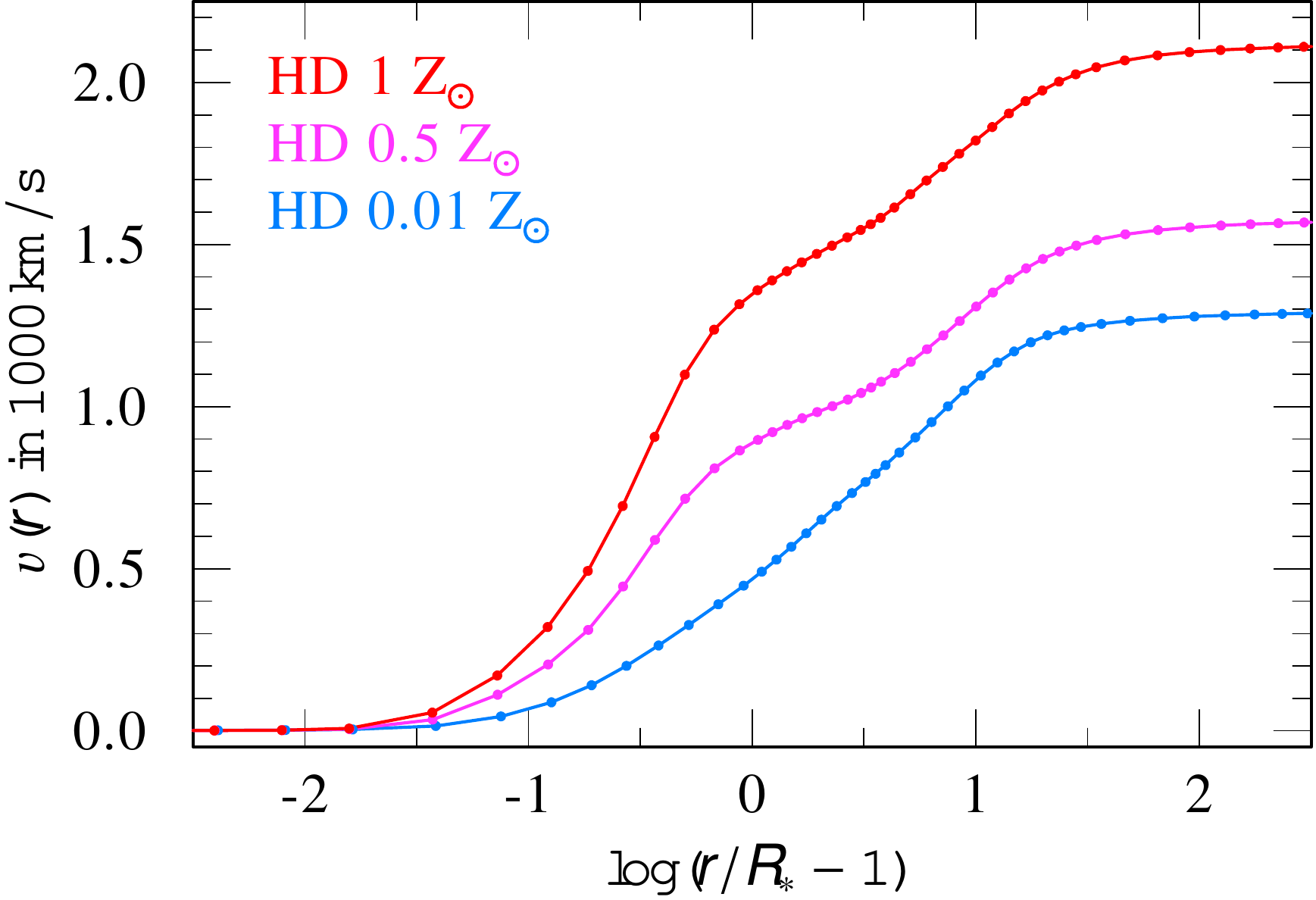}
  \caption{Normalized velocity profile $\varv(r)/\varv_\infty$ for a WC model with $\log \dot{M} = -5.15$ at $Z_\odot$ compared to non-HD models using different $\beta$-laws (left panel) and for WN models with different $\dot{M}$ at $Z_\odot$ (middle panel). The right panel depicts the absolute $\varv(r)$ for WC stars with the same $L$ and $\dot{M}$ at different $Z$}
  \label{fig:cmp-wr-velofields}
\end{figure*}

The velocity profile obtained in our hydrodynamically consistent models reflect the complex shape of $\Gamma_\text{rad}$. An example for a WC model is shown in the left panel of Fig.\,\ref{fig:cmp-wr-velofields}, where we compare the consistent solution to different $\beta$-laws smoothly connected to a hydrostatic part. In the outer wind, the HD solution reprises a $\beta$-law, but the increase in the inner wind is steeper than any $\beta$-type velocity law and the decrease of the radiative acceleration between the two main bumps leads to a bump in $\varv(r)$ as well. These results are in line with the findings of \citetalias{GH2005}, stating that none of the currently used analytic approximations accurately reflect the velocity field of their model for WR\,111. We can now extend this conclusion to the whole regime of classical WR stars, as we illustrate in the central panel of Fig.\,\ref{fig:cmp-wr-velofields}, where the velocity profiles from various WN models with different mass-loss rates are shown. With the exception of the lowest $\dot{M}$, all models show a structure with bumps or plateaus for $\varv(r)$. The closer $\Gamma_\text{rad}$ gets back to unity after the hot iron bump, the more plateau-like the velocity profile will become. Only when the ionization structure changes to the thin-wind situation discussed above in Sect.\,\ref{sec:mdotgamma}, $\varv(r)$ returns to a more standard, $\beta$-like behavior.

\begin{figure}
  \includegraphics[width=\columnwidth]{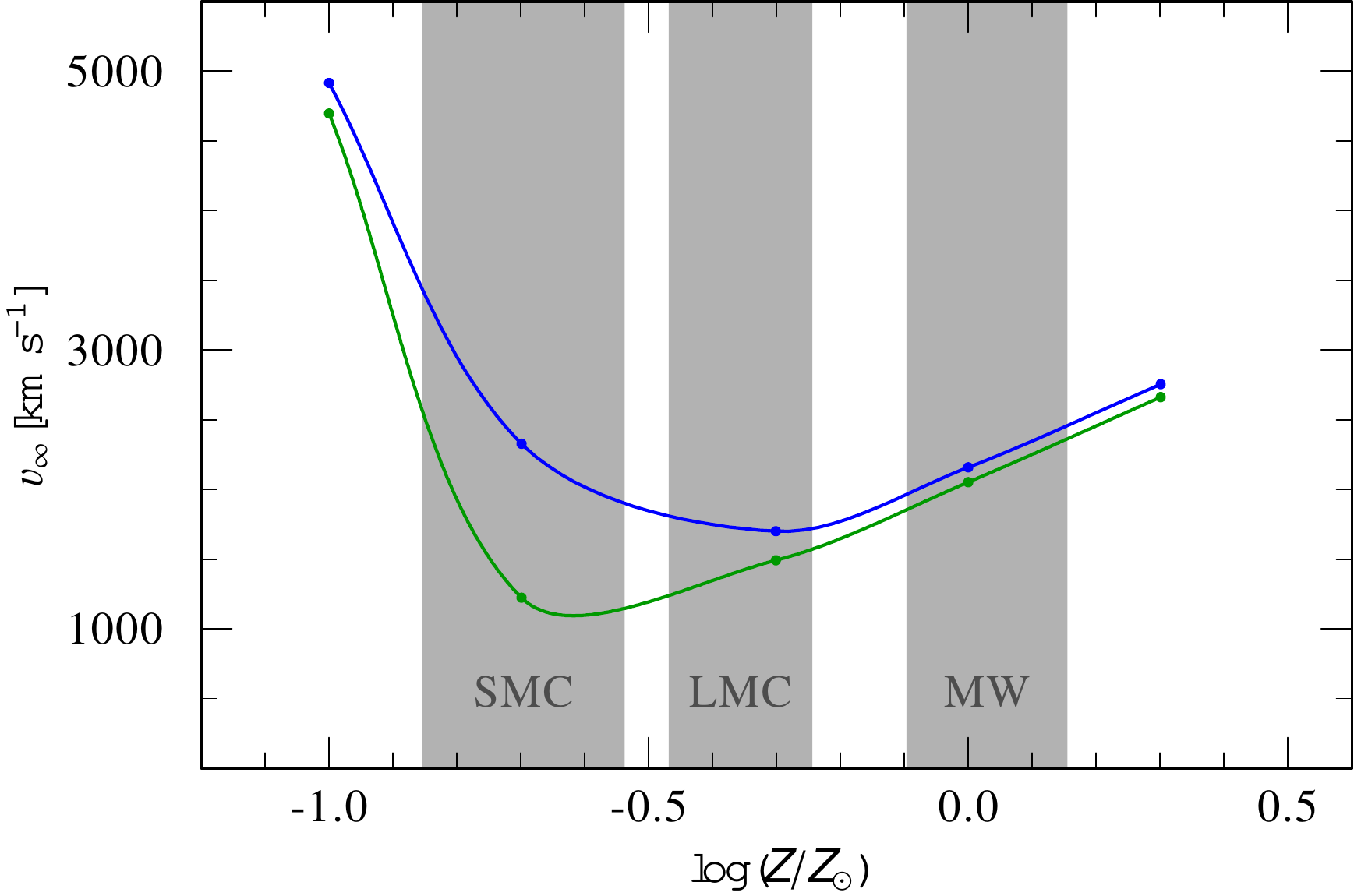}
  \caption{$\varv_\infty(Z)$ for WN (green) and WC (blue) models at different metallicity $Z$ with the same $L$ and $M_\ast$}
  \label{fig:cmp-vinf-z}
\end{figure}

Aside from lower wind densities, the `bumpy' structure of the velocity profile also becomes weaker and eventually vanishes when transitioning to lower metallicities. This is a straight-forward consequence of the lower Fe abundance since the first increase of $\varv(r)$ is caused by the opacities of \ion{Fe}{ix} to \ion{Fe}{xvi} (cf.~Figs.\,\ref{fig:aradelemillu} and \ref{fig:leadions-wne}). For models with the same $\dot{M}$, the imprint of the hot iron bump is noticeable in the velocity structure until $Z < 0.1\,Z_\odot$ (cf.~right panel of Fig.\,\ref{fig:cmp-wr-velofields}). Another consequence of the lower $Z$ is the decrease of $\varv_\infty$, which eventually stalls when the wind acceleration is maintained by continuum driving (cf.~Sect.\,\ref{sec:eddprox}). From the perspective of constant $L/M$, we see a different trend for $\varv_\infty$ in Fig.\,\ref{fig:cmp-vinf-z}. Our test model series with $\log L/L_\odot = 5.45$ and $M_\ast = 11.5\,M_\odot$ (cf.~Sect.\,\ref{sec:mdotz}) predicts only a decrease in $\varv_\infty$ down to LMC metallicities for WCs and down to SMC metallicities for WNs. At even lower $Z$, where $\dot{M}$ drops by orders of magnitude, we find a sharp increase in $\varv_\infty$. With lower Z, less radiative cooling is available, and higher ions such as e.g.\ \ion{Fe}{ix} or \ion{Ne}{vii} stay significantly populated throughout the wind, boosting the acceleration despite their lower abundances. A higher $\varv_\infty$ would also provide a first, qualitative, explanation to why \citet{VdK2005} find a less steep decrease in $\dot{M}$. In their global Monte Carlo approach, they fix $\varv_\infty$ and derive
\begin{equation}
  \label{eq:mdotmc}
  \dot{M} = \frac{2 \Delta L}{\varv_\infty^2 + \varv_\text{esc}^2} 
\end{equation}
with $\Delta L$ being the luminosity removed between the hydrostatic layers and infinity due to the acceleration of the stellar wind. From Eq.\,(\ref{eq:mdotmc}) it is immediately clear that an underestimation of $\varv_\infty$ would lead to an overestimation in $\dot{M}$. A quick scaling shows that this would not be able to account for the full difference, but as the models from \citet{VdK2005} are for late-type WR stars, the quantitative comparison might be biased. Newer calculations with Monte Carlo models based on the approach by \citet{MV2008} have been performed for a series of He star models in \citet{Vink2017} include a prediction of $\varv_\infty$ based on a semi-analytic expression for the line acceleration, which we discussed in Sect.\,\ref{sec:arad}. There are no direct matches in terms of stellar parameters between the sets in \citet{Vink2017} and our $\dot{M}(Z)$ sequence, but we can use their results for the $15\,M_\odot$ star at $0.33\,Z_\odot$ and $0.1\,Z_\odot$ for a rough comparison. At $0.33\,Z_\odot$, the mass-loss rates are lower than in our case, but at $0.1\,Z_\odot$ they are considerably higher. Given that \citet{Vink2017} used a temperature of $50\,$kK in the Monte Carlo calculations, this does not tell much, apart from reflecting that the metallicity trends for helium stars are highly uncertain. This is even more true for the terminal velocities, where none of the sequences in \citet{Vink2017} shows an increase for $\varv_\infty$ at sub-LMC metallicity. Recent models by \citet{Vink2018} for very massive and rather cool ($15\,$kK) stars also showed that $\varv_\infty$ is not simply decreasing with lower $Z$, but their minimum is located at way lower metallicities ($0.03\,Z_\odot$) than in our series.

Given the complex shapes of $\varv(r)$, an in-depth analysis of its behaviour and a physically motivated analytical recipe for $\varv(r)$ is -- despite the high scientific demand, for example for the accurate time-dependent modelling of WR winds -- a whole study of its own, which we have to leave as an important future follow-up.

\section{Conclusions}
  \label{sec:conclusions}
	
In this work we have introduced a series of next-generation stellar atmosphere models for classical Wolf-Rayet stars of WN and WC type. The velocity stratification is obtained by solving the hydrodynamic equation of motion, thus being locally dynamically consistent throughout the whole atmosphere in contrast to previous, e.g. Monte Carlo based, models, with the exception of the prototypical study by \citetalias{GH2005}.
Our examination of the resulting radiative acceleration yields a total breakdown of CAK-like descriptions. Neither the CAK optical depth parameter, nor the force multiplier itself turn out to be monotonic functions of radius or optical depth. Moreover, bound-free and free-free contributions cannot be neglected. 

All winds in our study are launched by the opacities of Fe M-shell ions. The complex ionization structure is imprinted in the velocity profile in the form of `bumps' or plateaus, causing significant deviations from $\beta$-type velocity laws. In line with the prototypical result for WR\,111 from \citetalias{GH2005}, we assume a depth-dependent clumping with $D_\infty = 50$ in the outer wind and derive that neither a $\beta$, nor a $2\beta$-law accurately reprises the derived $\varv(r)$, especially not in the inner part of the wind. At typical mass-loss rates, the winds of classical WR stars can be optically thick with regards to $\tau_\text{Ross}$ for a few stellar radii. The optical depth $\tau_F$ defined by the flux-weighted mean opacity is even larger, reflecting that the wind is `optically thick' with regards to driving. The radius where $\tau_F$ drops below $2/3$ is usually an order of magnitude larger than for $\tau_\text{Ross}$. Our assumed $D_\infty = 50$ is on the high end of what has been empirically derived. Compared to the more conservative $D_\infty = 10$, our $\varv_\infty$ is about $30\%$ higher. $\dot{M}$ however, is set in a region barely affected by $D_\infty$. Thus, we consider our derived trends robust against changes of $D_\infty$, but recommend further studies on the impact of clumping.

We obtain a steep dependence between $\dot{M}$ and $\Gamma_\text{e}$ or $\log L/M$ for both WN and WC stars. Tests show only a weak additional $L$-dependence, making $L/M$ the crucial quantity. For WC stars, larger C or O mass fractions lead to a decrease in $\dot{M}$ for constant $L$ and $M_\ast$. To maintain a certain $\dot{M}$ at lower metallicity, stars have to get closer to the Eddington limit as the line opacities diminish. All types of continuum opacities then take part in maintaining the wind acceleration, more pronounced in the case of WC stars.

Our finding that the optically thick winds of potentially all types of classical WR stars can be launched at the so-called `hot iron bump' matches with the conclusions from recent hydrodynamic stellar structure calculations \citep[][]{Grassitelli+2018} as long as $\tau_F(R_\text{crit}) \gg 1$, paving the way for an agreement between wind and structure models for He stars after decades of severe disagreement. However, we also do find solutions for winds launched below the $\dot{M}$-limit recently suggested on the basis of OPAL opacity tables. 
 For lower $\dot{M}$, the leading ionization in the outer stellar wind switches to higher Fe ions and the winds become optically thin, making the approach of using Rosseland opacities as a proxy for flux-mean opacities invalid. Given the steep slope of our $\dot{M}$-predictions for low $L/M$, we cannot exclude an abrupt transition in the observed properties of He stars and their conditions at the base of the wind. In the larger context, the caveats of hydrostatic structure models and our results derived in this work call for a unified approach where the subsonic stellar structure and the supersonic wind are treated consistently.
We plan to study next-generation models of He stars in a more coherent context in a follow-up study, including their emergent spectra and detailed ionizing fluxes. So far, our results indicate that the presently assumed \ion{He}{ii} ionizing fluxes for stripped He stars might still be underestimated due to an overestimation of their mass-loss rate $\dot{M}$.

We did a first comparison between the $\dot{M}$-predictions based on our new models with empirically derived $\dot{M}$ values using traditional stellar atmospheres with prescribed $\beta$-laws for the wind velocity. We obtain a considerable scatter, illustrating that the present approaches of analysing classical WR stars likely have severe shortcomings with regards to the derived stellar parameters. As their spectra are originating in the rapid wind layers, hydrodynamically consistent models are the only way to break existing parameter degeneracies.

We also calculated a test series of models to study the mass-loss rate of WR stars with constant $\Gamma_\text{e}$ at different $Z$. The results indicate a breakdown of $\dot{M}$ by about two orders of magnitude between $0.2$ and $0.1\,Z_\odot$. This is significantly more than deduced by \citet{VdK2005} and would allow for massive stars below $0.2\,Z_\odot$ to form black holes with higher masses than currently assumed, potentially having considerable effects on gravitational wave statistics.

\section*{Acknowledgements}

The authors would like to thank the referee, L.\ Grassitelli, for fruitful comments and suggestions that helped to improve the manuscript.
A.A.C.S.\ and J.S.V. are supported by STFC funding under grant number ST/R000565/1. W.-R.H.\ acknowledges support from the Deutsche Forschungsgemeinschaft
under grant HA\,1455/26. We further acknowledge helpful discussions with G. Gräfener, L.M. Oskinova, H. Todt, R. Hainich, and T. Shenar.




\bibliographystyle{mnras}
\bibliography{literatur} 



%

\appendix

\section{The breakdown of CAK-like descriptions for WR stars}
  \label{asec:cakfail}

Theoretical modeling of the winds of hot massive O and WR stars has a long history. Following up on the suggestion
that radiation intercepted by spectral lines could be sufficient to explain the driving of hot star winds \citep[][]{LS1970},
the path-breaking work in this area was the CAK paper \citep*[after][]{Castor+1975} which developed 
the so-called CAK force-multiplier framework, describing the line force as an amplification of the electron
scattering (Thomson) force by a force multiplier $\mathcal{M}$, involving both a line-strength parameter $k$, and 
a parameter $\alpha$ which reflects the distribution of optically thin to thick lines \citep[e.g.][]{Puls+2000}.

Neglecting bound-free and free-free contributions, the radiative acceleration in (m)CAK is written as
\begin{equation}
  \Gamma_\text{rad}^\textsc{cak} = \left(1 + \mathcal{M}\right) \Gamma_\text{e}\text{.}
\end{equation}
with the force multiplier $\mathcal{M}$ (cf.~Eq.\,\ref{eq:fmdef}) parametrized in the form
 \begin{equation}
  \label{eq:fmcak}
  \mathcal{M}(t) = \hat{k} t^{-\alpha} \hat{n}^{\delta} 
\end{equation}
with the definitions
\begin{align}
        t := & n_\text{e}(r) \sigma_\text{e} \varv_\text{th,H} \left(\frac{\mathrm{d}\varv}{\mathrm{d}r}\right)^{-1}\text{,}\\
	\hat{n} := & \frac{n_\text{e}(r)}{10^{-11}\,\text{cm}^{-3}}\frac{1}{W(r)}\text{,} \mbox{\hspace{1cm}} \text{and} \\
	   W(r) := & \frac{1}{2} \left( 1 - \sqrt{ 1 - \left(\frac{R_\ast}{r}\right)^2 } \right)\text{.}
\end{align}
All quantities in Eq.\,(\ref{eq:fmcak}) are dimensionless and ideally, $\alpha$ and $\delta$ should be constant, at least throughout the wind. It is known for decades already that the latter is not the case, even in OB supergiants \citep[e.g.][]{Kudritzki+1998,Vink2000Thesis,Kudritzki2002,Muijres+2012}. When plotting e.g. $\log \mathcal{M}$ versus $\log t$, the resulting curve is not a straight line, but instead requires a non-linear fit. While this makes the formula slightly less elegant, it still provides an efficient and fast way to properly describe the radiative force. 

With our detailed calculation of the radiative acceleration in the CMF, we can test the validity of a description like in Eq.\,(\ref{eq:fmcak}) using the previous definitions and the obtained values for $\mathcal{M}$ via Eq.\,(\ref{eq:fmdef}). The parameters $\alpha$ and $\delta$ can then be obtained via
\begin{align}
  \alpha(r) &= - \frac{\partial \left(\log \mathcal{M}\right)}{\partial \left(\log t\right)} & \text{and} & & \delta(r) &= - \frac{\partial \left(\log \mathcal{M}\right)}{\partial \left(\log \hat{n}\right)}
\end{align}
with $\hat{k}(r)$ then being determined via the obtained quantities and Eq.\,(\ref{eq:fmcak}). Such defined quantities are termed `effective' force multiplier parameters as they are not based on an underlying line-strength distribution \citep[cf.][]{Puls+2000}. The original CAK concept treats $\hat{k}$, $\alpha$, and $\delta$ as constant throughout the wind. This was relaxed in later extensions \citep[e.g.][]{Kudritzki2002} to account for the fact that no lines with infinite strengths exist and thus $\mathcal{M}(t)$ eventually saturates. Since this depth-dependence can already be important for O stars, we do expect a depth-dependency for these parameters for WR stars, as already hinted by the results from \citetalias{GH2005}.

\begin{figure}
  \includegraphics[width=\columnwidth]{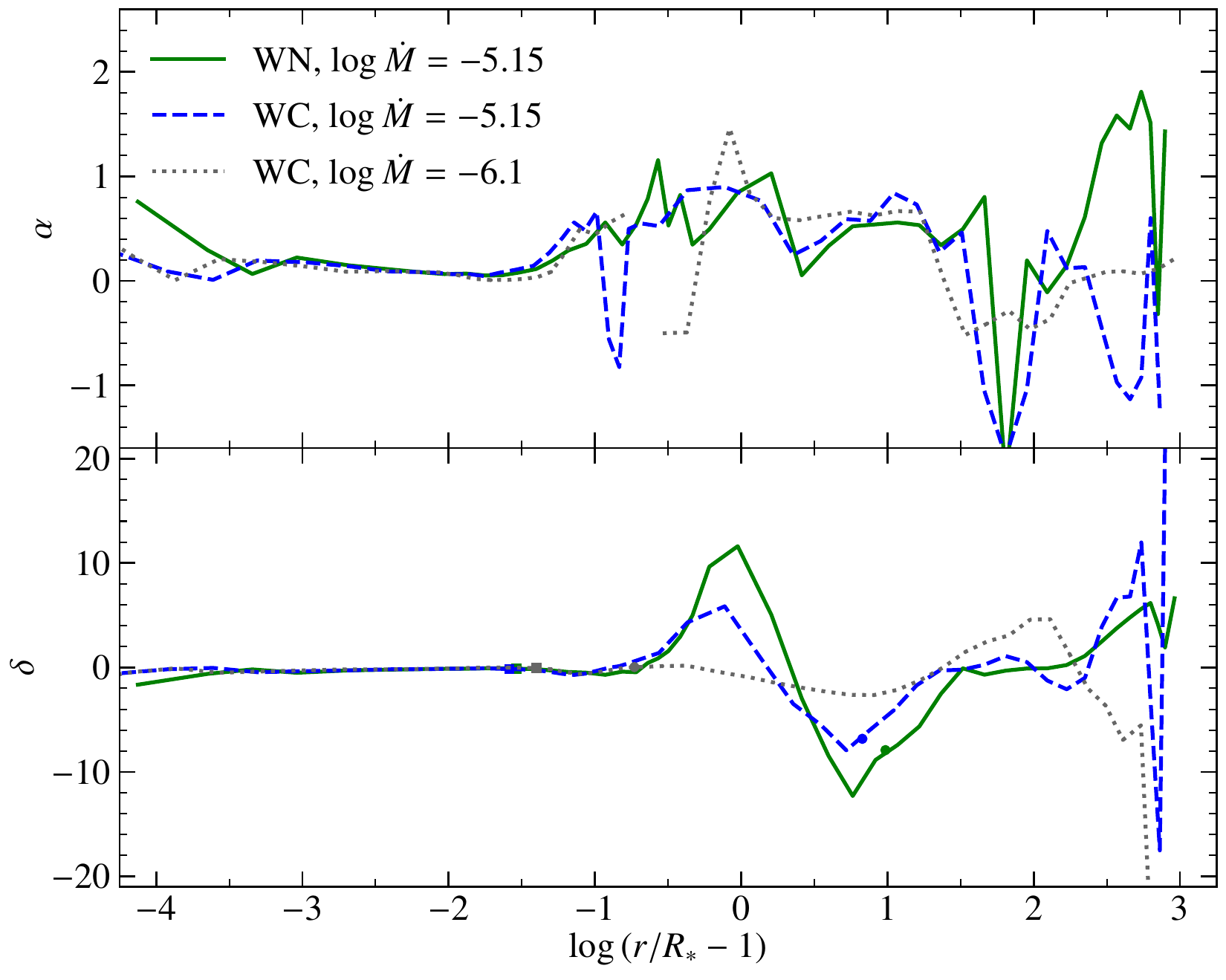}
	\vspace{-1em}
  \caption{Effective force multiplier parameters $\alpha$ (upper panel) and $\delta$ (lower panel) as a function of radius for three different WR models (see Table\,\ref{tab:wcwnesolcmp} for detailed parameters). The squares denote the location of the critical point in the models while the dots mark $\tau_\text{F} = 2/3$.}
  \label{fig:fmpara-alpha-delta-r}
\end{figure}

The obtained behavior of $\alpha(r)$ and $\delta(r)$ for typcial Galactic WN and WC models are depicted in Fig.\,\ref{fig:fmpara-alpha-delta-r}. All curves in both graphs look very different from what is obtained for OB-type stars. In particular the part outwards from $R_\text{crit}$ shows wild jumps for $\alpha$. The $\delta$ parameter reflecting the ionization change has a smoother shape, but with a huge positive and negative amplitudes it by far exceeds the region of $0.0\dots0.2$ known in usual mCAK-descriptions.

\begin{figure}
  \includegraphics[width=\columnwidth]{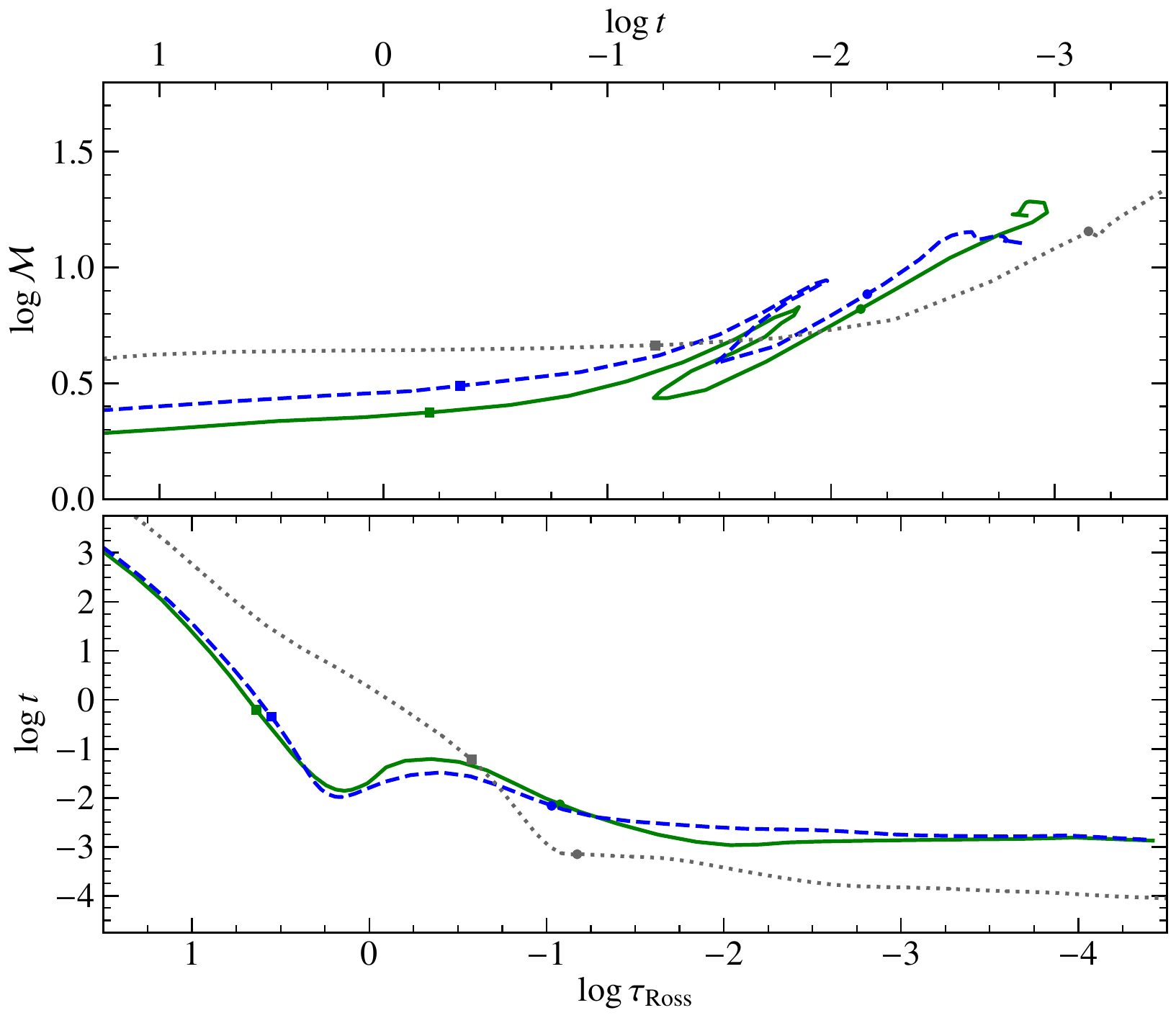}
	\vspace{-1em}
  \caption{Upper panel: Force multiplier $\mathcal{M}$ versus the Sobolev optical depth parameter $t$.
	         Lower panel: Sobolev optical depth parameter $t$ as a function of the Rosseland optical depth. (Same models as in Fig.\,\ref{fig:fmpara-alpha-delta-r})}
  \label{fig:fmpara-m-t-taur}
\end{figure}

A close inspection of the Sobolev optical depth parameter $t$ in Fig.\,\ref{fig:fmpara-m-t-taur} reveals that $t$ is not a monotonic representation of the optical depth scale (here $\tau_\text{Ross}$) and thus $\mathcal{M}(t)$ is not a proper mathematical function, leading to the complete breakdown of the CAK theory. For a lower $\log \dot{M} = -6.1$, we find a monotonic relation. However, this does not lead to simple slopes for $\alpha$ and $\delta$ as in mCAK, but instead only reduces the amplitude of the scatter.

\begin{figure}
  \includegraphics[width=\columnwidth]{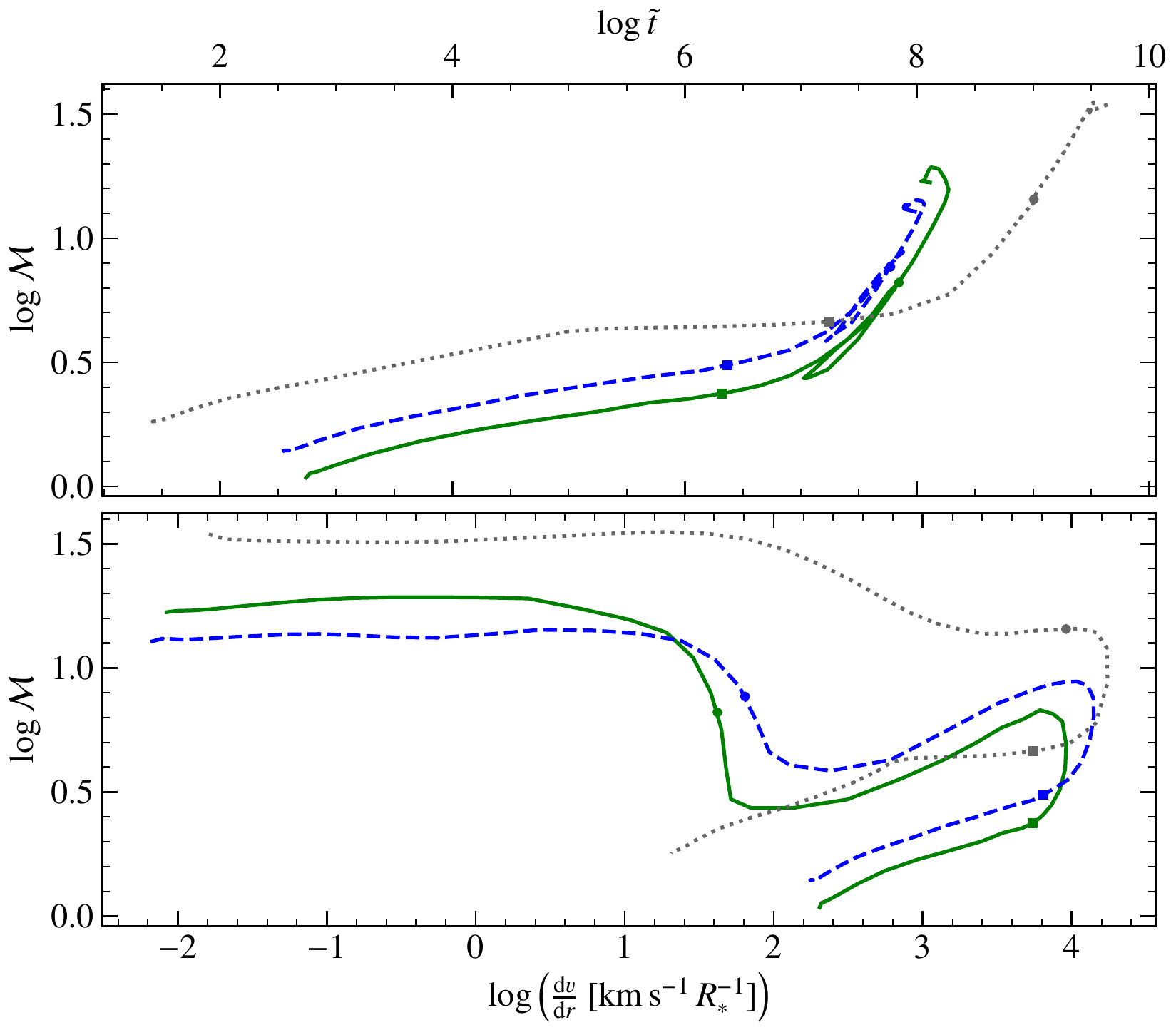}
	\vspace{-1em}
  \caption{Force multiplier $\mathcal{M}$ versus the depth parameter $\tilde{t}$ from \citetalias{GH2005} (upper panel) and velocity gradient (lower panel) for the three different WR models specified in Table\,\ref{tab:wcwnesolcmp}.}
  \label{fig:fmpara-m-tgoetz-dvdr}
\end{figure}

Alternatively to a straight-forward mCAK description, one could try to apply a general term containing the density and the velocity gradient following \citetalias{GH2005}, i.e.
\begin{equation}
  \mathcal{M} \propto \left(\frac{1}{\rho}\frac{\mathrm{d}\varv}{\mathrm{d}r}\right)^{\tilde{\alpha}} \equiv \tilde{t}^{~\tilde{\alpha}}\text{.}
\end{equation}
The resulting curve for $\mathcal{M}(\tilde{t})$ is shown in the upper panel of Fig.\,\ref{fig:fmpara-m-tgoetz-dvdr}, revealing a non-monotonic behaviour of $\tilde{t}$ for dense WR winds, similar to what we obtained for the mCAK description. More fundamentally, the force multiplier $\mathcal{M}$ itself is not a monotonically increasing function, not even over the radius, not even in the outer wind where WR atmospheres behave more like OB atmospheres. In fact, we even see this in OB models, such as the HD calculations for $\zeta$\,Pup \citep{Sander+2017} and Vela X-1 \citep{Sander+2018}. It is also hinted in the slope of $\Gamma_\text{rad}$ shown in our demonstration model from \citet{Sander+2015}. While the $\beta$-law implies that $\Gamma_\text{rad}$ asymptotically reaches a maximum value, the CMF calculations show that $\Gamma_\text{rad}$ can decrease in the outer wind. This is more subtle in the OB models, but quite prominent in the WR models presented in this work. Plotting the force multiplier versus the velocity gradient in the lower panel of Fig.\,\ref{fig:fmpara-m-tgoetz-dvdr} reveals how complex the interplay between the two quantities really is. Even if we account only for the section outwards of $\tau_F = 2/3$, i.e. the area we concern as optically thin for driving, $\log \mathcal{M}$ is clearly a non-monotonic function of $\log \frac{\mathrm{d}\varv}{\mathrm{d}r}$, thus spoiling any mCAK description of the radiative force.

\section{The challenge to describe the radiative acceleration}
  \label{asec:aradfit}
	
In this section we illustrate the complexity of deriving an analytical approximation of the radiative acceleration for cWR stars using our WNE model from Table\,\ref{tab:wcwnesolcmp}. We aim at reproducing the general shape of $\Gamma_\text{rad}(r)$ as a kind of blueprint for dealing with WR-type winds in situations where a detailed CMF-based calculation is not feasible, e.g. due to computational limitations in multi-dimensional or time-dependent models. Caution is advised when adopting these recipes as their parameters depend significantly on the individual stellar parameters such as $T_\ast$, $L$, $M$, and the chemical composition.

In order to not be too dependent on the particular dimensions of our archetypical model, we do not fit $\Gamma_\text{rad}$ directly on the standard radius scale, but instead introduce the dimensionless quantity
\begin{equation}
  \label{eq:dimlessx}
	x := 1 - \frac{R_\ast}{r}
\end{equation}
which is well known as the kind of `radikand' from the $\beta$-velocity law (i.e. $\varv(x) = \varv_\infty x^\beta$). It essentially maps our interval $r \in \left[R_\ast,\infty\right[$ to $x \in \left[0,1\right[$. We then fit $\Gamma_\text{rad}(x)$ with a series of shifted Legendre polynomials, which provide an orthogonal basis on our interval. While \citet{KK2011} used this technique to constrain an improved description for the wind velocity profile of O-type stars (including central stars of planetary nebulae), we apply it here to the radiative force itself. Thus we can write $\Gamma_\text{rad}$ as a series
\begin{equation}
  \Gamma_\text{rad}(x(r)) = \sum\limits_{i=0}^{n}\gamma_i \tilde{P}_i(x(r)) 
\end{equation}
with the shifted Legendre polynomials $\tilde{P}_i(x)$ and the coefficients $\gamma_i$. The shifted Legendre Polynomials follow from the ordinary Legendre Polynomials $P_i(x)$ via $\tilde{P}_i(x) = P_i (2x-1)$.

\begin{figure}
  \includegraphics[width=\columnwidth]{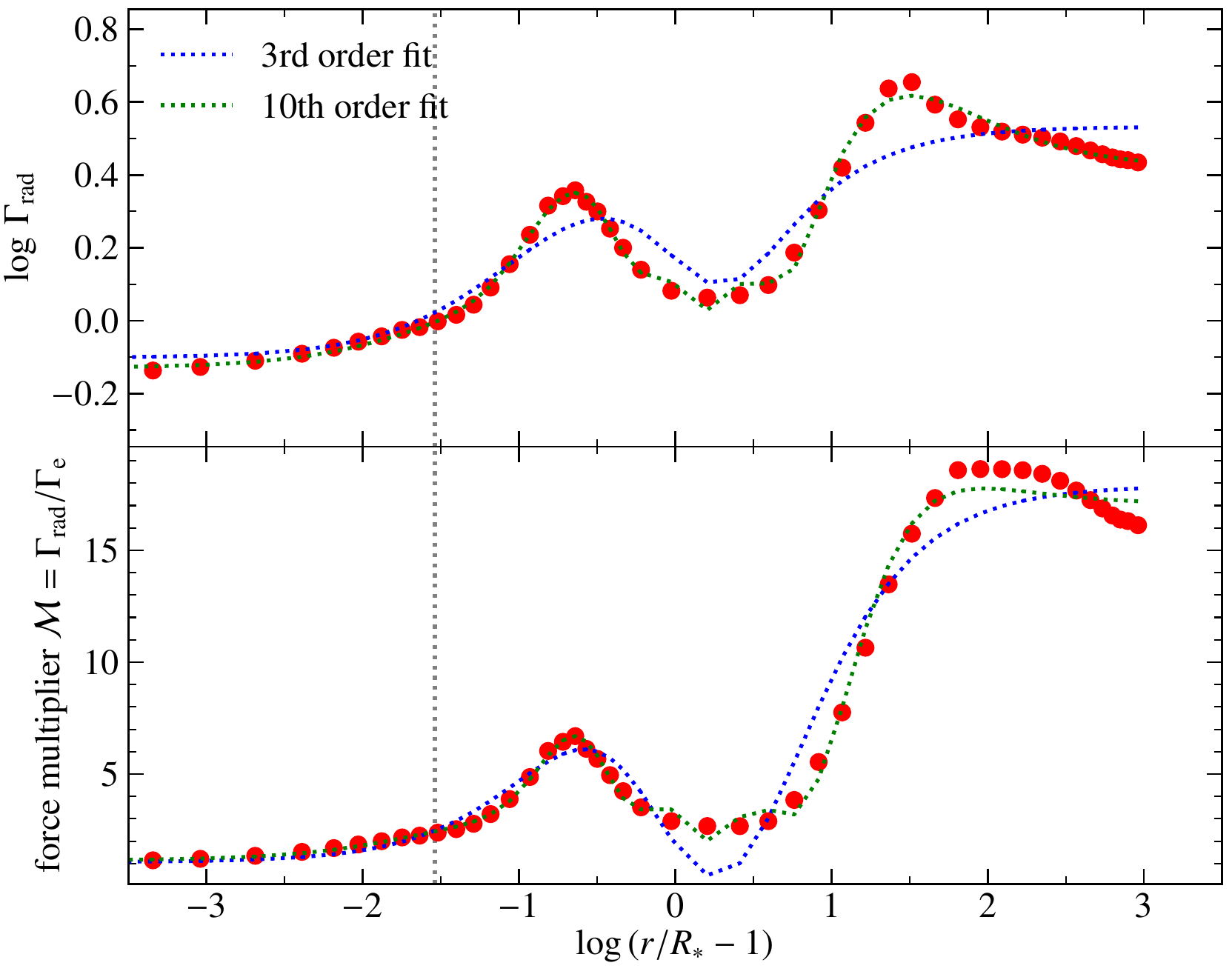}
  \vspace{-1em}\caption{Stratification for $\Gamma_\text{rad}$ (upper panel) and the force multiplier $\mathcal{M}$ (lower panel) in the consistent WN model at $Z_\odot$ (see Table\,\ref{tab:wcwnesolcmp} for detailed parameters). In each panel, two fits with shifted Legendre Polynomials are shown, giving either a rough representation of the curve or a detailed fit.}
  \label{fig:gammarad-fm-fit}
\end{figure}

\begin{table}
  \caption{Analytic fit for the radiative acceleration and the force multiplier}
  \label{tab:radfitparams}

  \centering
  \begin{tabular}{lcccc}
      \hline
                             & \multicolumn{2}{c}{$\Gamma_\text{rad}$} & \multicolumn{2}{c}{$\mathcal{M}$} \\
                             &  shape & detailed & shape & detailed \\[0.5mm]
      \hline		
		 $\gamma_0$    &   $1.6860$   &   $1.6397$   &   $4.3956$   &    $4.5850$    \\
		 $\gamma_1$    &   $0.3206$   &   $0.3596$   &   $1.5917$   &    $1.6429$    \\
		 $\gamma_2$    &   $0.4139$   &   $0.8629$   &   $5.0803$   &    $3.9346$    \\
		 $\gamma_3$    &   $0.9880$   &   $1.3064$   &   $6.8170$   &    $5.7556$    \\
		 $\gamma_4$    &              &  $-0.0481$   &              &    $1.2600$    \\
		 $\gamma_5$    &              &   $0.4051$   &              &    $2.8576$    \\
		 $\gamma_6$    &              &  $-0.0570$   &              &    $1.1911$    \\
		 $\gamma_7$    &              &  $-0.5899$   &              &   $-0.5528$    \\
		 $\gamma_8$    &              &  $-0.2478$   &              &    $0.2333$    \\
		 $\gamma_9$    &              &  $-0.5414$   &              &   $-1.7546$    \\
		 $\gamma_{10}$ &              &  $-0.4671$   &              &   $-2.1131$    \\
    \hline
  \end{tabular}
\end{table}

Instead of fitting $\Gamma_\text{rad}$, one could also take a look at the so-called \emph{force multiplier} introduced by \citet{Castor+1975}
\begin{equation}
  \label{eq:fmdef}
	\mathcal{M} := \frac{\Gamma_\text{lines}}{\Gamma_\text{e}} = \frac{\Gamma_\text{rad} - \Gamma_\text{cont}}{\Gamma_\text{e}}\text{,}
\end{equation}
describing the effect of line-driving as a depth-dependent `boost factor' to $\Gamma_\text{e}$. Stratifications for both, $\Gamma_\text{rad}$ and $\mathcal{M}$ are shown in Fig.\,\ref{fig:gammarad-fm-fit} including fits with shifted Legendre Polynomials. It becomes clear that the stratifcations are quite complex and a high number of orders is required to successfully reproduce the shapes. The 10th order fits given in the figures have coefficients listed in Table\,\ref{tab:radfitparams}. In addition, a much simpler 4th-order representation is provided as well. While this does not represent the precise shape of the curves, it is the lowest order fit that contains all the major features of a WR-like behavior and could be useful for qualitative studies.

\begin{figure}
  \includegraphics[width=\columnwidth]{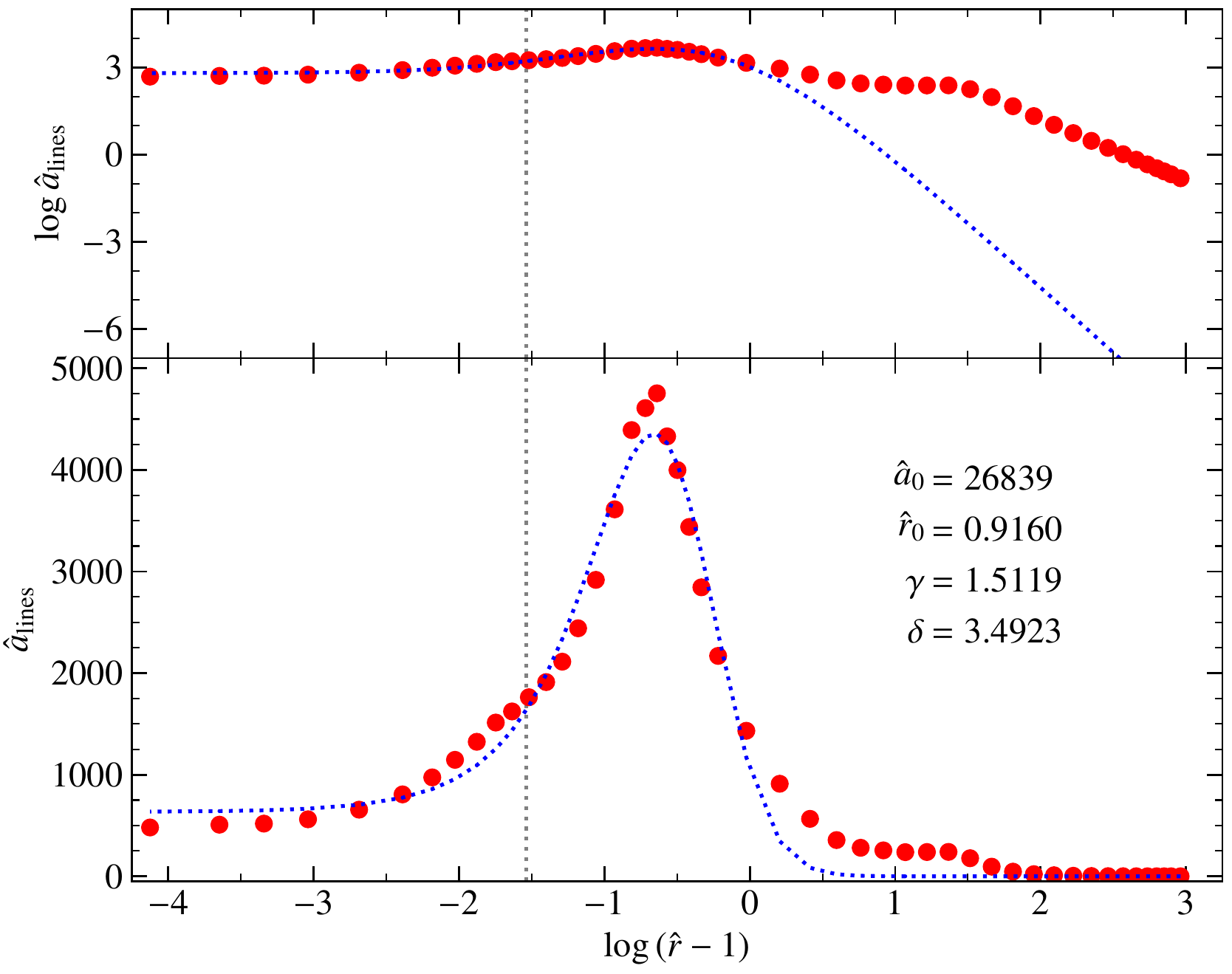}
	\vspace{-1em}
  \caption{ Line acceleration of the WN model from Table\,\ref{tab:wcwnesolcmp} in the dimensionless from suggested by \citet{MV2008}, shown on linear (lower panel) and logarithmic (upper panel) scale. The best fit according to Eq.\,(\ref{eq:alinemv}) is shown as a blue dashed line, compared to the HD model in red.}
  \label{fig:aradmvfit}
\end{figure}

Using the Monte Carlo approach, \citet{MV2008} introduced a semi-analytical description of the radiative acceleration that is more physically motivated than a polynomial. \citep[See Sect.\,2.4 in][for a detailed discussion of the motivation.]{MV2008} Their description for the line contribution has the form
\begin{equation}
  \label{eq:alinemv}
	\hat{a}_\text{lines}(\hat{r}) = \frac{\hat{a}_0}{r^{1+\delta}} \left(1 - \frac{\hat{r}_0}{\hat{r}^\delta}\right)^\gamma
\end{equation}
with the four free parameters $\hat{a}_0$, $\hat{r}_0$, $\gamma$, and $\delta$. $\hat{a}_\text{lines} = a_\text{lines}\,R_\ast \,a_\text{s}^{-2}$ is the total line acceleration, normalized into a dimensionless form, and $\hat{r} = r/R_\ast$ is simply the radius in units of $R_\ast$. In their models, the wind is isothermal, i.e. $a_\text{s} = \mathrm{const.}$, which is not the case here. Nonetheless, we can calculate the corresponding values for $\hat{a}_\text{lines}$ and perform a numerical fit for our WNE model in order to check how accurate the line acceleration of a cWR star can be reproduced by a description in the form of Eq.\,(\ref{eq:alinemv}). The result in shown in Fig.\,\ref{fig:aradmvfit}, where we also display the parameter set of the best fit. Looking at the lower panel, the result is quite promising at first glance, but the logarithmic scale in the upper panel reveals that the description is off by orders of magnitude in the outer wind. The reason is that Eq.\,(\ref{eq:alinemv}) permits only one local maximum and no local minima for $\hat{a}_\text{lines}$, meaning it cannot handle more than one bump. A closer inspection also shows that the increase of the acceleration just below the critical point is underestimated.

Despite the shortcomings, our fit result with the recipe from \citet{MV2008} is promising. Their recipe has been successfully applied for predictions of OB star winds \citep[e.g.][]{Muijres+2012}, where not as many different ionization stages contribute as in WR winds. A full physically meaningful analytic representation of the radiative acceleration will likely require an extension of the model from \citet{MV2008} as well as a detailed mapping of the whole WR parameter range with hydrodynamically consistent models. This is beyond the scope of the present paper, but a sufficient description of the wind acceleration of cWR and other He stars marks an important scientific goal with implications to other astrophysical fields on a likely mid- to long-term timescale.

\section{Additional tables}
  \label{asec:tables}

\begin{table}
  \caption{Parameters for the WR models with $\log \dot{M} = -5.15$ at different metallicities}
  \label{tab:wrgeddzmodels}

  \centering
  \begin{tabular}{cccccc}
      \hline 
         $\log Z/Z_\odot$ & 
         $\log L/M$ &
         $\Gamma_\text{e}(R_\ast)$  &  
         $\varv_\infty$ &
         $T_{2/3}^\text{Ross}$ &
         $T_\text{e}(R_\text{crit})$ \\[0.5mm]
      \hline
 	\multicolumn{6}{c}{\textit{WN models}} \\
     1.0 &  4.07 & 0.17 & 4746 & 122 & 214  \\
 0.7 &  4.10 & 0.19 & 3579 & 119 & 204  \\
 0.3 &  4.20 & 0.24 & 2375 & 109 & 200  \\
 0.0 &  4.27 & 0.29 & 1640 &  92 & 195  \\
 -0.3 &  4.37 & 0.36 & 1391 &  80 & 194  \\
 -0.7 &  4.45 & 0.43 & 1165 &  75 & 193  \\
 -1.0 &  4.51 & 0.50 & 1063 &  74 & 191  \\
 -1.3 &  4.56 & 0.56 & 1029 &  75 & 192  \\
 -2.0 &  4.66 & 0.70 & 1048 &  81 & 180  \\
 -3.0 &  4.72 & 0.81 & 1099 &  83 & 169  \\
 -4.0 &  4.74 & 0.85 & 1106 &  83 & 165  \\
 -5.0 &  4.74 & 0.86 & 1113 &  83 & 165  \\
 \medskip
 -6.0 &  4.75 & 0.86 & 1117 &  83 & 165  \\

 	\multicolumn{6}{c}{\textit{WC models}} \\
		     1.0 &  4.09 & 0.15 & 4351 & 123 & 212  \\
 0.7 &  4.16 & 0.17 & 3446 & 122 & 205  \\
 0.3 &  4.25 & 0.21 & 2277 & 114 & 199  \\
 0.0 &  4.34 & 0.27 & 2120 & 113 & 197  \\
 -0.3 &  4.42 & 0.32 & 1573 &  98 & 191  \\
 -1.0 &  4.51 & 0.39 & 1169 &  75 & 189  \\
 -1.3 &  4.55 & 0.43 & 1151 &  77 & 188  \\
 -2.0 &  4.65 & 0.54 & 1291 &  87 & 181  \\
 -2.7 &  4.69 & 0.59 & 1262 &  86 & 178  \\
 -3.0 &  4.70 & 0.60 & 1282 &  87 & 178  \\
 -3.3 &  4.70 & 0.61 & 1228 &  87 & 177  \\
 -3.7 &  4.71 & 0.62 & 1214 &  88 & 176  \\
 -4.0 &  4.72 & 0.64 & 1325 &  90 & 174  \\
 -4.3 &  4.71 & 0.62 & 1228 &  87 & 177  \\
 -5.0 &  4.71 & 0.63 & 1313 &  88 & 174  \\
   \hline
  \end{tabular}
\end{table}
	
  \begin{table}\centering
\caption{Overview of the number of levels and line transitions used in the radiative transfer calculations. Nitrogen was only included for the WN models and is thus marked by an asterisk. The number of lines refers to the transitions considered in the radiative transfer. For collisions, also radiatively forbidden transitions are taken into account. The atom listed as Fe here is a generic element that also includes Sc, Ti, V, Cr, Mn, Co, and Ni. For details, see \citet{GKH2002}\label{tab:datom}}
 \begin{tabular}{lccc c lccc}
 \hline &&&\\[-2.0ex]
 \multicolumn{2}{l}{Ion}    & Levels                & Lines   &  \mbox{\hspace{2mm}} &
 \multicolumn{2}{l}{Ion}    & Levels                & Lines\\
 \hline &&&&\\[-2.0ex]
  He   &      I   &   35   &   271             &   &              P    &    III   &   10   &    11     \\          
  He   &     II   &   26   &   325             &   &              P    &     IV   &   10   &     8     \\           
  He   &    III   &    1   &     0             &   &              P    &      V   &   10   &    20     \\         
  C    &     II   &    3   &     2             &   &              P    &     VI   &    1   &     0     \\         
  C    &    III   &   40   &   226             &   &              S    &    III   &   10   &     7     \\      
  C    &     IV   &   25   &   230             &   &              S    &     IV   &   25   &    54     \\      
  C    &      V   &    1   &     0             &   &              S    &      V   &   10   &    13     \\      
  N$^{\ast}$ &   I  &   3    &    2            &   &              S    &     VI   &   22   &    75     \\        
  N$^{\ast}$ &  II  &  38    &  201            &   &              S    &    VII   &    1   &     0     \\        
  N$^{\ast}$ & III  &  30    &   94            &   &              Cl   &    III   &    1   &     0     \\        
  N$^{\ast}$ &  IV  &  38    &  154            &   &              Cl   &     IV   &   24   &    34     \\          
  N$^{\ast}$ &   V  &  20    &  114            &   &              Cl   &      V   &   18   &    29     \\          
  N$^{\ast}$ &  VI  &   1    &    0            &   &              Cl   &     VI   &   23   &    46     \\          
  O    &     II   &   36   &   146             &   &              Cl   &    VII   &    1   &     0     \\        
  O    &    III   &   33   &   121             &   &              Ar   &     II   &   10   &     9     \\        
  O    &     IV   &   29   &    76             &   &              Ar   &    III   &   30   &    63     \\       
  O    &      V   &   36   &   153             &   &              Ar   &     IV   &   13   &    20     \\      
  O    &     VI   &   16   &   101             &   &              Ar   &      V   &   10   &    11     \\   
  O    &    VII   &    1   &     0             &   &              Ar   &     VI   &    9   &    11     \\   
  Ne   &     II   &   10   &     9             &   &              Ar   &    VII   &   20   &    34     \\   
  Ne   &    III   &   18   &    18             &   &              Ar   &   VIII   &   11   &    24     \\   
  Ne   &     IV   &   35   &   159             &   &              Ar   &     IX   &   10   &    10     \\   
  Ne   &      V   &   54   &   263             &   &              Ar   &      X   &    3   &     1     \\   
  Ne   &     VI   &   49   &   239             &   &              K    &    III   &   10   &    12     \\
  Ne   &    VII   &   36   &   229             &   &              K    &     IV   &   23   &    27     \\
  Ne   &   VIII   &   77   &   517             &   &              K    &      V   &   19   &    33     \\
  Ne   &     IX   &   23   &    51             &   &              K    &     VI   &   28   &    38     \\
  Ne   &      X   &   15   &    25             &   &              K    &    VII   &    8   &     6     \\
  Ne   &     XI   &    1   &     0             &   &              K    &   VIII   &    1   &     0     \\
  Na   &    III   &    1   &     0             &   &              Ca   &    III   &   10   &     7     \\ 
  Na   &     IV   &   10   &     9             &   &              Ca   &     IV   &   24   &    43     \\ 
  Na   &      V   &   10   &    14             &   &              Ca   &      V   &   15   &    12     \\ 
  Na   &     VI   &   10   &    13             &   &              Ca   &     VI   &   15   &    17     \\  
  Na   &    VII   &    1   &     0             &   &              Ca   &    VII   &   10   &     9     \\  
  Mg   &    III   &   10   &    12             &   &              Ca   &   VIII   &    1   &     0     \\
  Mg   &     IV   &   10   &     9             &   &              Fe   &    III   &    1   &     0     \\
  Mg   &      V   &   10   &     9             &   &              Fe   &     IV   &   18   &    77     \\
  Mg   &     VI   &   10   &    19             &   &              Fe   &      V   &   22   &   107     \\
  Mg   &    VII   &    1   &     0             &   &              Fe   &     VI   &   29   &   194     \\
  Al   &    III   &   10   &    18             &   &              Fe   &    VII   &   19   &    87     \\
  Al   &     IV   &   10   &    10             &   &              Fe   &   VIII   &   14   &    49     \\
  Al   &      V   &   10   &     9             &   &              Fe   &     IX   &   15   &    56     \\
  Al   &     VI   &   10   &     9             &   &              Fe   &      X   &   28   &   170     \\
  Al   &    VII   &    1   &     0             &   &              Fe   &     XI   &   26   &   161     \\
  Si   &     II   &   10   &    13             &   &              Fe   &    XII   &   13   &    37     \\
  Si   &    III   &   10   &     9             &   &              Fe   &   XIII   &   15   &    50     \\
  Si   &     IV   &   10   &    22             &   &              Fe   &    XIV   &   14   &    49     \\
  Si   &      V   &   10   &     9             &   &              Fe   &     XV   &   10   &    25     \\
	Si   &     VI   &   10   &     9             &   &              Fe   &    XVI   &    9   &    20     \\
  Si   &    VII   &    1   &     0             &   &              Fe   &   XVII   &    1   &     0     \\
  Si   &   VIII   &    1   &     0             &   &                   &          &        &           \\
\hline
 \end{tabular}
\end{table}

\bsp	
\label{lastpage}
\end{document}